\documentclass[aps,prb,twocolumn,floatfix,showpacs,superscriptaddress]{revtex4-1}
\usepackage{graphicx}
\usepackage{amsmath}
\usepackage{bm}
\usepackage{verbatim}
\usepackage{color}
\usepackage{ulem}
\usepackage{subfig}
\usepackage{hvfloat, subfig}
\usepackage{graphicx}
\usepackage{color}
\usepackage{xcolor}
\usepackage{float}
\usepackage{graphicx}
\usepackage{silence}
\usepackage[utf8]{inputenc}
\usepackage{epsf}
\usepackage{float}
\usepackage{subfig} 
\usepackage[font=small,skip=0pt]{caption}
\usepackage[export]{adjustbox}

\def\be{\begin{equation}}
\def\ee{\end{equation}}

\def\rv{{\bf r}}
\def\vv{{\bf v}}
\def\Ev{{\bf E}}
\def\Rv{{\bf R}}
\def\Av{{\bf A}}
\def\sv{{\bf s}}
\def\Sv{{\bf S}}
\def\Jv{{\bf J}}

\def\sigmav{{\bm \sigma}}

\begin{document}
 
\title{The Anomalous Hall Effect in Magnetic Topological Insulators}

\author{Amir Sabzalipour}
\affiliation{University of Antwerp, Department of Physics, Groenenborgerlaan 171, 2020 Antwerp, Belgium}
\affiliation{School of Physics, Institute for Research in Fundamental Sciences (IPM), Tehran 19395-5531, Iran}
\author{Bart Partoens}
\affiliation{University of Antwerp, Department of Physics, Groenenborgerlaan 171, 2020 Antwerp, Belgium}

\begin{abstract}
The anomalous Hall effect (AHE) is studied on the surface of a 3D magnetic topological insulator. By applying a modified semi-classical framework, all three contributions to the AHE, the intrinsic Berry phase curvature effect, the side-jump effect and the skew scattering effects are systematically treated, and analytical expressions for the conductivities are obtained in terms of the Fermi level, the spatial orientation of the surface magnetization and the concentration of magnetic and non-magnetic impurities. 
 We demonstrate that the AHE can change sign by altering the orientation of the surface magnetization, the concentration of the impurities and also the position of the chemical potential, in agreement with recent experimental observations. Hence, each contribution to the AHE, or even the whole AHE, can be turned off by properly adjusting the given parameters. For example, one can turn off the anomalous hall conductivity in a system with in-plane magnetization by pushing the system into the fully metallic regime.

\end{abstract}
\maketitle

%%%%%%%%%%%%%%%%%%%%%%%%%%%%%%%%%%%%%%%%%%%%%%%%%
\section{Introduction}\label{sec:intro}
%%%%%%%%%%%%%%%%%%%%%%%%%%%%%%%%%%%%%%%%%%%%%%%%%
Topological insulators are a new class of matter that resemble band insulators in the bulk while capable of conducting along gapless states on the surfaces or edges ~\cite{kane_1,zhang_1,kane_2,bernevig_1,Hassan_m}. 
Topological properties of the band structure in these materials protect the metallic surface or edge states, as long as time reversal or crystalline symmetry is present~\cite{fu_tci}. Surface states of a 3D topological insulator can be described by an effective 2D massless Dirac Hamiltonian, within a certain energy range~\cite{TI_book}. Spin-momentum locking of these massless Dirac fermions prohibits backscattering of the itinerant electrons off non-magnetic impurities and consequently results in anti-weak localization~\cite{roushan_nat_2009, weak_1}. All these exotic features of topological insulators have attracted a lot of interests theoretically and experimentally~\cite{moore_1, Qi_1, hish_1}. 
 Revealing these topological features in surface transport is an important direction of research, and the dependency of the surface charge transport on the type of disorder and the range of disorder-electron interaction has been extensively studied theoretically~\cite{culcer_prb_2010,culcer_review}.

The anomalous Hall effect (AHE) as one of the most fundamental transport properties of magnetic materials, is the manifestation of the Hall effect in systems without time-reversal symmetry. This effect has been an enigmatic problem for almost a century and still remains a poorly understood phenomenon. A magnetic topological insulator with strong spin orbit coupling is a valuable host medium for realizing both the quantized version of the anomalous Hall effect~\cite{rui, scalQAHE, beyond2d}, and the unquantized version~\cite{signchangeex, sensitivity}.   The collective behavior of the randomly distributed point-like magnetic impurities on the surface and in the bulk of a topological insulator can break time reversal symmetry and drive the system into a gapped system. This introduced gap in spin space influences, via the spin-orbit coupling, the charge dependent properties of the massive Dirac fermions, like the anomalous charge conductivity. 

While this time reversal symmetry breaking makes the system topologically trivial, the chiral nature of the surface states could still play an important role in the transport properties of the system. In the current effort, the magnetic and non-magnetic impurities on the surface of the TI are the source of scattering for the itinerant massive Dirac Fermions on the surface. In fact, we ignore the likely induced small gap by the magnetic scattering on the surface, compared to the already existing gap induced by the magnetic impurities in the bulk or by any other mechanism. In this work, we comprehensively investigate the three different contributions to the AHE  arising from the intrinsic berry-phase curvature, the extrinsic side-jump and skew scattering.

%TI's are often materials with a large spin-orbit coupling (SOC). The SOC also lies at the origin of the anomalous Hall effect (AHE)~\cite{karplus_1}, the focus of this work, but which occurs in materials with broken time-reversal symmetry%.

% Interplay of the exchange field and the spin-orbit coupling via simultaneously breaking time reversal symmetry and the chiral symmetry leads to emergence of AHE in ferromagnet states. By going beyond the first ordered quantum Boltzmann equation, Kohn and Luttinger built a comprehensive transport theory for this effect~\cite{Lutinger_1,Lutinger_2}. Building further on their work, Smit and Berger identified various contributions to this AHE: the intrinsic Berry phase contribution, and the extrinsic skew scattering and side jump effects originating from a simple Bloch band structure and a spin-orbit interaction effect due to the external scattering potential~\cite{smit_1,smit_2,Berger1,Berger2,Berger3}. 

%The strength of the SOC for the states at the surface of a magnetically doped TI make it a valuable host medium for realizing the AHE~\cite{rui, zhangscience}.%
Exerting an external electric field determines the momentum direction of the charge carriers which is locked to their spin. Therefore, novel phenomena can be expected in the spin dependent interaction of itinerant electrons with these surface magnetic impurities. %The SOC induces an anomalous deflection in the dynamics of surface itinerant electrons during scattering off disorder. %This deflection or side jump has a noticeable effect after averaging over all scattering events.  Simultaneously, the magnetic impurity influences the spin texture of massive Dirac fermions via transferring a torque. 
Furthermore, by altering the orientation of TI's surface magnetization by an applied field, the strength of the scattering potential can be changed, and by changing the Fermi level, the spin orientation can be altered. As a consequence, the relative importance of the different contributions to the AHE will also depend on the surface magnetization direction and the Fermi level. Therefore we also address in this work how the orientation of the surface magnetization, the material's Fermi level and also the impurity concentration control the side jump, skew scattering and anomalous velocity contributions to the AHE.
 
In this study we rely on the semi-classical Boltzmann formalism. The semiclassical description of transport through a Boltzmann equation does lead to a Hall contribution if skew scattering is accounted for in the collision term, but the other contributions, the anomalous velocity and the side-jump effects, are ignored. Therefore we included the anomalous velocity to the formalism. Furthermore, in Ref.~\onlinecite{firstcoordinate} it was also shown how the ad hoc addition of a gauge invariant expression for the coordinate shift leads to a semi-classical description of all contributions to the anomalous Hall
effect and that the same final results can be obtained as in a more complicated approach using the microscopic perturbative approach
of Luttinger\cite{lutinger1,lutinger2, lutinger3}. In this work we make use of this physically transparent semi-classical description~\cite{topical}. 
 
We have organized the rest of this paper as follows. In Sec.~\ref{sec:model}, we introduce the effective model of massive Dirac fermions on the surface of a magnetically doped three-dimensional topological insulator. In addition, we present the semi-classical approach to correctly incorporate the side jump and skew scattering contributions in the AHE dynamics of the charge carriers. The obtained results are shown in Sec.~\ref{sec:results}. In Sec.~\ref{sec:summary} we summarize our findings and conclude with our main results. Finally, some important relations are derived in the appendix to ease tracing some results presented in the previous sections.
 
\section{Model and Approach}\label{sec:model}

\subsection{Model Hamiltonian}
The \textit{minimal} effective Hamiltonian describing massive Dirac fermions on the surface of a 3D TI is given by
\begin{equation} \label{eq:1}
H _{D}=\hbar v_{F} \left(\textbf{\textit{k}} \times \sigma\right)_z+ M \sigma_{z},
\end{equation}
 where the $\hat{z}$-direction is chosen normal to the surface of the TI. Here, $v_{\rm F}$, $\bm k=(\textit{k}_x, \textit{k}_y)$, and $M$ are respectively the Fermi velocity, the wave vector, and the mass of the surface Dirac electrons, and $\sigma=(\sigma_x,\sigma_y,\sigma_z)$ is the vector of Pauli matrices acting on the spin of the electrons. The eigenvalues and eigenvectors of $H _{D}$ are
\begin{equation}\label{eq:eigenvector}
\psi_{\textbf{\textit{k}},\alpha}(r)=\frac{e^{i \textbf{\textit{k}} \cdot \textbf{r}} } {\sqrt{A(1+\xi_k^{2\alpha})}}
	\left(
	\begin{array}{c}
	e^{-i \phi_{\textbf{\textit{k}}}/2} \\
	i \alpha \xi_k^{\alpha} e^{i \phi_{\textbf{\textit{k}}}/2}
	\end{array}
	\right)~,
\end{equation}

\begin{equation}\label{eq:eigenvalue}
\varepsilon_{k,\alpha}=\alpha\varepsilon_k= \alpha \sqrt{(\hbar v_{\rm F} k)^2+M^2}~,
\end{equation}
where $\alpha$ labels the conduction $(\alpha=+1)$ and valence $(\alpha=-1)$ bands, $k=|\textbf{\textit{k}}|$, $\xi_{\textit{k}}=\sqrt{(1-\gamma_{\textit{k}})/(1+\gamma_{\textit{k}})}$, with $\gamma_{\textit{k}}=M/\varepsilon_{\textit{k}}$, and $\phi_{\textbf{\textit{k}}}=\arctan({\frac{\textit{k}_y}{\textit{k}_x}})$ refers to the direction of the wave vector of the surface electrons. In the following we will also label the eigenstates and energies with the index $l\equiv(\textbf{\textit{k}},\alpha)$ as the combined (momentum, band) index. 

The presence of dilute and randomly placed magnetic impurities on the surface of a 3D TI, scatter electrons and influence the transport properties of the system. 
We model the interaction between an electron located at $\rv$ and a single magnetic impurity at $\Rv_{m}$ as 
\begin{equation}\label {eq:vsc_single}
V^{m}(\textbf{r}-\textbf{R}_{m})=J \delta (\rv-\textbf{R}_{m})~\textbf{S}_{m}\cdot \sv,
\end{equation}
where $\Sv_{m}$ and $\sv=\hbar \sigmav/2$ are the spins of the impurity and the electron, respectively. $J$ is the exchange coupling and the Dirac delta function refers to the short-range nature of the electron-impurity interaction we have considered in this study. 
In the regime of large magnetic spin $|S|\rightarrow \infty $, weak interaction $J\rightarrow 0$ and $J |S|$=constant, we can treat the spin of the magnetic impurities classically. We assume that the magnetic impurities are all aligned in the same direction and lie in the $yz$-plane, with the $z$-axis perpendicular to the surface of the magnetic TI. Since the system is usually not pure and often also contains non-magnetic impurities, we also consider, in addition to the term $V^{m}$, 
\begin{equation}
V^{nm}(\textbf{r}-\textbf{R}_{nm})=V^{nm}_{0}\delta(\textbf{r}-\textbf{R}_{nm})
\end{equation}
 as another source for scattering of itinerant electrons off non-magnetic impurities located at $\Rv_{nm}$. We relied on the modified Boltzmann formalism~\cite{Jairo_method} separately for these two kinds of impurities to obtain analytical results for the different contributions to the AHE in a magnetic TI.

\subsection{Semi-classical approach}
Within the semi-classical Boltzmann approach, the rate of change of the surface electrons' distribution function $f$, due to the scattering from impurities, can be obtained from
\begin{equation}\label{eq:boltzmann}
\left(\frac{\partial f(\varepsilon_{l})}{\partial t}\right)_{sc}=
|e| \mathbf{E}\cdot\mathbf{v}_{0l}
\left(-\frac{\partial f^0(\varepsilon_{l})}{\partial \varepsilon_{l}} \right)~,
\end{equation}
where $\mathbf{v}_{0l}=\dfrac{\partial \varepsilon_{l}}{\partial \bm{k}}$ is the velocity of the incident wave packet, ${\bf E}$ is the external applied electric field, and $f^0(\varepsilon_{l})$ is the Fermi-Dirac distribution function. Considering only elastic scattering events, and using the detailed balance principle, we obtain 
\begin{equation}\label{eq:balance}
\left(\frac{\partial f(\varepsilon_{l})}{\partial t}\right)_{ sc}
=\sum_{l'}   w_{ll'}\left(f_{l'}- f_{l}\right)~,
\end{equation}
where $w_{l l'}$ is the transition rate between states $l$ and $l'$. Combining Eq.~(\ref{eq:boltzmann}) with Eq.~(\ref{eq:balance}) gives us
\begin{equation}\label{eq:boltzmann2}
-|e|\mathbf{E}\cdot\mathbf{v}_{0l} \left(-\frac{\partial f^0}{\partial \varepsilon_{l}} \right)=
\sum_{l'}   w_{ll'}\left(f_{l}- f_{l'}\right)~.
\end{equation}
The semi-classical Eq.~(\ref{eq:boltzmann2}) deals only with gauge invariant quantities, such as the scattering rate, band velocity and the distribution function. Nevertheless, since in this equation the only role of the electric field is to accelerate wave packets, and the only role of impurities is to produce incoherent instantaneous  events, it is clear
that this approach must often be insufficient. In studying the AHE, more than ever, we need to modify the semi-classical framework to incorporate all the relevant phenomena correctly. This can be done by staying within the semi-classical framework. We now discuss separately the corrections that are added to the velocity of the electrons, the transition rates and also the distribution function of the electrons in the semi-classical approach, in order to correctly include all phenomena -skew scattering as well as the side jump and anomalous velocity effects- during the scattering time of electrons off impurities under the presence of an external electric field. 

\subsubsection{Transition rate}
 To study the transport of electrons in a quantum regime, we need to find the scattering matrix (or $T$-matrix) of the electrons. Within the semi-classical framework, the  scattering rate, as a classical object, can be obtained by its connection to the scattering matrix through Fermi's golden rule. However, it should be noted that only the absolute value of the $T$-matrix elements are present in the scattering rate. Consequently, all the phase information of the $T$-matrix elements is lost. In this section we forget about this insufficiency of the golden rule, but in the following sections we will discuss how we can restore all the missing phase information.     
The scattering rate between two different quantum states is connected to the $T$-matrix elements and is given by
\begin{equation}\label{eq:rate}
w_{ll'}=\frac{2 \pi}{\hbar}\rvert T_{ll'}\rvert^{2}\delta(\varepsilon_{l'}-\varepsilon_{l}),
\end{equation}
in which the scattering $T$-matrix is defined as
\begin{equation}\label{eq:tmatrix}
T_{ll'}=\langle \l|\ V_{sc}|\psi_{l'}\rangle,
\end{equation}
where $| l\rangle $ is an eigenstate of the Hamiltonian $H_{D}$, $V_{sc}$ is the scattering potential operator and $|\psi_{l'}\rangle$ is an eigenstate of the full Hamiltonian $H=H_{D}+V_{sc}$, that satisfies the Lippmann-Schwinger equation
\begin{equation}\label{eq:limann}
| \psi_{l'}\rangle=|l'\rangle+\frac{V_{sc}}{\varepsilon_{l'}-H_{0}+i \eta}| \psi_{l'}\rangle.
\end{equation}
For weak disorder, $|\psi_{l'}\rangle$ can be approximated by a truncated series in powers of $V_{ll'}=\langle l |V_{sc}|l'\rangle$. By applying Eq.~(\ref{eq:limann}) and Eq.~(\ref{eq:tmatrix}), $T_{ll'}$ up to third order in $V_{sc}$ is given by 
\begin{equation}\label{eq:T-series}
\begin{split}
T_{l l'}= &V_{ll'}+\sum_{l''}\frac{V_{ll''} V_{l''l'}}{\varepsilon_{l}-\varepsilon_{l''}+i\eta}+\\
&\sum_{l'''} \sum_{l''}\frac{V_{ll''}V_{l''l'''}V_{l'''l'}}{(\varepsilon_{l}-\varepsilon_{l''}+i\eta)(\varepsilon_{l}-\varepsilon_{l'''}+i\eta)}.
\end{split}
\end{equation}
Substituting this expansion for $T_{ll'}$ into Eq.~(\ref{eq:rate}) leads to the following scattering rate up to fourth order in the scattering potential
\begin{equation}\label{eq:allw}
w_{ll'}=w^{(2)}_{ll'}+w^{(3)}_{ll'}+w^{(4)}_{ll'}~,
\end{equation}
where $w^{(2)}_{ll'}$ is symmetric under changing $l$ $\longleftrightarrow$ $l'$, and is given by 
\begin{equation}\label{eq:w}
w^{(2)}_{ll'}=\frac{2\pi}{\hbar}\langle | V_{ll'}|^{2}\rangle_{dis}\, \delta(\varepsilon_{l}-\varepsilon_{l'})~,
\end{equation}
where $dis$ denotes averaging over all possible distributions of impurities in our system.
For dilute and randomly placed impurities, it has been shown that $\langle | V_{ll'}|^{2}\rangle_{dis}\sim n_{im} V^{2}_{0}$, with $n_{im}$ the impurity concentration\cite{lutinger1}.  
 As already indicated, one of the main contributions to the AHE originates from Skew scattering. In order to investigate this contribution, we need to calculate the asymmetric part of the transition rate, $w^{(a)}_{ll'}=\dfrac{w_{ll'}-w_{l'l}}{2}$. Since $w^{(2)}_{ll'}$ is symmetric, the first asymmetric term in the transition rate $w_{ll'}$ appears at the order of $V^{3}_{0}$. Now $w^{(3)}_{ll'}$ is given by
\begin{equation}\label {eq:10}
w^{(3)}_{ll'}=\frac{2\pi}{\hbar}  \left(\sum_{l''}\frac{\left\langle V_{ll'} V_{l'l''} V_{l'' l}\right\rangle_{dis}}{\varepsilon_{l}-\varepsilon_{l''}-i\eta}+ c.c.\right) \delta(\varepsilon_{l}-\varepsilon_{l'})~. 
\end{equation}
$w^{(3)}_{ll'}$ itself can be written as a sum of a symmetric term $w^{(3s)}$ and an asymmetric term $w^{(3a)}$. Then, $w^{(3)}_{ll'}=w^{(3a)}_{ll'}+w^{(3s)}_{ll'}$, where $w^{(3s/a)}_{ll'}=\dfrac{w^{(3)}_{ll'}\pm w^{(3)}_{l'l}}{2}$. Since the symmetric part of  $w^{(3)}_{ll'}$
\begin{equation} \label{eq:w3s}
w^{(3s)}_{ll'}=\frac{4\pi}{\hbar} P\left(\sum_{l''}\dfrac{Re[\langle V_{ll'} V_{l'l''}V_{l''l}\rangle_{dis}]}{\varepsilon_{l}-\varepsilon_{l''}}\right),
\end{equation}
 just renormalizes $w^{(2)}_{ll'}$, it does not introduce a new physical contribution to the scattering and is further not considered. $P$ in the above equation refers to the \textit{principal value}. The remaining asymmetric term $w^{(3a)}_{ll'}$ can be expressed as 
 
 \begin{equation}\label{eq:w3a}
\begin{split}
w^{(3a)}_{ll'}&=\frac{-(2\pi)^{2}}{\hbar}\delta(\varepsilon_{l}-\varepsilon_{l'}) \\ &\times \sum_{l''} \langle
{\rm Im}  \left( V_{ll'}V_{l'l''}V_{l''l}\right)\rangle_{dis}
\delta(\varepsilon_{l}-\varepsilon_{l''} )~,
\end{split}\end{equation}
This contribution scales (for a so-called non-Gaussian disorder model~\cite{lutinger1}) as $\left( V_{ll'}V_{l'l''}V_{l''l}\right)_{dis}\sim n_{im} V^{3}_{0}$ with the impurity concentration. Consequently we can expect that the transverse conductivity associated to this term will be inversely proportional to $n_{im}$ \cite{lutinger1}. This contribution to the conductivity of the system  dominates in dilute systems.

Two different scattering processes contribute to the fourth order expression for the scattering rate. A fourth order scattering process can occur at a single defect, but also two second order scattering processes can occur at two different defects. As the sequence of scatterings that lead to these two second order pair scattering events is arbitrary, this process leads to three contributions in the expression for $w^{(4)}_{ll'}$:\cite{fourth}
\begin{equation}\label{w4}
\begin{split}
w^{(4)}_{ll'}=(&\sum_{l''}\sum_{l'''}\left[\dfrac{\langle V_{l'''l} V_{ll''}\rangle_{dis}}{\varepsilon_{l}-\varepsilon_{l''}+i\eta}\dfrac{\langle V_{l''l'} V_{l'l'''}\rangle_{dis}}{\varepsilon_{l}-\varepsilon_{l'''}-i\eta}+c.c\right]\\
&+\sum_{l''}\sum_{l'''}\left[\dfrac{\langle V_{l''l} V_{ll'}\rangle_{dis}}{\varepsilon_{l}-\varepsilon_{l''}-i\eta}\dfrac{\langle V_{l'l'''} V_{l'''l''}\rangle_{dis}}{\varepsilon_{l}-\varepsilon_{l'''}-i\eta}+c.c\right]\\
&+\sum_{l''}\sum_{l'''}\left[\dfrac{\langle V_{ll'} V_{l'l'''}\rangle_{dis}}{\varepsilon_{l}-\varepsilon_{l''}-i\eta}\dfrac{\langle V_{l'''l''} V_{l''l}\rangle_{dis}}{\varepsilon_{l}-\varepsilon_{l'''}-i\eta}+c.c\right] ) \\
\times &~~\delta(\varepsilon_{k}-\varepsilon_{k'}),
\end{split}
\end{equation}
The factors like $\langle V_{l'''l} V_{ll''}\rangle_{dis}$ are all proportional to $n_{im}$, and therefore these contributions to $w^{(4)}_{ll'}$ are proportional to $ n^{2}_{im}$. The fourth order contribution due to a scattering event at a single impurity contains factors like $\langle V_{ll''} V_{l''l'''} V_{l'''l'} V_{l'l}\rangle_{dis}$ which are proportional to $n_{im}$. This contribution is therefore physically similar to $w^{(3a)}_{ll'}$ (with respect to the concentration of impurities) but much smaller. Therefore we only consider the contribution of the two second order pair scattering events in $w^{(4)}_{ll'}$. This contribution due to the fourth-order $V_{0}$ in skew scattering leads to an intrinsic, disorder-independent term in the conductivity of the system, as will be shown later.

In this work, we are interested in the zero temperature regime. Furthermore, in the weak disorder limit that we consider, the energy width of the Bloch state spectral peaks is smaller than the gap, allowing us to ignore
direct interband scattering. Therefore we will only consider intraband transitions in calculating $w^{(2)}_{ll'}$ and $w^{(3)}_{ll'}$. We consider electron transport in electron doped systems, thus the chemical potential $\mu$ lies inside the conduction band, thus $\alpha=+1$ for all states in Eqs.~(\ref{eq:w}) and~(\ref{eq:w3a}). 
 However, for $w^{(4)}_{ll'}$ we also incorporate the off-diagonal scattering matrix elements as they produce virtual transitions that mix states in the two bands in a way which is ultimately crucial~\cite{mac}. Thus for the calculation of $w^{(4)}_{ll'}$, also interband transitions with $\alpha,\alpha'=\pm 1$ are taken into account.

\subsubsection{Electron velocity}

To obtain the current density of the system $\textbf{J}=\sum_{l} \textbf{\textbf{v}}_{l} f_{l}$, we need to calculate the velocity $\textbf{v}_l$ of the itinerant electrons and also their distribution function $f_{l}$ in the presence of an external electric field and randomly placed dilute magnetic and non-magnetic impurities. The conventional semi-classical approach just studies electrons at scattering events and ignores the evolution of the  
wave packets during the scattering time interval where a side jump can occur. Furthermore, in a system with broken either time reversal or inversion symmetries, an additional term should be added to the velocity expression of electrons to incorporate properly the effect of the non-zero Berry curvature in the electron dynamics. If we incorporate both extra effects, which are missing in the conventional semi-classical approach, the velocity can be written as
 \be\label{eq:velo}
\vv_{l}=\vv_{0l}+\vv^{an}_{l}+\vv^{sj}_{l}~,
\ee
in which
 $\vv^{an}_{l}=- \dot{\textbf{\textit{k}}}\times(\nabla_{\textbf{\textit{k}}}\times \Av_{l})$ is the anomalous velocity, with $\Av_{l}=i \langle u_{l}| \nabla_{\textbf{\textit{k}}}|u_{l} \rangle$ the Berry connection where $u_{l}(\textbf{r})= e^{-i \textbf{\textit{k}}\cdot\textbf{r}}\psi_{ l }(\bm r)$, and $\vv^{sj}_{l}=\sum _{l'} \delta \rv_{ll'} w_{ll'}$ is the side jump velocity.
Here, $\delta \rv_{ll'}$ denotes the anomalous deflection which electrons experience during scattering time. The gauge invariant expression of this anomalous displacement is given by \cite{firstcoordinate}
\begin{equation}\label {eq:deltar}
F= \Av_{l'}-\Av_{l} -\left(\nabla_{\textbf{\textit{k}}}+\nabla_{\textbf{\textit{k}}}\right)  \arg(V_{ll'})~,
\end{equation}
where $\arg(V_{ll'})$ is the argument of $V_{ll'}$.
While the phase information of the scattering amplitude is absent in the first-order Born approximation, the third term on the right hand side of Eq.~(\ref{eq:deltar}) is responsible for restoring this information to the dynamics of the charge carriers.

\subsubsection{Distribution function} \label{distribution}

After obtaining all terms for the velocity expression of the electrons, the next step is to calculate the distribution function of the electrons. Therefore we write the electron distribution function as follows    
\begin{equation}\label{eq:all distribution}
f_{l} = f^{0}+ g^{s}_{l}+g^{a1}_{1}+g^{a2}_{l}+g^{ad}_{l}~.
\end{equation}
The largest deviation from the Fermi-Dirac distribution is given by $g^{s}_{l}$. It arises from the symmetric part of the scattering rate $w^{(2)}_{ll'}$ and also describes the longitudinal conductivity. $g^{a1}_{l}$ is defined as the deviation due to the asymmetric part of the scattering rate $w^{(3a)}_{ll'}$, and $g^{a2}_{l}$ due to $w^{(4)}_{ll'}$. Finally, $g^{ad}_{l}$ is responsible for capturing the effect of the side jump which changes the energy of the scattered electrons and consequently their distribution function.
Substituting the transition rate $w_{ll'}$ expressed in Eq.~(\ref{eq:allw}) along with the above non-equilibrium distribution function into Eq.~(\ref{eq:boltzmann2}), we obtain the following self consistent  time-independent integral equations \begin{align}
&- e \mathbf{E} \cdot \vv_{0l} \left(-\frac{\partial f^{0}}{\partial \varepsilon_l}\right) 
=\sum_{l'} w^{(2)}_{ll'} \left(g^{s}_{l} - g^{s}_{l'} \right)~, \label{eq:longconduc}\\
& \sum_{l'} w^{(3a)}_{ll'}\left(g^{s}_{l}-g^{s}_{l'}\right)+\sum_{l'} w_{ll'}\left(g^{a1}_{l}-g^{a1}_{l'}\right)=0~  \label{eq:skew1conduc},\\
& e \mathbf{E} \cdot \vv^{sj}_l \left(-\frac{\partial f^{0}}{\partial \varepsilon_l}\right) 
=\sum_{l'}w_{ll'}\left(g^{ ad}_{l}-g^{ ad}_{l'}\right)~,\label{eq:sideconduc}\\
&\sum_{l'} w^{(4)}_{ll'}(g^{s}_{l}-g^{s}_{l'})+\sum_{l'}w_{ll'}(g^{a2}_{l}-g^{a2}_{l'})=0. \label{eq:skew2conduc}
\end{align}
In the presence of an external electric field ${\mathbf E}$, electrons acquire an extra potential energy $\Delta U_{ll'}= -e \mathbf{E}\cdot \delta \mathbf{r}_{ll'}$ during the side jump $\delta \mathbf{r}_{ll'}$. Since the energy of the electrons is conserved during elastic scattering, this change in potential energy during a side jump event should be compensated by a change in the kinetic energy of the electrons given by $\Delta \varepsilon_{ll'}=\varepsilon_{l'}-\varepsilon_{l}= e \mathbf{E}\cdot \delta \mathbf{r}_{ll'}$. Therefore, based on conservation of energy, one obtains $\sum_{l'} w_{ll'} [(f^{0}(\varepsilon_{l})- f^{0}(\varepsilon_{l'})]=- e \mathbf{E} \cdot \vv^{sj}_{l} \left(-\frac{\partial f^{0}}{\partial \varepsilon_{l}}\right)$, which reduces to Eq.~(\ref{eq:sideconduc}) based on $\sum_{l'} w_{ll'} [(f^{0}(\varepsilon_{l})+g^{ad}_{l})-(f^{0}(\varepsilon_{l'})+g^{ad}_{l'})] =0$. As $g^{a1}_{l}, g^{a2}_{l}$ and $g^{ad}_{l}$ are small compared to $g^{s}_l$, we approximate $w_{ll'}$ in Eqs.~(\ref{eq:skew1conduc}), (\ref{eq:sideconduc}) and (\ref{eq:skew2conduc}) by $w^{(2)}_{ll'}$.

It is now interesting to deduce how each contribution to the distribution function scales with the impurity concentration. Since $w^{(2)}_{ll'} \sim n_{im}$, we find, based on equation Eq.~(\ref{eq:longconduc}), that $g^{s}_{l}\sim n^{-1}_{im}$. Like $w^{(2)}_{ll'}$, $w^{(3a)}_{ll'} \sim n_{im}$, therefore referring to Eq.~(\ref{eq:skew1conduc}) shows that $g^{a1}_{l}\sim n^{-1}_{im}$. As it is clear that $\textbf{v}^{sj}_{l} \sim w^{(2)}_{ll'} \sim n_{im}$, then based on Eq.~(\ref{eq:sideconduc}) one can conclude that $g^{ad}_{l}\sim n^{0}_{im}$. Finally, let us consider $g^{a2}_{l}$. Since $w^{(4)}_{ll'}\sim n^{2}_{im}$ and $g^{s}_{l}\sim n^{-1}_{im}$ we come to the conclusion that $g^{a2}_{l}\sim n^{0}_{im}$. 

\subsubsection{Current density} The next step is to calculate the relevant terms in the current density. Using Eqs.~(\ref{eq:velo}) and~(\ref{eq:all distribution}), the charge current density is given by
\begin{equation}\label {eq:current}
\begin{split}
\Jv &=-e\sum_{l} f_{l}\vv_{l} \\
&\simeq \Jv^{in}+\Jv^{s}+\Jv^{ ad}+\Jv^{sj}+\Jv^{sk1}+ \Jv^{sk2}~,
\end{split}
\end{equation}   
where $\Jv^{in}=-e\sum_{l} f^{0} (\varepsilon_l) \vv^{an}_{l}$ is the intrinsic current density,
$\Jv^{s}=-e\sum_{l} g^{s}_{l} \vv_{0l}$ is the regular contribution to the charge current, arising from impurity scattering events within the first-order Born approximation,
$\Jv^{ad}=-e \sum_{l} g^{ ad}_{l} \vv_{0l}$ and $\Jv^{sj}=-e \sum_{l} g^{s}_{l} \vv^{sj}_{l}$ are consequences of the side jump effect on the distribution function and the electron velocity, respectively. $\Jv^{sk1}=-e\sum_{l}g^{a1}_{l}  \vv_{0l}$ and $\Jv^{sk2}=-e\sum_{l}g^{a2}_{l}  \vv_{0l}$ result from skew scattering. In the second line of Eq.~(\ref{eq:current}), among the 15 terms we just consider 6 terms non-negligible. It is obvious that $\sum_{l} f^{0}_{l}(\vv_{0l}+\vv^{sj}_{l})=0$ for the equilibrium distribution function. In addition, we have ignored the small contributions $\sum_{l} (g^{a1}_{l}+g^{a2}_{2}+g^{ad}_{l})\vv^{sj}_{l}$. Moreover, as $\vv_{an}$ is already linear in the electric field, the non-linear contributions to the current $\sum_{l} (g^s_{l}+g^{a1}_{l}+g^{a2}_{l}+g^{ad}_{l})\vv^{an}_{l}$ are also omitted.

\subsection{Generalized relaxation time approximation}
In order to solve the integral equations~(\ref{eq:longconduc}), (\ref{eq:skew1conduc}), (\ref{eq:sideconduc}) and~(\ref{eq:skew2conduc}), we rely on the generalized relaxation-time approach introduced first in Ref.~\onlinecite{Jairo_method}. When both the energy spectrum and the scattering potential are isotropic, the transition probability $w_{ll'}$ will depend only on the angle between $\textbf{\textit{k}}$ and $\textbf{\textit{k}}'$, and one can employ the standard relaxation time approach~\cite{mahan_nutshell}. This is indeed the case when the spins of the magnetic impurities in Eq.~(\ref{eq:vsc_single}) are aligned perpendicular to the surface, \textit{i.e.}, $\Sv_m=S_{m} {\hat z}$.
On the other hand, for an arbitrary orientation of the spins of the aligned magnetic impurities, the scattering of helical electrons becomes anisotropic and the transition probability depends on the directions of both $\textbf{\textit{k}}$ and $\textbf{\textit{k}}'$. Consequently, the relaxation time is strongly anisotropic and depends on the magnitude and direction of $\textbf{\textit{k}}$, and on the orientation of the magnetic impurities. The generalized relaxation time approximation captures the effects of this anisotropy in the conductivity ~\cite{Jairo_method}. In this approach, the different contributions to the non-equilibrium distribution function are written as
\begin{equation}\label {eq:generalg}
g^{p}_{l}=e E \left[\lambda^{p}_{1l}\cos\chi+\lambda^{p}_{2l}\sin\chi\right]
\frac{\partial f^0}{\partial  \varepsilon_l}~.
\end{equation}  
Here, $p$ stands for $s$, $a1$, $a2$ and $ad$. $\chi$ is the angle of $\textbf{E}$ with the $\hat{x}$-direction, $\lambda^{p}_{il}$ ($i=1, 2$) are the generalized mean free paths of the charge carriers~.

Considering now an electric field in the $\hat{x}$ or $\hat{y}$ direction  $\Ev=E \hat{x_{i}}$ $(\hat{x}_{1}=\hat{x}, \hat{x}_{2}=\hat{y})$ and substituting $g^{p}_{l}$ from Eq.~(\ref{eq:generalg}) into Eqs.~(\ref{eq:longconduc}), (\ref{eq:skew1conduc}), (\ref{eq:sideconduc}) and (\ref{eq:skew2conduc}), we arrive at
\begin{equation}\label {eq:gs}
\vv_{0l}\cdot \hat{x_{i}} =\sum_{l'} w^{(2)}_{ll'} \left[\lambda^{s}_{il} - \lambda^{s}_{il}\right],
\end{equation}
\begin{equation} \label{eq:gadist}
\vv^{sj}_{l}\cdot \hat{x_{i}} =\sum_{l'}w^{(2)}_{ll'}\left[\lambda^{ad}_{il}-\lambda^{ad}_{il'}\right]~,      
\end{equation}
\begin{equation}\label {eq:ga}
\sum_{l'} w^{(2)}_{ll'}\left[\lambda^{a1}_{il}- \lambda^{a}_{il'}\right]+ \sum_{l'} w^{(3a)}_{ll'}\left[\lambda^{s}_{il} - \lambda^{s}_{il'}\right]=0, 
\end{equation}

\begin{equation}\label {eq:ga2}
\sum_{l'} w^{(4)}_{ll'}\left[\lambda^{s}_{il}-\lambda^{s}_{il'}\right]+
\sum_{l'}w^{(2)}_{ll'}\left[\lambda^{a2}_{il}-\lambda^{a2}_{il}\right]=0. 
\end{equation}
To solve the above equations, all mean free paths $\lambda_i^p$ are expanded in Fourier series. Finally we obtain the Fourier coefficients of $\lambda^{m.p} _{i}$, the general mean free path of the Dirac fermions due to scattering off magnetic impurities, and also the Fourier coefficients of $\lambda^{nm.p} _{i}$, the general mean free path of the Dirac fermions due to scattering off non-magnetic impurities.

\section{Results and Discussions}\label{sec:results}
In this section  we present our results for the different contributions to the AHE on the surface of a 3D topological insulator. In subsection~\ref{subsec:intrinsic}, we first investigate the contribution to the charge conductivity of massive Dirac fermions arising from the non-zero anomalous velocity. Next in subsection~\ref{subsection:sidejump}, we include the corrections in the velocity and the electron distribution function that are responsible for the side jump effect and consequently calculate their associated contributions to the conductivity of the system. In subsection~\ref{subsection:skew} we discuss the skew scattering contribution to the charge conductivity of the system. Finally, the sum of all these contributions, the total conductivity is discussed in subsection~\ref{subsection:total}.

\subsection{Intrinsic contribution} \label{subsec:intrinsic}
To calculate the intrinsic contribution to the current density $\Jv^{in}=-e\sum_{l} f^{0} (\varepsilon_l) \vv^{an}_{l}$, one has to know $\mathbf{v}^{an}_{l}$. In contrast to the two other contributions associated to the side jump and skew scattering effects, one has to consider all electrons residing in the whole Fermi sea instead of just in the conduction band in order to calculate this anomalous velocity. Using the eigenstates in Eq.~(\ref{eq:eigenvector}), one arrives at the following expression for the anomalous velocity
\begin{equation}
\textbf{v}^{an}_{\bm k,\alpha}=-\mathbf{\dot{\bm k}}\times \Omega_{\textit{ k},\alpha}= \frac{e \mathbf{E}}{\hbar} \times  \frac{ -\alpha M v_{F}^2 \hbar^2 }{2 \left(\textit{k}^2 v^2_{F} \hbar ^2+M^2\right)^{3/2}}\hat{z}.
\end{equation}
Therefore, the intrinsic contribution to the velocity of the electrons due to this anomalous correction is given by 
\begin{equation}
\begin{split}
\sigma^{in}_{xy}= -e(\sum^{\textit{k}_{F}}_{\textit{k}} \mathbf{v}^{an}_{k,+}+\sum^{\infty}_{k} \mathbf{v}^{an}_{k,-})= -\frac{1}{2 m}.
\end{split}
\end{equation}
with $\sigma^{in}_{yx}=-\sigma^{in}_{xy}$ and $m=\frac{\mu}{M}$. Note that all contributions to the AHE, like the expression above, are given in the unit of $\frac{e^{2}}{h}$. This contribution can be regarded as an “unquantized” version of the quantum Hall effect which is given by $ \sigma^{in}_{xy}=-\frac{1}{2}$. 

\subsection{Side Jump}\label{subsection:sidejump}
As we indicated before, there are two distinct effects due to the anomalous coordinate shift: the side jump $\delta \rv_{\bm k \bm k'}$ and a change in the energy of the electron. After averaging over many scattering events, side jumps do not cancel out and give rise to a non-zero contribution $\mathbf{v}^{sj}_{\bm k}$ to the velocity of the electrons given in Eq.~(\ref{eq:velo}). This correction to the velocity of the electrons itself changes the conductivity of the electrons and we call this contribution $\sigma^{sj}_{ij}$. The second effect is the energy change of an electron when it makes a deflection $\delta \rv_{\bm k \bm k'}$ in the presence of an external electric field $\Ev$. This change in its potential energy is given by $e \delta \rv_{\bm k \bm k'}\cdot \textbf{E}$, which eventually leads to the deviation of the distribution function of the electrons that we indicate as $g^{ad}_{l}$ in Eq.~(\ref{eq:all distribution}). We now separately discuss the resultant conductivities $\sigma^{sj}_{ij}$ and $\sigma^{ad}_{ij}$.

Using the eigenvectors of the conductive massive Dirac fermions given in Eq.~(\ref{eq:eigenvector}) and applying Eq.~(\ref{eq:deltar}), we obtain the deflections $\delta \rv_{\bm k \bm k'}^{m}$ and $\delta \rv_{\bm k \bm k'}^{nm}$ due to the scattering off magnetic and non-magnetic impurities, and they are given by 

\begin{widetext}
\begin{equation}\label{19}
\delta \rv_{\bm k \bm k'}^{m}[\frac{\hbar v_{F}}{ \varepsilon_{\textit{k}}}]= \frac{\gamma (1/2-c)}{\sqrt{(1-\gamma^{2})}} [\hat{\bm \phi}_{\bm k'}-\hat{\bm \phi}_{\bm k}]+ c \tan \theta [\cos \phi_{\bm k'}~\hat{\phi_{\bm k}}-\cos \phi_{\bm k}~\hat{\bm{\phi}_{\bm k'}}]+ (2 c\gamma^{2} \tan\theta \sin\phi_{-} \cos\phi_{+} - c \sqrt{ \gamma^{2} -\gamma^{4} }\sin 2\phi_{-}) [\hat{\textbf{\textit{k}}}+\hat{{\textbf{\textit{k}}}'}],
\end{equation}

\begin{equation}\label{20}
\delta \rv_{\bm k \bm k'}^{nm}[\frac{\hbar v_{F}}{\varepsilon_{\textit{k}}}]=\frac{\gamma (1-\gamma^{2})^{1/2}} {4[\cos^{2}\phi_{-}+\gamma^{2}\sin^{2}\phi_{-}]}\{\sin 2\phi_{-} [\hat{\bm k}+\hat{{\bm k}'}]- 2\sin^{2}\phi_{-}[\hat{\bm \phi}_{{\bm k}'}-\hat{\bm \phi}_{\bm k}]\},
\end{equation}
\end{widetext}
where the two vectors $\hat{\bm k}$ and $\hat{\bm \phi _{\bm k}}$ are unit vectors in spherical coordinates, respectively in the radial and polar direction, $\theta$ is the tilting angle of the randomly placed magnetic impurities on the surface of magnetic Ti, $\gamma=\gamma_{\textit{k}}=\dfrac{M}{\varepsilon_{\textit{k}}}$, $\phi_{\pm}=\dfrac{\phi_{\bm k}\pm \phi_{{\bm k}'}}{2}$ and finally $c=\left(2 \sin^{2}\phi_{-}+2[\gamma_{\textit{k}}\cos\phi_{-}+ \sqrt{1-\gamma_{\textit {k}}^{2}}\tan\theta \cos\phi_{+}]^{2}\right)^{-1}$. 
\begin{figure*}[t]
	\centering
	{\includegraphics[width=.94\textwidth,center]{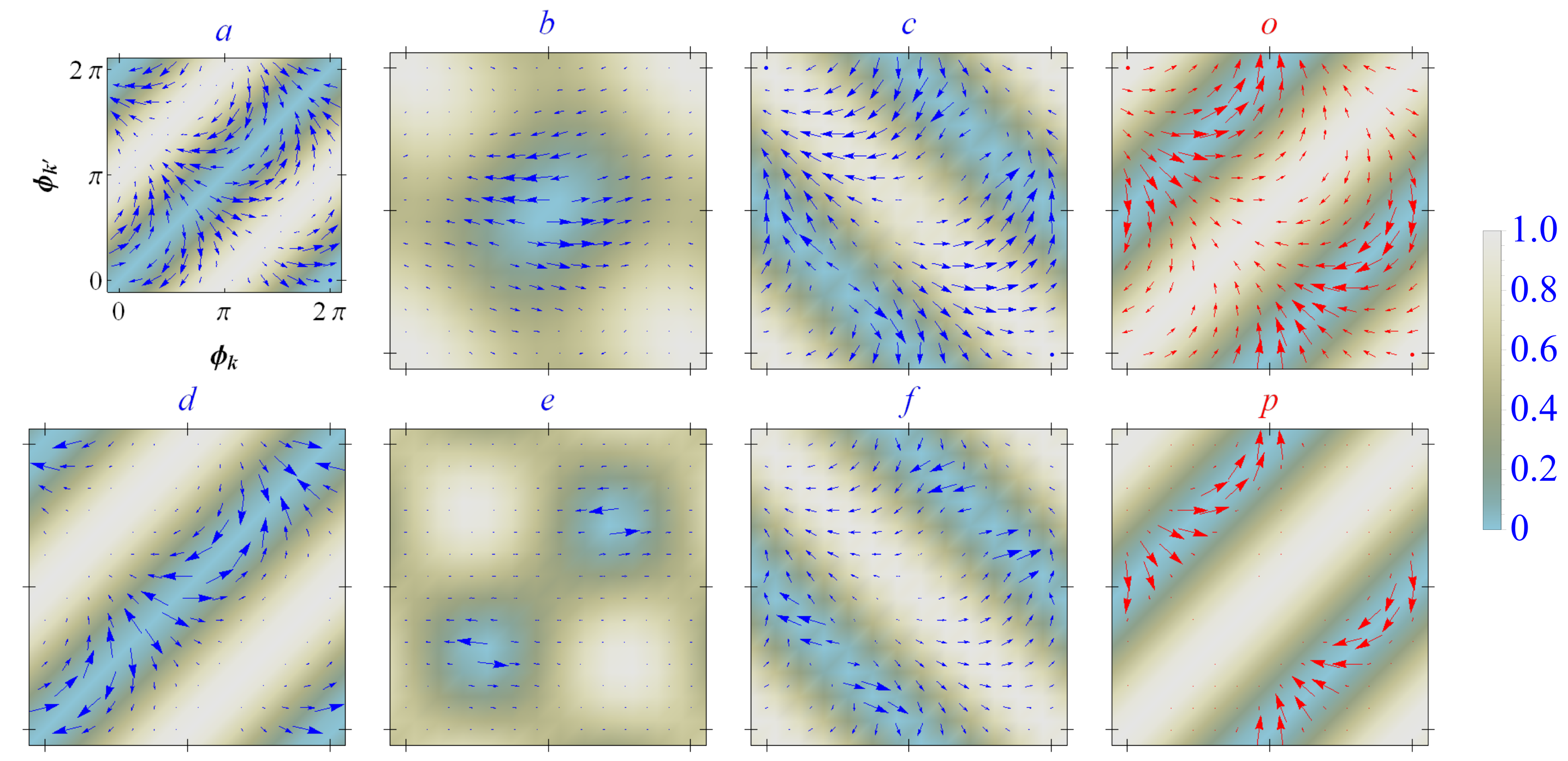}}
	\caption{(Color online) Side jump vectors of electrons during scattering off magnetic impurities $\delta \rv_{\bm k \bm k'}^{m}$ are shown in panels $(a)-(f)$ and for scattering off non-magnetic impurities $\delta \rv_{\bm k \bm k'}^{nm}$ in panels $(g)-(h)$, as function of the incident angle of the Fermi electrons $\phi_{\bm k}$ and the scattering angle of the electrons $\phi_{{\bm k}'}$, for different values of $\theta$ and $m= \frac{\varepsilon_{F}}{M}$. In panel $a$, $(\theta, m)=(0, 1.1)$, in $b$ $(\theta, m)=(\frac{\pi}{4}, 1.1)$ and in $c$, $(\theta, m)=(\frac{\pi}{2}, 1.1)$. In the second row, $(\theta, m)=(0, 8)$, $(\frac{\pi}{4}, 8)$ and $(\frac{\pi}{2}, 8)$ in panels $d$, $e$ and $f$, respectively. In addition, $\delta \rv_{\bm k \bm k'}^{nm}$ is shown in terms of $\phi_{\bm k }$ and $\phi_{\bm k'}$ for $m=1.1$ and $8$, in panel $o$ and $p$, respectively.   } \label{fig:delta}
\end{figure*}
Since electrons undergo two distinctive and independent scattering events, magnetic and non-magnetic, we treat them separately. As Eq.~(\ref{19}) shows, the side jump of an electron during a magnetic scattering strongly depends on its incident angle $\phi_{\bm k}$, scattering angle $\phi_{\bm k'}$ and also $\theta$, the tilting angle with respect to the $\hat{z}$-direction of the magnetic orientation of the surface impurities. In Fig.~\ref{fig:delta}, the deflection of a Fermi electron during scattering off magnetic impurities (illustrated by blue vectors) and non-magnetic impurities (illustrated by red vectors) is shown with respect to $\phi_{\bm k}$ and $\phi_{\bm k'}$, for different values of $m=\frac{1}{\gamma_{\textit{k}}} = \frac{\varepsilon_{\textit{k}}}{M}$ (with $\varepsilon_{\textit{k}_{F}}$ the Fermi level) and $\theta$. Just as an example here we assume that the mass of the Dirac Fermions is caused by doping the surface of $Bi_{2}Se_{3}$ with \textit{Fe}, resulting in $v_{F} \simeq 5 \times 10^{5} \textit{m} \textit{s}^{-1}$ and $M=25 m eV$, motivated by the experimental results presented in Ref.~\onlinecite{size of gap}. The length of the arrows indicate the relative size of that deflection. The background color in all panels of Fig.~\ref{fig:delta} shows the scattering probability $|T^{m}_{\bm k {\bm k}'}|^{2}$ (with $J^2 S_{m}^2=1 $), given by 

\begin{equation}\label{eq:w2}
\begin{split}
	\left|T^{m}_{\bm k{\bm k}'}\right|^2=&  \left|
	\frac{2\gamma_{\textit {k}}}{1+\gamma^2_{\textit{ k}}} \sin \theta \cos\phi_+ \right. \\
	&\left.+\frac{1-\gamma^2_{\textit {k}}}{1+\gamma^2_{\textit {k}}} \cos\theta \cos\phi_- + i \cos \theta \sin\phi_- \right|^2.
\end{split}
\end{equation}
Blue corresponds to zero probability, beige with the highest probability.

 In the first row in Fig.~\ref{fig:delta}, $m$ is taken equal to 1.1, the Fermi energy is thus just above the lowest surface conduction band state. In panel $a$ the magnetization is chosen to be perpendicular to the surface of the TI ($\theta= 0$). From this panel it is clear that the side jump is maximal when $\phi_{{\bm k}'}\approx\phi_{\bm k}\pm \frac{n\pi}{2}$, with $n$ an odd number, and is minimal when $\phi_{{\bm k}'}\approx\phi_{\bm k}\pm n\pi$, with $n$ an integer.
 
 In panel $b$ of this figure the side jump and the corresponding probabilities for a magnetic scattering event are shown for a magnetization direction rotated in the $yz$-plane with $\theta=\frac{\pi}{4}$. Note that the probability for many scattering events increases, however just in a small region of the $(\phi_{{\bm k}},\phi_{{\bm k}'})$ space the electrons feel a considerable side jump coordinate shift. Increasing $\theta$ further to $\frac{\pi}{2}$ (thus ending up in a magnetization in the $\hat{y}$-direction) makes that electrons undergo a considerable coordinate shift especially in the region with maximum scattering probability, as shown in panel $c$ of Fig.~\ref{fig:delta}. The second row of the figure (panels $d$, $e$ and $f$) shows what happens with the side jump if the Fermi level is increased up to $m=\frac{\varepsilon_{\bm k}}{M}=8 $, again for the same $\theta$ values. Note that for a large number of scattering events with different incident and scattering angle $(\phi_{\bm k},\phi_{{\bm k}'})$, the size of the anomalous coordinate shift decreases (in comparison to the upper row), though based on Eq.~(\ref{eq:w2}) its scattering probability in general increases. Thus, the side jump effect will be maximal for a surface magnetization in the plane as well as for low Fermi level values. Furthermore, as increasing $\theta $ from 0 to $\frac{\pi}{2}$ increases the scattering probability as well as the size of the side jump, it can be inferred, based on $\textbf{v}^{sj}_{\bm k}=\sum_{\bm k'} w_{\bm k \bm k'} \delta \rv_{\bm k \bm k'}$, that the side jump velocity also increases with increasing $\theta$. 

In panel $g$ and $h$ of Fig.~\ref{fig:delta} the anomalous coordinate shifts $\delta \rv_{\bm k \bm k'}$ of the Fermi electrons involved in a non-magnetic scattering event are shown as a function of $\phi_{\bm k}$ and $\phi_{{\bm k}'}$ for the same two values of $m$. The background profiles show again the scattering probability (with $V_{0}^2$=1), now given by 
\begin{equation}
\left|T^{nm}_{\bm k{\bm k}'}\right|^2 =( 1+ [\gamma_{\textit {k}}^{2}-1]\sin^{2}\phi_{-}),
\end{equation}
with $V_{0}=V^{nm}_{0}$. We can deduce that the general trend for $\delta \rv_{\bm k \bm k'}^{nm}$ is that the size of the side jump and the corresponding scattering event probability are $\pi$ out of phase. 

  \begin{figure*}[t]
 	\centering
 	{\includegraphics[width=0.85\textwidth,center]{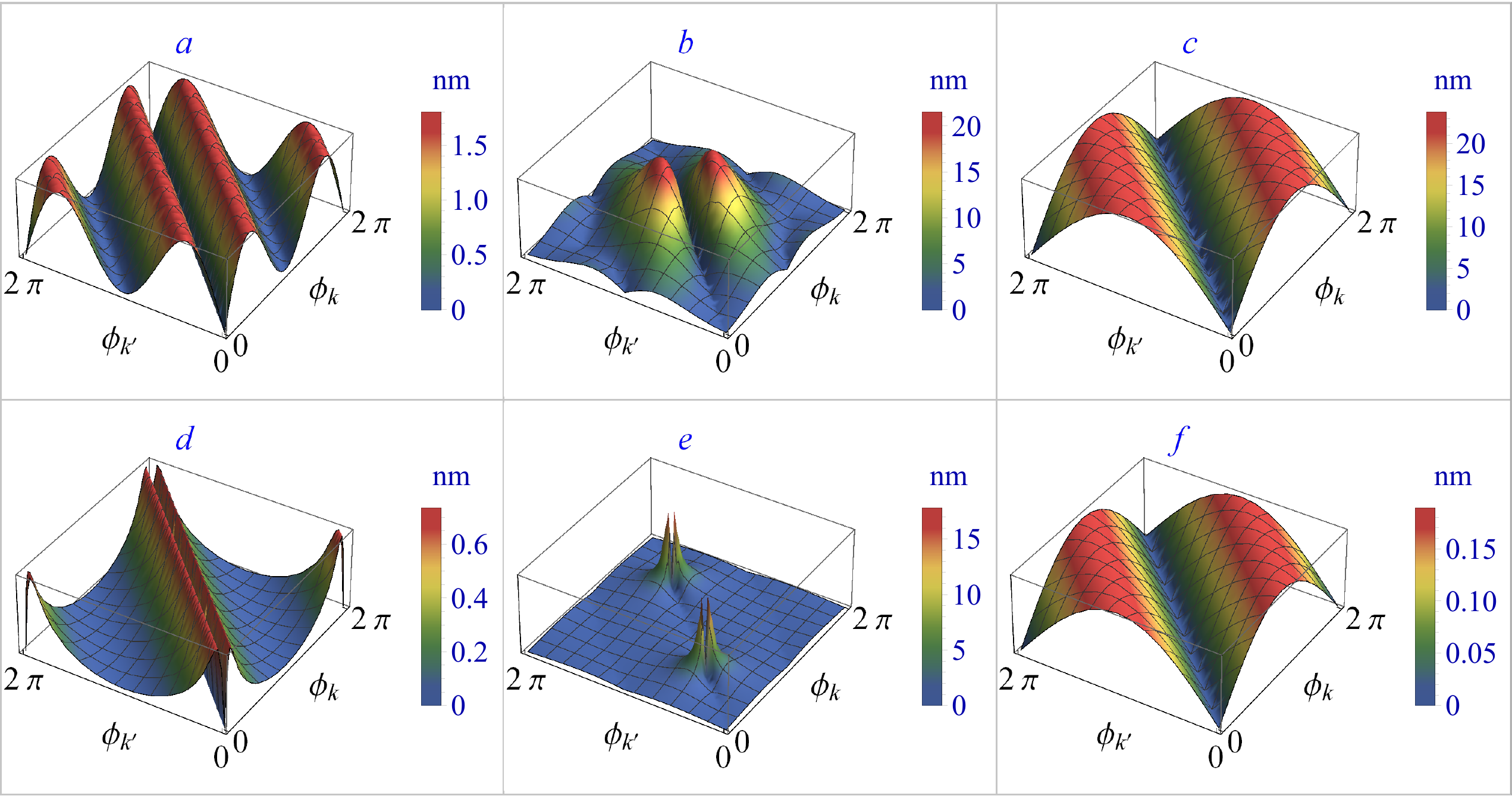}}
 	 	\caption{Numerical value of $\delta \rv_{\bm k \bm k'}^{m}$ for an $Fe$-doped topological insulator $Bi_{2}Se_{3}$ with $v_{F} \simeq 5 \times 10^{5} \textit{m} \textit{s}^{-1}$ and $M=25$ meV are shown as function of the incident angle of the electrons $\phi_{\bm k}$ and the scattering angle of the electrons $\phi_{{\bm k}'}$, for different values of $\theta$ and $m= \frac{\varepsilon_{\bm k_{F}}}{M}$. In panel $\textit{a}$, $(\theta, m)=(0, 1.1)$, in panel $\textit{b}$, $(\theta, m)=(\frac{\pi}{4}, 1.1)$ and in panel \textit{c}, $(\theta, m)=(\frac{\pi}{2}, 1.1)$. In the second row, $(\theta, m)=(0, 8)$, $(\frac{\pi}{4}, 8)$, $(\frac{\pi}{2}, 8)$ in panels $d$, $e$ and $f$, respectively. \label{fig:numericr}} 
 \end{figure*}

Fig.~\ref{fig:delta} only qualitatively shows how the side jump vectors of the electrons behave. Fig.~\ref{fig:numericr} shows the numerical information for exactly the same $(\theta,m)$ values as in Fig.~\ref{fig:delta}, to complete the discussion.
This figure shows that magnetic side jump events can undergo a one order of magnitude change in their numerical value, by changing the spatial orientation of the surface magnetization from $\theta=0$ to $\frac{\pi}{4}$. Also, even though increasing $\theta$ from $\frac{\pi}{4}$ to $\frac{\pi}{2}$ does not impose a large change in the magnetic side jump, it provides a suitable regime in which electrons will experience a significant side jump.
 
Using the derived side jumps $\delta \rv_{\bm k \bm k'}$ and scattering rates $w^{(2)}_{\bm k \bm k'}$ for magnetic and non-magnetic scattering events, we can obtain the following side jump velocities:
  \begin{eqnarray}\label{eq:vsidm}
\textbf{v}^{m.sj}_{\bm k}= \vartheta^{m}_{\textit{k}}  \left[ (1+2\sin^{2}\theta)\sin\phi_{\bm k}~\hat{x} -\cos\phi_{\bm k} ~\hat {y}\right],
\end{eqnarray}
 
  \begin{eqnarray}\label{eq:vsidnm}
\textbf{v}^{nm.sj}_{\bm k}= \vartheta^{nm}_{\textit{k}} \hat{\phi}_{\bm k},
\end{eqnarray}

 where $\vartheta^{m}_{\textit{k}}=\dfrac{S_{m}^{2}J^{2} n_{im}}{8\hbar^{2} v_{F}} \Lambda_{k}$ and $\vartheta^{nm}_{\textit{k}}=\dfrac{ V_{0}^{2} n_{inm}}{2\hbar^{2} v_{F}} \Lambda_{k}$ with $\Lambda_{k}= \gamma_{k} \sqrt{1-\gamma^{2}_{k}}$ and $n_{inm}$ the concentration of the non-magnetic impurities. Both of these velocity expressions are only non-zero for gapped systems (i.e. $\gamma_{\bm k}\neq0$) and therefore this effect is a consequence of the gap opening.
Also note that Eq.~(\ref{eq:vsidm}) and Eq.~(\ref{eq:vsidnm}) show that ${\textbf{v}_{\bm k}}^{nm.sj}$ is always perpendicular to the band velocity $\textbf{v}_{0 \bm k}$ (directed in the $\hat{\bm k}$ direction), in contrary to ${\textbf{v}_{\bm k}}^{m.sj}$. Only when all magnetic impurities are aligned perpendicular to the surface of the TI (i.e. $\theta = 0$) or when the electrons move on the surface of the TI in the direction perpendicular to the in-plane component of the magnetization (i.e. $\phi_{\bm k}= 0$) we find  ${\textbf{v}}^{m.sj}_{\bm k} \parallel {\textbf{v}}^{nm.sj}_{\bm k}$.

\begin{figure}
  	 	{\includegraphics[width=0.4\textwidth,center]{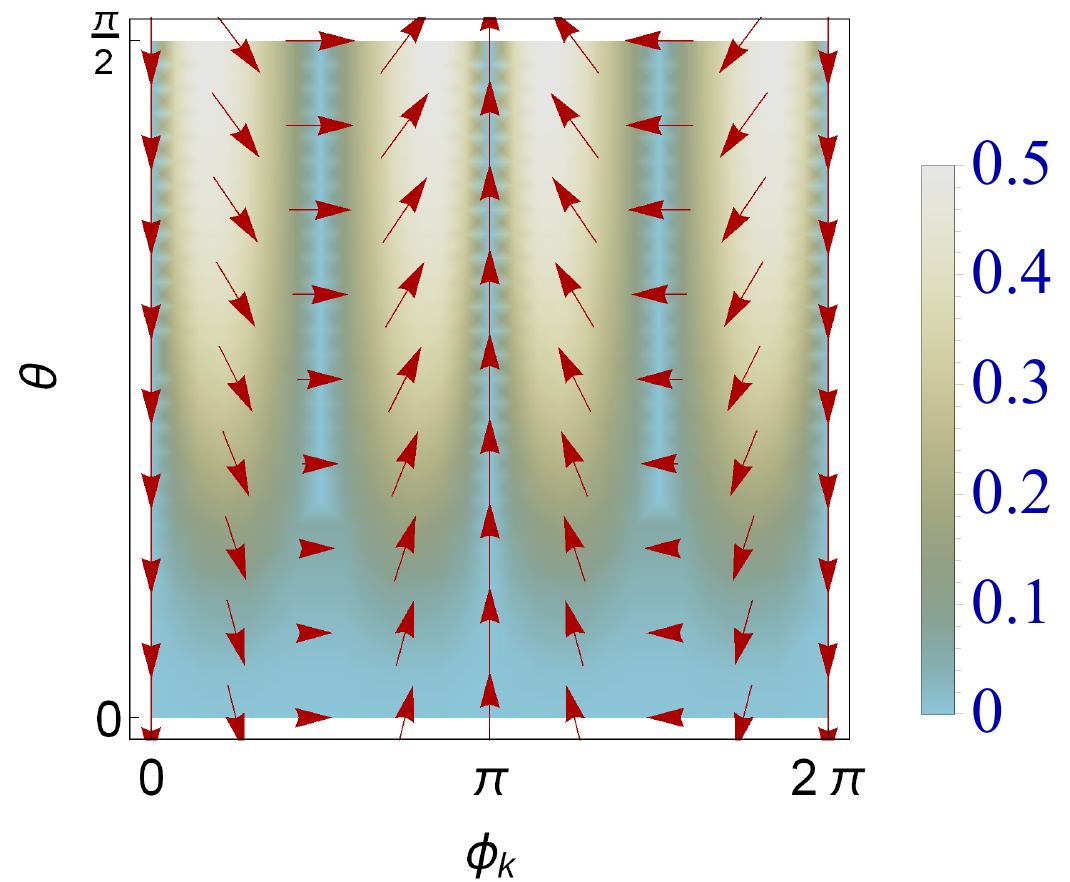}}
 	\caption{ $\textbf{v}^{m.sj}_{\bm k}$ is shown as a function of $\phi_{\bm k}$ and $\theta$, with $\vartheta^{m}_{\textit{k}}$=1 . The background color shows $\cos \varphi$, with $\varphi$ the angle between  $\textbf{v}^{m.sj}_{\bm k}$ and $ \textbf{v}_{0k}$.\label{fig:sidevelocity} } 
 \end{figure}

To better trace the behavior of the side jump velocity of the electrons during a magnetic scattering event, Fig.~\ref{fig:sidevelocity} is provided to show $\textbf{v}^{m.sj}_{\bm k}$ as function of $\theta$ and $\phi_{\bm k}$. Here $\vartheta ^{m}_{\textit{k}}$ was set to 1 and background color representing $\cos \varphi$, with $\varphi$ the angle between  $\textbf{v}^{m.sj}_{\bm k}$ and $ \textbf{v}_{0\bm k}$. This figure now reveals when the side jump contribution to the transverse conductivity is largest.

As we previously predicted by looking at the profile of side jump of electrons in Fig.~\ref{fig:delta}, by increasing $\theta$ the side jump velocity ${\textbf{v}}^{m.sj}_{\bm k}$ increases (indicated by the increase in length of the red arrows in Fig.~\ref{fig:sidevelocity} and consequently a larger transverse conductivity can be expected. The averaged band velocity of itinerant electrons in the presence of an external electric field along the $\hat{x}/\hat{y}$ direction would be along the $-\hat{x}$ /$- \hat{y}$ direction, corresponding to $\phi_{\bm k} = \pi$ and $\phi_{\bm k} = \frac{3\pi}{2}$, respectively. Fig.~\ref{fig:sidevelocity} shows that the side jump velocity ${\textbf{v}}^{m.sj}_{\bm k}$ of electrons for $\phi_{\bm k}=\pi$ is along the $\hat{y}$ direction and for $\phi_{\bm k} = \frac{3\pi}{2}$ is along the $-\hat{x}$ direction. Therefore, we can expect that $\sigma_{yx}^{m.sj}<0 $ and $\sigma_{xy}^{m.sj}> 0 $ for the corresponding transverse conductivities. Furthermore, from the background color of Fig.~\ref{fig:sidevelocity} which shows $\cos \varphi$ with $\varphi$ the angle between  $\textbf{v}^{m.sj}_{\bm k}$ and $ \textbf{v}_{0\bm k}$, we can deduce that the area for which ${\textbf{v}}^{m.sj}_{\bm k} \cdot {\textbf{v}}_{0\bm k}\sim 0$ is larger around $\phi_{\bm k}=\frac{3\pi}{2}$ than for $\phi_{\bm k}=\pi$. Therefore we expect $\rvert \sigma_{xy}^{m.sj}\rvert > \rvert \sigma_{yx}^{m.sj}\rvert$.

Now we are ready to derive all side jump contributions in the charge conductivity of the massive Dirac fermions due to the magnetic impurities ${\bm J}^{tot.m.sj}={\bm J}^{m.sj}+{\bm J}^{m.ad}$ and the non-magnetic impurities ${\bm J}^{tot.nm.sj}={\bm J}^{nm.sj}+{\bm J}^{nm.ad}$.  In order to find ${\bm J}^{m.ad}$ and  ${\bm J}^{nm.ad}$, we need to solve the corresponding Eq.~(\ref{eq:gadist}). We obtain all the mean free paths $\lambda^{m.ad}_{i}$ and $\lambda^{nm.ad}_{i}$ by relying on their Fourier expansions (see appendix) and obtain the following corresponding charge conductivities

\begin{equation}\label{eq:28}
\sigma^{m.ad}_{xy}=\frac{2(1-m^2) }{4m(4m^2\cos^2\theta+g(m^2+\cos2\theta))},
\end{equation}
\begin{equation}
\sigma^{m.ad}_{yx}=\dfrac{(2-\cos 2\theta)(m^2+\cos 2 \theta)(g-2)}{4 m(\cos 4 \theta-1+ (m^2+\cos2\theta)(g-2))},
\end{equation}
\begin{equation}\label{eq:26}
\sigma^{nm.ad}_{xy}=-\sigma^{nm.sj}_{yx}=\frac{1-m^{2}}{m(m^{2}+3)},
\end{equation}
with $ g =\dfrac{\left(4[m^4+1]+2m^2[4\cos2\theta+\cos4\theta-1]\right)^{1/2}}{\left|m^2+\cos2\theta\right|}$. Because $\textbf{J}^{sj}=-e \sum_{\bm k} g^{s}_{\bm k} \textbf{v}^{sj}$ and using the already reported distribution function $g^{s}_{\bm k}$ in Ref.~\onlinecite{amir_2015}, we come to the conclusion that $\sigma^{m.sj}_{ij}=\sigma^{m.ad}_{ji}$ and $\sigma^{nm.sj}_{ij}=\sigma^{nm.ad}_{ij}$, for $i\neq j$.

How a contribution to the correction of the transverse charge conductivity of a system is linked to the longitudinal conductivity has always been a vital question in this context. The following connections are found
\begin{align}
&\sigma ^{m.sj}_{xy}=\sigma ^{m.ad}_{yx}=\dfrac{2-\cos 2 \theta}{4 m }~\tilde{\sigma}^{m.s}_{yy}\label{msjxy},\\
&\sigma ^{m.sj}_{yx}=\sigma ^{m.ad}_{xy}=-\dfrac{1}{4 m}~\tilde{\sigma}^{m.s} _{xx}\label{msjyx},\\
&\sigma^{nm.ad}_{ij}=\sigma^{nm.sj}_{ij}=-\frac{1}{m}\tilde{\sigma}^{nm.s}_{ii}\label{nmsjyx},
\end{align}
with $\tilde{\sigma}^{m.s} _{ii}=\sigma^{m.s} _{ii}/\sigma^{m}_{0}$, $\tilde{\sigma}^{nm.s}_{ii}=\sigma^{nm.s}_{ii}/\sigma^{nm}_{0}$,  $\sigma^{m}_{0}=2\hbar^2 v^2_\textrm{F}/(n_{\rm im}J^2 S_{m}^2)$ and $\sigma^{nm}_{0}=2\hbar^2 v^2_\textrm{F}/(n_{\rm inm}V_{0}^2 )$.

\begin{figure}
	\centering
	\subfloat{\includegraphics[width=0.42\textwidth,center]{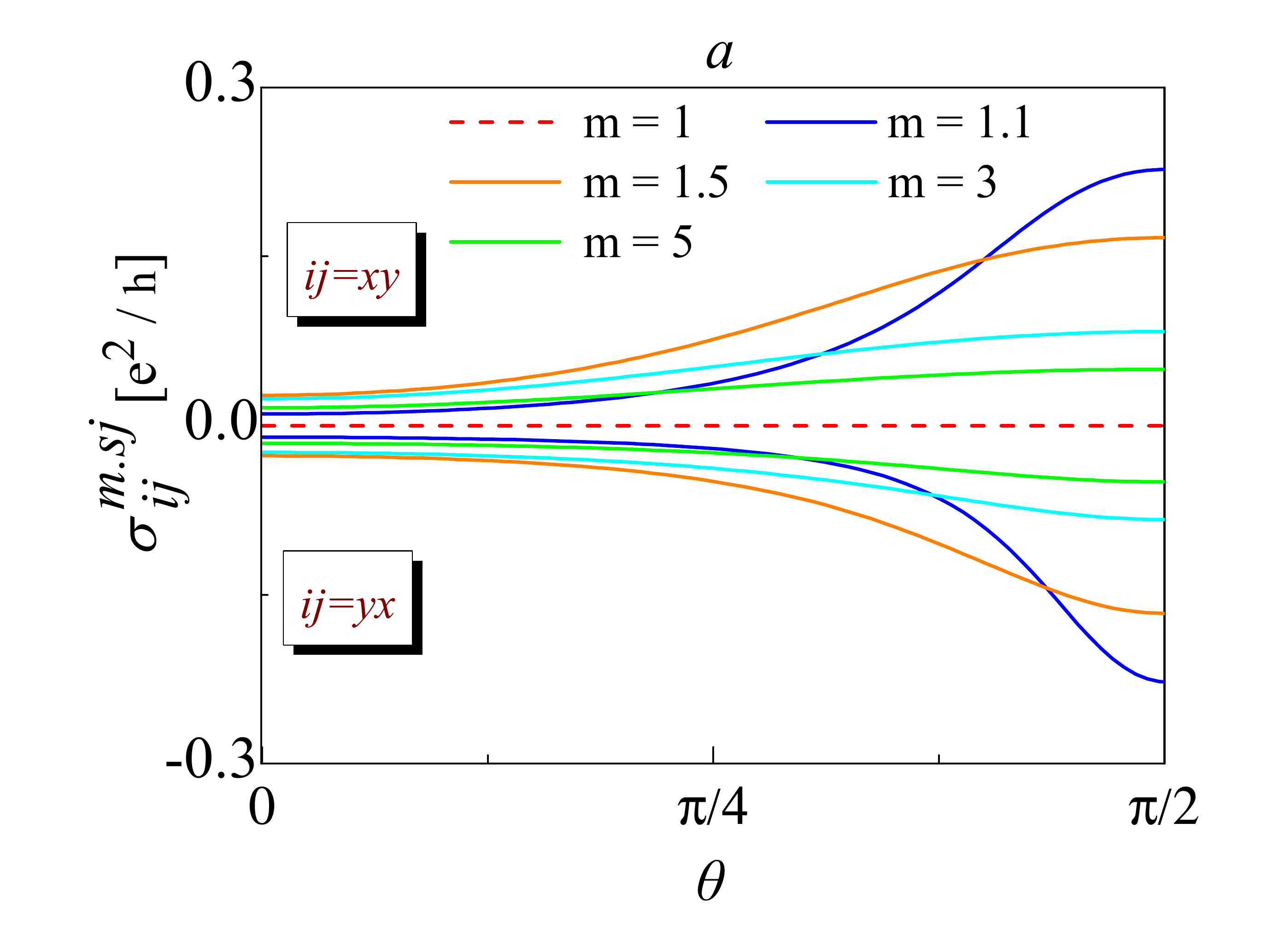}}\\
	\subfloat{\includegraphics[width=0.42\textwidth, center]{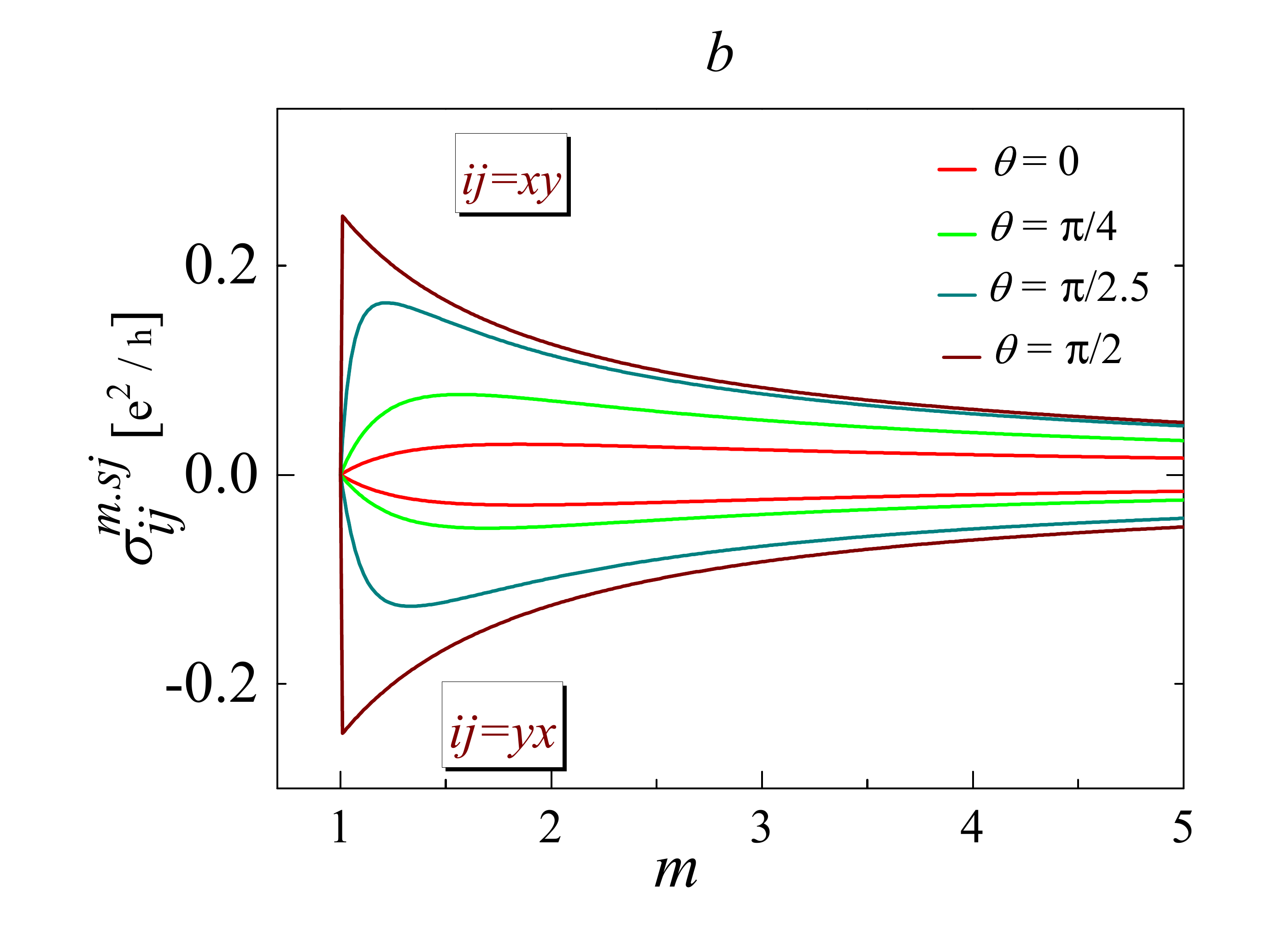}}
	\caption{$\sigma_{xy}^{m.sj}$ and $\sigma_{yx}^{m.sj}$ are plotted in terms of $\theta$ for some different values of $m$ in panel $a$, and in terms of $m$ for some different values of $\theta$ in panel $b$, respectively. \label{fig:sj}}
\end{figure}

\begin{figure}
  	 	{\includegraphics[width=0.41\textwidth,center]{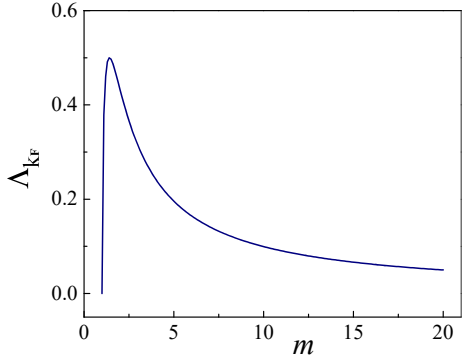}}
 	\caption{ $\Lambda_{k_{F}}$ is shown as function of $m$.\label{fig:velocitygama}} 
 \end{figure}

$\sigma^{m.sj}$ is plotted against $\theta$ for some different values of $m$ in panel $a$ of Fig.~\ref{fig:sj} and also in terms of $m$ for some different values of $\theta$ in panel $b$ of this figure. In case of $\theta=0$, $\rvert \sigma^{m.sj}_{xy}\rvert=\rvert \sigma^{m.sj}_{yx}\rvert$ and putting aside the sign of these conductivities, the system behaves isotropically relative to the external electric field direction. Increasing $\theta$ increases the magnitude of the transverse conductivities, and this is caused by the interplay between two factors. Firstly, the backscattering probability is the main mechanism which suppresses both longitudinal and transverse conductivity. By increasing $\theta$, the backscattering probability $w^{(2.m)}(\bm k,-\bm k)\sim [(1-\gamma_{\textit{k}}^2) \sin^{2}\theta \sin^{2}\phi_{\textit{k}} + \cos^{2} \theta]$ decreases, so the transverse conductivity will increase. Secondly, angular part of the magnetic side jump velocity $[~1+ 4 \sin^{2}\phi_{\bm k} (\sin^{2}\theta +\sin^{4}\theta ) ~]^{1/2}$ increases with increasing $\theta$ and subsequently the conductivity increases. As it is clear from panel $b$ of the figure, $\sigma^{m.sj}$ is zero if we put the chemical potential exactly on the lowest state of the surface band structure ($m=1$ or $\mu=M$). Beyond $m=1$, $\sigma^{m.sj}$ experiences a peak close to $m=1$, and thereafter decreases by increasing $m$ for all values of $\theta$. This non-monotonic feature of the conductivity arises from the non-angular part of the side jump velocity ($\Lambda_{\textit{k}}$) which has the same trend against $m$ (as shown in Fig.~\ref{fig:velocitygama}). Therefore, deviating the system a bit from the insulating state and being far enough away from the perfect metallic state $m\gg 1$ or $\mu\gg M$, can produce a large value for the side jump conductivity. 

Based on Eq.~(\ref{msjxy}) and Eq.~(\ref{msjyx}) we come to the conclusion that $\sigma^{tot.m.sj}_{xy}=\sigma^{m.sj}_{xy}+\sigma^{m.ad}_{xy}=\sigma^{tot.m.sj}_{yx}$, with
\begin{equation}\label{sjtot}
\sigma^{tot.m.sj}_{xy}=\sigma^{tot.m.sj}_{yx}=\frac{(2-\cos 2 \theta ) \sigma^{m.s}_{yy}-\sigma^{m.s}_{xx}}{4 m\sigma _0}.
\end{equation}
\begin{figure}
	\includegraphics[width=0.43\textwidth,center]{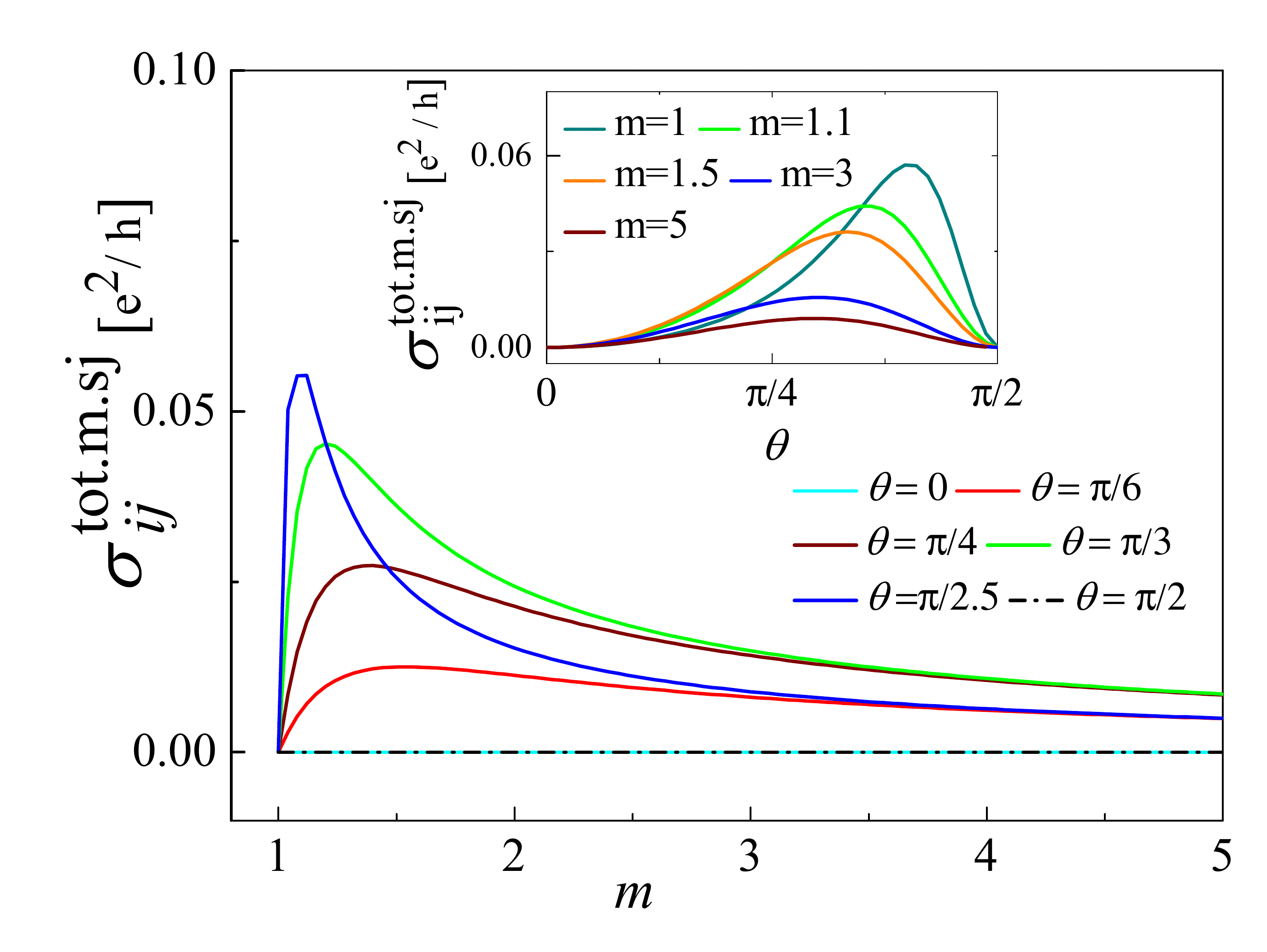}	
	\caption{$\sigma_{ij}^{tot.m.sj}$ is plotted in terms of $m$ for some values of $\theta$ in the main panel, and in terms of $\theta$ for some different values of $m$ in the inset.
	}\label{fig:sjtot}
\end{figure}
Therefore, while in the presence of magnetic scatterers $\sigma^{tot.m.sj}_{xy}=\sigma^{tot.m.sj}_{yx}$ and hence the side jump itself has an isotropic feature relative to the direction of the external electric field, this transverse conductivity somehow measures the anisotropy of the system when $\theta=0$, being the difference between the two components of the longitudinal conductivities $\sigma^{m.s}_{xx}$ and $\sigma^{m.s}_{yy}$.

\begin{figure}
	\centering
	\subfloat{\includegraphics[width=0.44\textwidth,center]{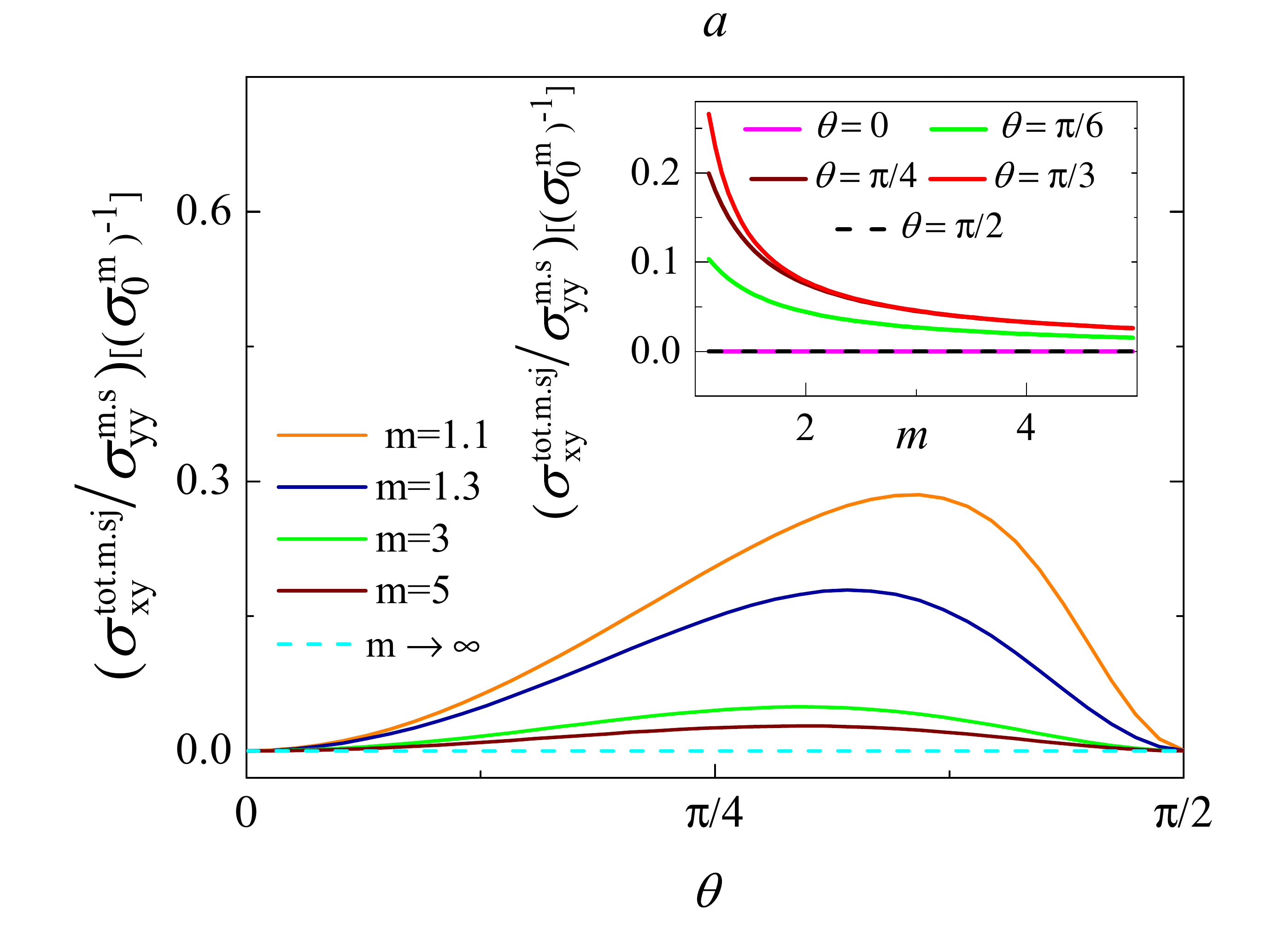}}\\
	\subfloat{\includegraphics[width=0.44\textwidth,center]{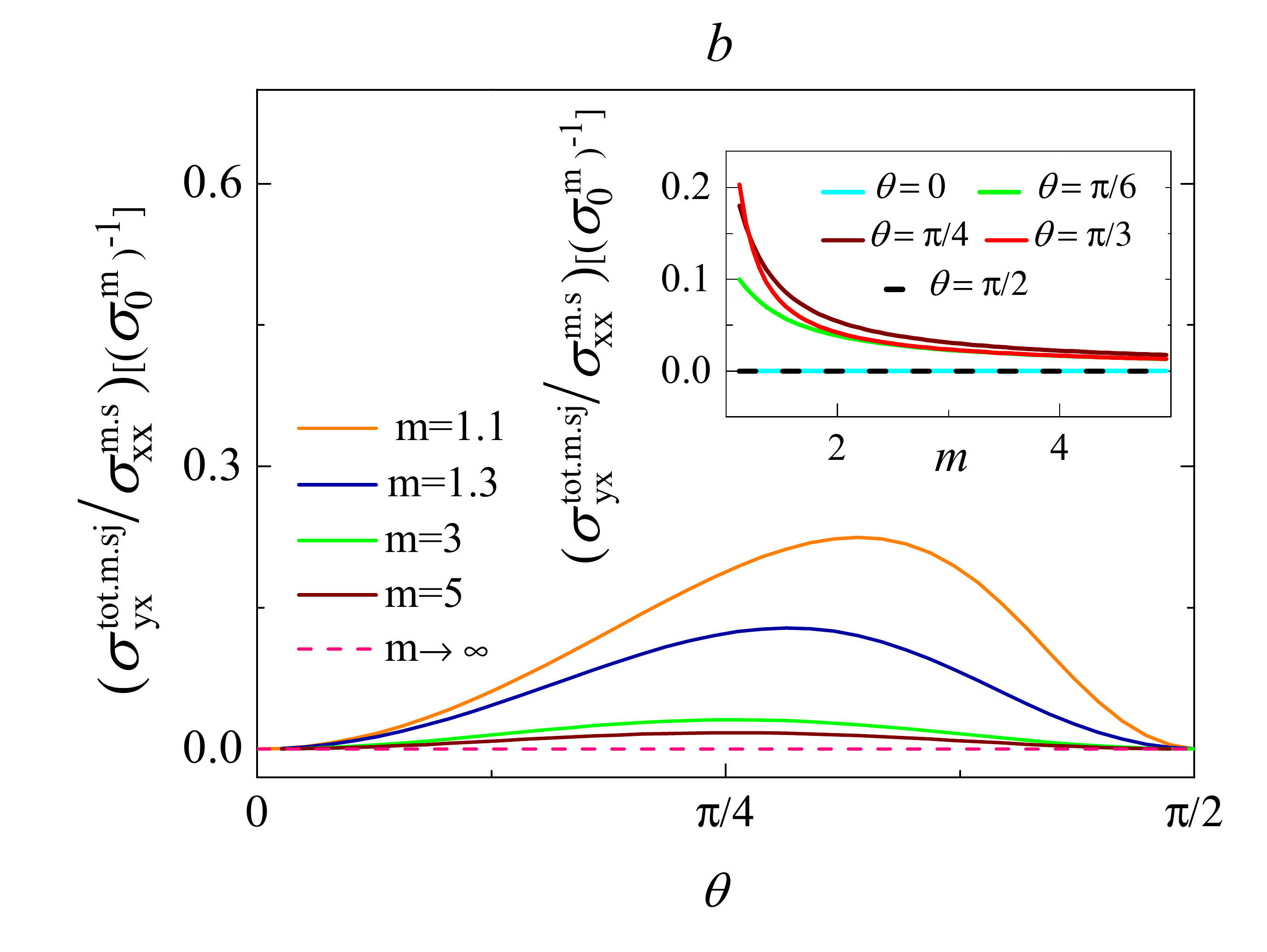}}
	\caption{$\sigma_{xy}^{tot.m.sj}$ / $\sigma_{yy}^{m.s}[(\sigma^{m.s}_{0})^{-1}]$ is plotted in terms of $\theta$ for some different values of $m$ in the main window of panel $a$, and against $m$ for some different values of $\theta$ in the inset of this panel.  $\sigma_{yx}^{m.sj}$ / $\sigma_{xx}^{to.m.s}[(\sigma^{m.s}_{0})^{-1}]$ is plotted in panel $b$ for the same choice of parameters $\theta$ and \textit{m}.  \label{fig:sjlongg}} 
\end{figure}

 In addition, using Eq.~(\ref{msjxy}) and Eq.~(\ref{msjyx}) we find $\sigma^{tot.m.sj}_{xy}= \sigma^{m.sj}_{xy} +\sigma^{m.sj}_{yx}$. Furthermore, as Fig.~\ref{fig:sj} shows $\sigma^{m.sj}_{yx} < 0 $, then we can rewrite $\sigma^{tot.m.sj}_{xy}= \rvert\sigma^{m.sj}_{xy}\rvert -\rvert\sigma^{m.sj}_{yx}\rvert$. Finally since always $\sigma^{tot.m.sj}_{xy}\geq 0 $, we can conclude that $\rvert \sigma^{m.sj}_{xy}\rvert \geq \rvert \sigma^{m.sj}_{yx}\rvert$. This inequality surprisingly proves our already made prediction just based on looking at the profile of $\textbf{v}^{m.sj}_{\bm k}$. Fig.~\ref{fig:sjtot} shows $\sigma^{tot.m.sj}_{ij}$ in terms of $m$ for some values of $\theta$ in the main window and against $\theta$ for some different values of $m$ in the inset. In this figure for two cases $\theta$=0 and $\theta =\frac{\pi}{2}$, $\textbf{J}^{m.ad}$ and $\textbf{J}^{m.sj}$ cancel out each other and there is no net conductivity due to the side jump effect. Within $0<\theta<\frac{\pi}{2}$, $\sigma^{tot.m.sj}_{xy}$ has a minimum value at $m=1$, thereafter increases sharply within interval of $1\leq m \lesssim 1.5$, until it reaches its maximum value. By further increasing $m$, the net side jump conductivity decreases until it reaches zero in the limit of $m \rightarrow \infty$ or a gapless system. In short, in order to have a maximal total side jump conductivity arising from just magnetic impurities, we need to put the chemical potential just above the lowest state in the surface band structure and also tune the orientation of the surface magnetization within the interval $\frac{\pi}{3}<\theta<\frac{\pi}{2}$.\\
    In order to make a comparison between  the mean free path of the electrons during a magnetic side jump and their magnetic longitudinal mean free path, Fig.~\ref{fig:sjlongg} is provided. In panel $a$ of this figure $\sigma^{tot.m.sj}_{xy} /\sigma^{m.sj}_{yy}$ (in units of $(\sigma^{m}_{0})^{-1})$ is plotted against $\theta$ for some values of $m$, and in the inset against $m$ for some values of $\theta$. As this panel illustrates, this ratio gradually increases by increasing $\theta$ until a specific $\theta$ (which varies with $m$), thereafter gradually decreases until it reaches zero at $\theta=\frac{\pi}{2}$. We can see that, if we align the surface magnetization at $\theta\sim \frac{\pi}{3}$ and in addition put the chemical potential just above the bottom of the conduction band, the maximum value of $\sigma^{tot.m.sj}_{xy} /\sigma^{m.sj}_{yy}$ is reached. The curves in the inset window show that this ratio is zero for the system with in plane magnetization or fully out of plane magnetization. Again we can conclude that adjusting $\theta\sim \frac{\pi}{3}$ and $m \sim 1$ lead to the maximum attainable value for $\sigma^{tot.m.sj}_{xy} /\sigma^{m.sj}_{yy}$. In panel $b$, $\sigma^{tot.m.sj}_{yx} /\sigma^{m.sj}_{xx}$ (in units of $(\sigma^{m}_{0})^{-1})$ is plotted in the same way. Comparing presented curves in the main and inset window of this panel indicates that they share their trends against $m$ and $\theta$ with their corresponding curves in panel $a$. In addition, even though the system is anisotropic with respect to the longitudinal conductivity and isotropic with respect to $\sigma^{tot.m.sj}_{ij}$, the numerical difference between two curves in panel $a$ and $b$ for the same parameters is insignificant.
    
\begin{figure}
	\centering
	\subfloat{\includegraphics[width=0.35\textwidth,center]{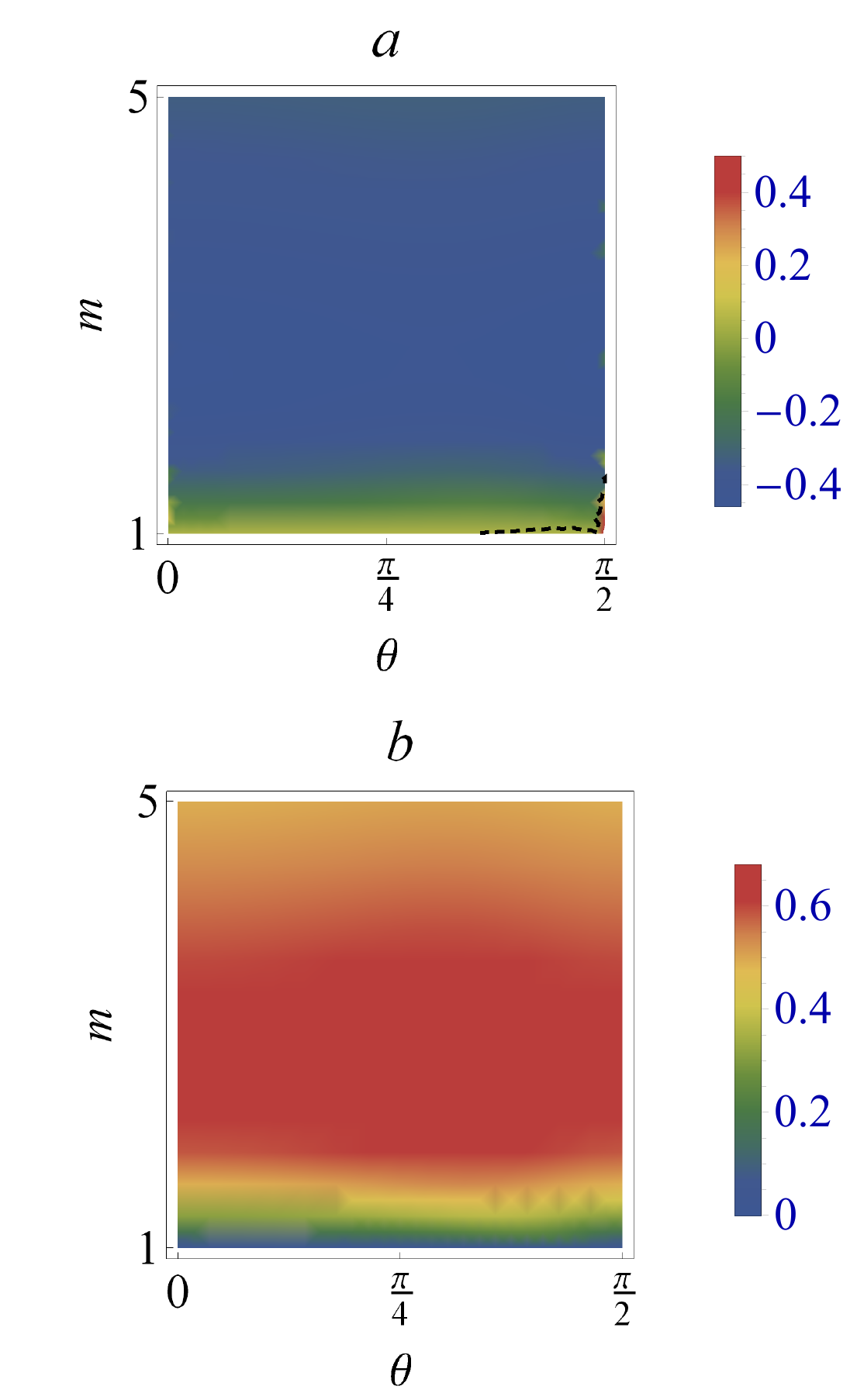} }
	\caption{$\sigma_{xy}^{tot.sj}$ (panel $a$) and $\sigma_{yx}^{tot.sj}$ (panel $b$) are plotted in terms of $\theta$ and $m$.	\label{fig:totsjxy}} 
\end{figure}

Finally, we consider the non-magnetic scattering events, and using Eq.~(\ref{nmsjyx}) and  Eq.~(\ref{eq:26}) we arrive at 
\begin{align}
&\sigma^{tot.nm.sj}_{xy}=\frac{2(1-m^{2})}{m(m^{2}+3)}\\
& \sigma^{tot.nm.sj}_{yx}=-\frac{2(1-m^{2})}{m(m^{2}+3)},
\end{align}
  where $\sigma^{tot.nm.sj}_{ij}=\sigma^{tot.nm.sj}_{ij}+\sigma^{tot.nm.ad}_{ij}$. Thus also this contribution to the side jump conductivity is isotropic (ignoring the sign difference).  
  Fig.~\ref{fig:totsjxy}(a) and Fig.~\ref{fig:totsjxy}(b) show the off diagonal elements of the total conductivity matrix $\sigma^{tot.sj}$ in terms of $\theta$ and $m$, where $\sigma^{tot.sj}_{ij}=\sigma^{tot.m.sj}_{ij}+\sigma^{tot.nm.sj}_{ij}$. As it is clear from this figure, $\sigma^{tot.sj}_{ij}$  has a negligible sensitivity against $\theta$. This fact implies that among the two different types of impurities, the non-magnetic impurities contribution to the side jump conductivity $\sigma^{tot.sj}_{ij}$ dominates. In addition, the black dashed line in  Fig.~\ref{fig:totsjxy}(a) specifies the $(\theta, m)$ combinations for which the corresponding conductivity is zero. In other words, tuning $\theta$ within $\frac{\pi}{3}\leq \theta \leq\frac{\pi}{2}$ and also the putting Fermi level close to the bottom of conduction band, enables us to turn off and on the side jump total conductivity in the presence of both magnetic and non-magnetic impurities, if we exert the external electric field parallel to in-plane component of the surface magnetization.       

\subsection{Skew Scattering}\label{subsection:skew}

\subsubsection{Conventional Skew Scattering}\label{subsec:conventional}
Finally we consider the last contribution to the AHE, the skew scattering. Recall that skew scattering gave rise to two terms in the current density. Let us first consider $\textbf{J}^{sk1}=\sum_{\bm k} \textbf{v}_{0\bm k}g^{a1}_{\bm k}$, due to magnetic impurities and non-magnetic impurities. This contribution is also called the conventional skew scattering. Since $g^{a1}_{\bm k}$ is inversely proportional to the concentration of the present impurities, this contribution dominates conductivity in a very dilute regime. From Eq.~(\ref{eq:w3a}) we obtain  $w^{(3a.m)}_{\bm k \bm k'}$ for a magnetic scattering event and $w^{(3a.nm)}_{\bm k \bm k'}$ for a nonmagnetic scattering event
\begin{align}
&w^{(3a.m)}_{\bm k \bm k'}=\dfrac{\pi ~n_{im}J^{3} S_{m}^{3} k^{2}}{2 ~\hbar~\varepsilon_{\textit{k}}} \cos\theta \sin(\phi_{\bm k}- \phi_{\bm k'}) \delta(\varepsilon_{\textit{k}}-\varepsilon_{\textit{k}'})\label{eq:w3am},\\
& w^{(3a.nm)}_{\bm k \bm k'}=\dfrac{\pi~ n_{inm}V_{0}^{3} k^{2} M}{2 ~\hbar ~\varepsilon^{2}_{\textit{ k}}} \sin(\phi_{\bm k}- \phi_{\bm k'}) \delta(\varepsilon_{k}-\varepsilon_{k'}).
\end{align}

Substituting $w^{(3a.m)}_{\bm k \bm k'}$ in Eq.~(\ref{eq:ga}) gives us the corresponding mean free paths $\lambda^{m.a1}_{i}(\bm k) (i=1,2)$, from which we obtain the distribution function $g^{m.a1}_{\bm k}$ using Eq.~(\ref{eq:generalg}), and consequently the following corresponding conductivity (for more details see the appendix): 
\begin{align}\label{eq:36}
&\sigma^{m.sk1}_{xy}=\frac{D (2-g)(\cos2\theta +m^2) }{(g\cos 4 \theta +2 (g-2) m^2 \cos 2 \theta +g-4)},\\
&\sigma^{m.sk1}_{yx}=-\sigma^{m.sk}_{xy},
\end{align}
with $D=\eta^{m}\dfrac{4 (m^2-1)\cos \theta}{(g+2) \cos 2 \theta +(g+4) m^2-2}$ and $\eta^{m}_{1}= \dfrac{\mu }{n_{im} S_{m} J}$. 
\begin{figure}
	\includegraphics[width=90mm,center]{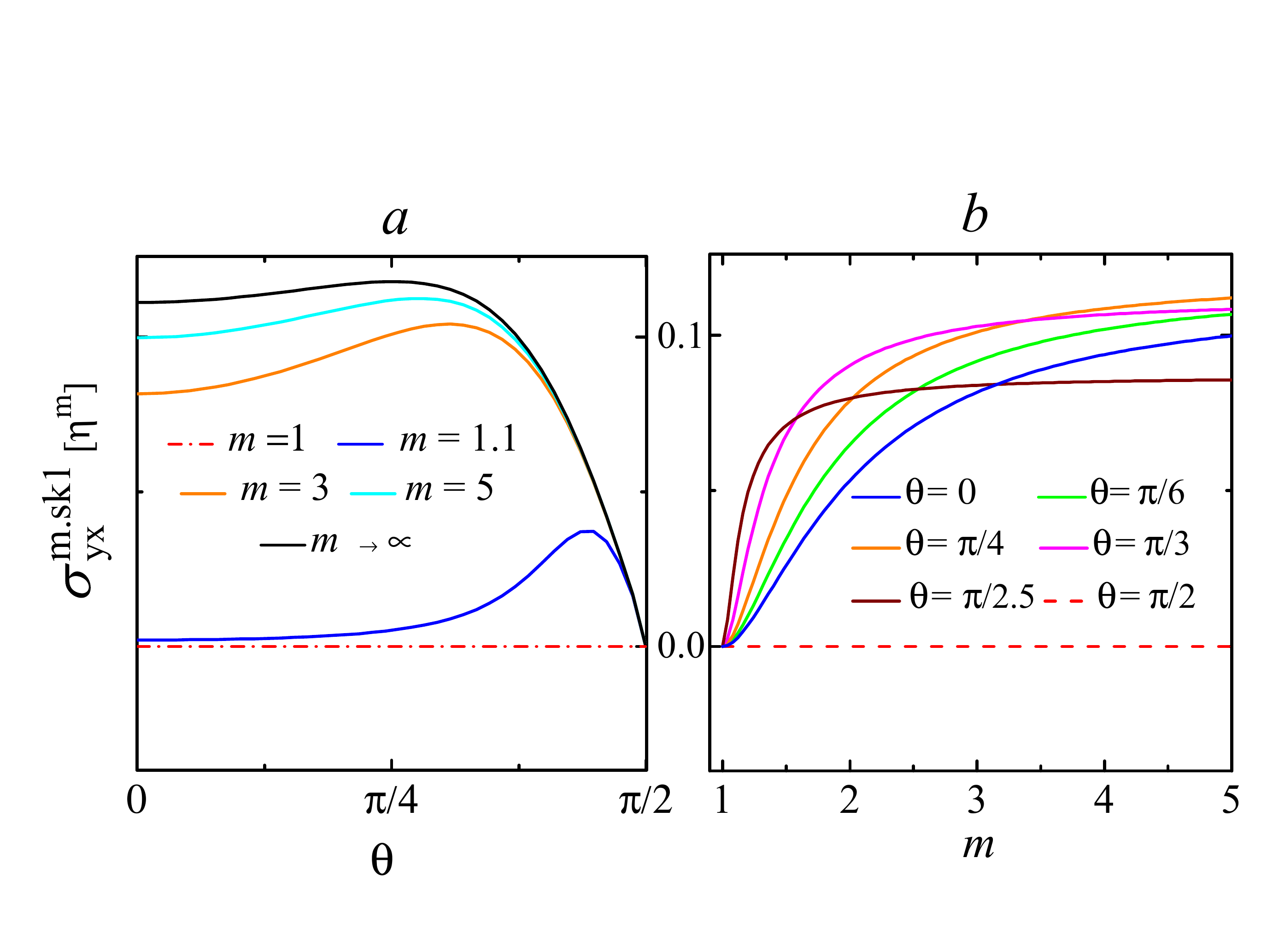}
	\caption{$\sigma_{yx}^{m.sk1}$ is plotted in terms of $\theta$ and \textit{m} in panel $a$ and $b$, respectively. 
	\label{fig:yxmsk} }
\end{figure}

In panel $a$ of Fig.~\ref{fig:yxmsk}, $\sigma^{m.sk1}_{yx}[\eta^{m}]$ is shown against $\theta$ for some values of \textit{m}. This figure shows that $\sigma^{m.sk1}_{yx}$ smoothly increases for increasing $\theta$, until it reaches a maximum in the interval $ \frac{\pi}{4}<\theta <\frac{\pi}{2}$. Thereafter it sharply decreases until it vanishes at $\theta=\frac{\pi}{2}$. Panel $b$ shows that putting the surface of the system into the insulator regime, i.e. $m=1$, turns off the transverse conductivity  $\sigma^{m.sk1}_{yx}$, regardless of the orientation of the magnetization. For larger values of $m$, $\sigma^{m.sk1}_{yx}$ increases with increasing $m$. In addition, as shown by the black curve in panel $a$, the conductivity saturates at $\sigma^{m.sk1}_{yx}(m\rightarrow \infty)$=$\eta^{m}\dfrac{\cos \theta }{6+ 3 \cos 2 \theta}$ for very large values of $m$. Thus by closing the gap or driving the system into a perfect metallic regime, skew scattering still has a non-zero contribution in the conductivity of the system. This feature reveals one of the main differences between this skew scattering contribution and the side jump contribution $\sigma^{to.m.sj}_{yx}$ which vanishes in a gapless system or in a perfect metallic regime.

\begin{figure}
	\centering
	\subfloat{\includegraphics[width=0.46\textwidth,center]{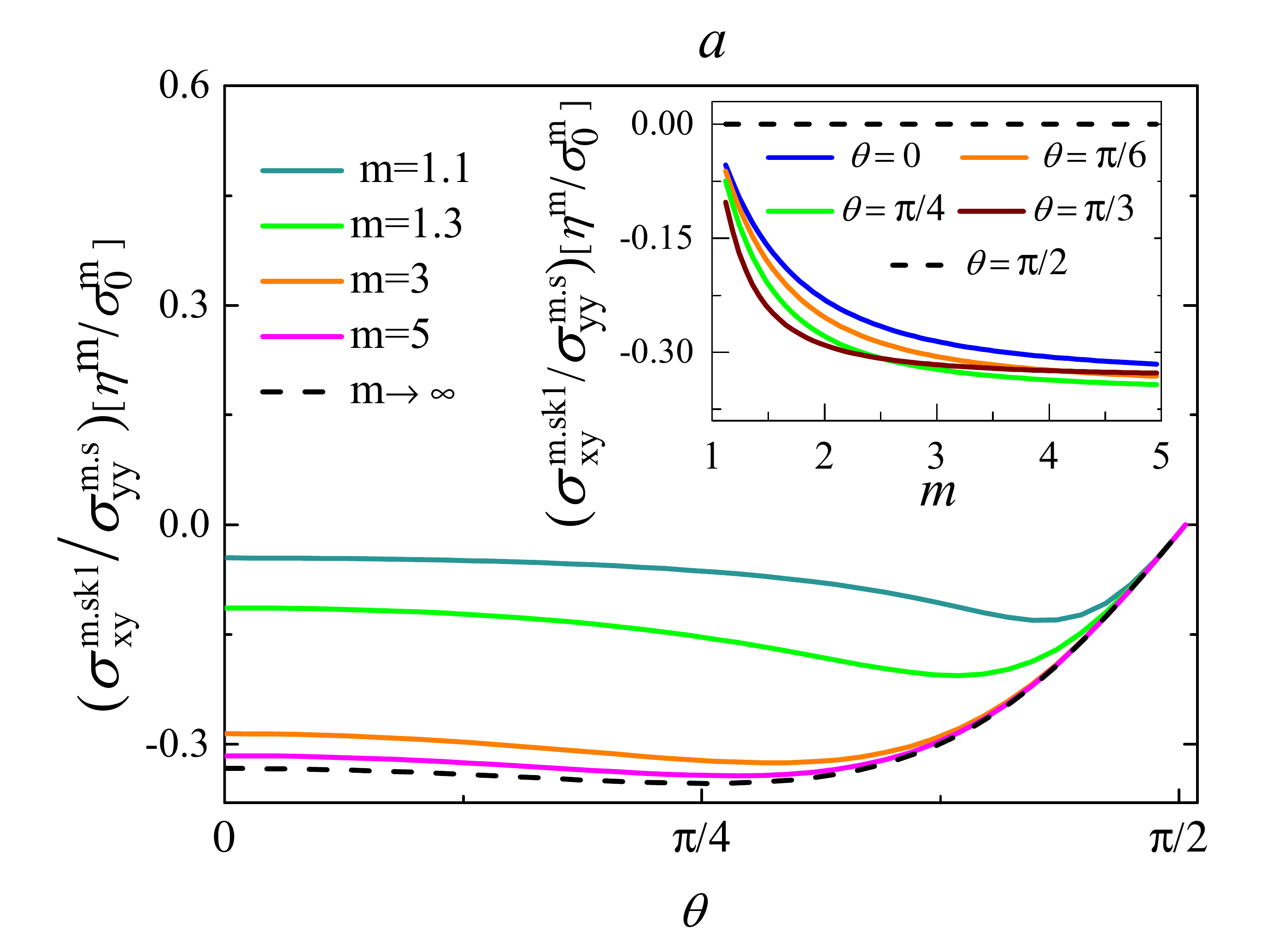}	\label{fig:msk1xyy} }\\
	\subfloat{\includegraphics[width=0.46\textwidth,center]{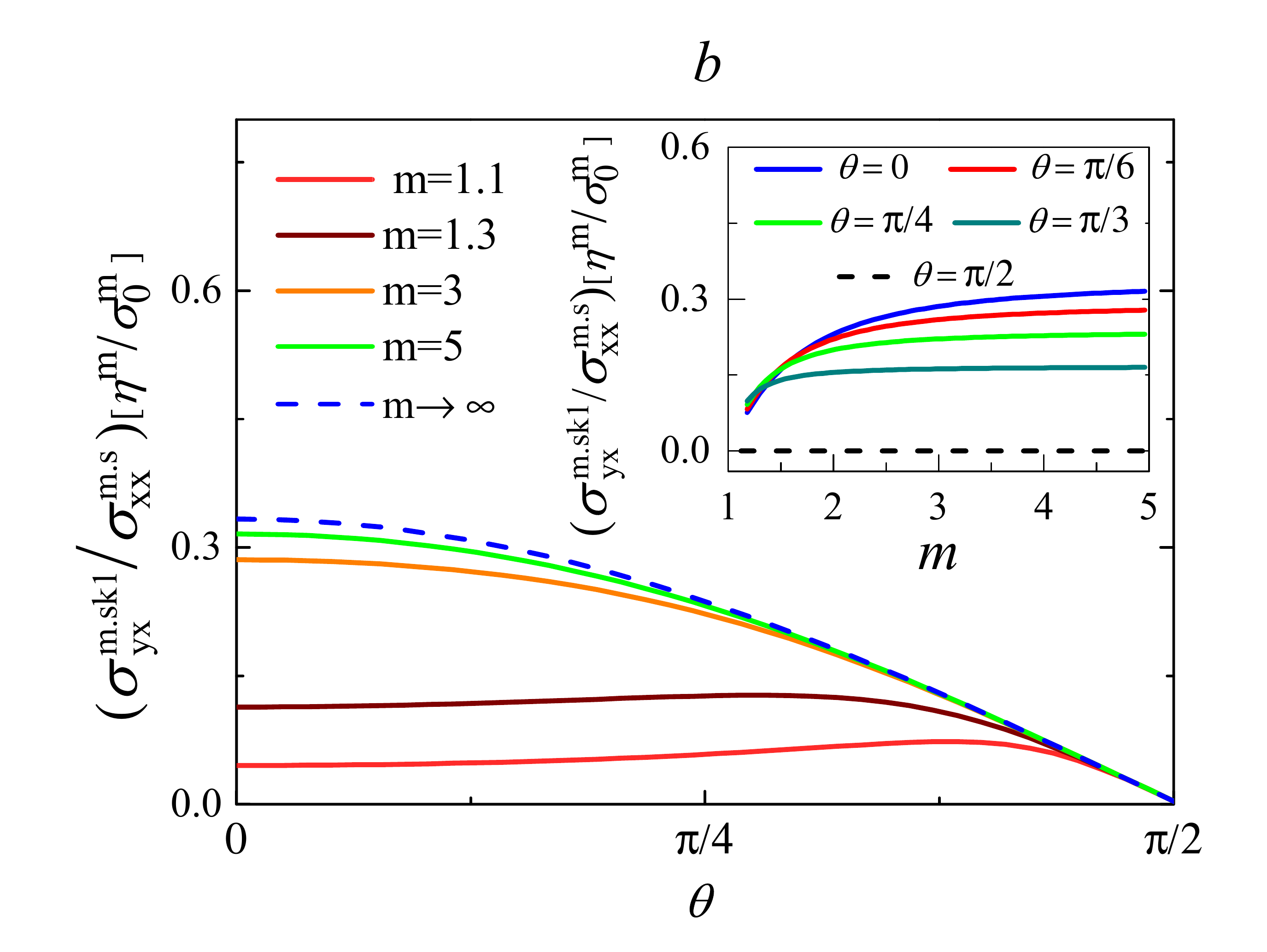}\label{fig:msk1yxxx}}
	\caption{$\sigma_{xy}^{m.sk1}$ / $\sigma_{yy}^{m.s}[(\sigma^{m.s}_{0})^{-1}]$ is plotted in terms of $\theta$ for some different values of $m$ in the main window of panel $a$, and against $m$ for some different values of $\theta$ in the inset of this panel.  $\sigma_{yx}^{m.sk1}$ / $\sigma_{xx}^{m.s}[(\sigma^{m.s}_{0})^{-1}]$ is plotted in panle $b$ for the same choice of parameters $\theta$ and \textit{m}. \label{fig:sklong}}
\end{figure}

Above we compared the side jump contribution with its corresponding longitudinal conductivity, here Fig.~\ref{fig:msk1xyy} and Fig.~\ref{fig:msk1yxxx} are provided to discuss how the ratio $\sigma_{ij}^{m.sk1}$/$\sigma_{ii}^{m.s}$ behaves with respect to $\theta$ (in the main window) and $m$ (in the inset). As is clear from panel $a$, the ratio $\sigma_{xy}^{m.sk1}$/$\sigma_{yy}^{m.s}[\sigma^{m}_{0}]$ is almost constant within the interval $ 0\leq\theta\leq \frac{\pi}{4}$ for all values of $m$. In the interval $\frac{\pi}{4}\leq\theta\leq \frac{\pi}{2}$, the absolute value of this ratio undergoes a significant decrease with increasing $\theta$, and finally becomes zero for $\theta=\frac{\pi}{2}$. The distinct behavior of this ratio within the interval $ 0\leq\theta\leq \frac{\pi}{4}$ and out of this interval can be traced back to the trends of the numerator and denominator in these two intervals. Since the numerator and denominator are both very smooth with respect to $\theta$ in the interval $0\leq\theta\leq \frac{\pi}{4}$, the ratio $\sigma_{xy}^{m.sk1}$/$\sigma_{yy}^{m.s}$ is almost constant. However, out of this interval, the absolute value of the numerator decreases while the denominator increases, leading to a decreasing behavior with respect to $\theta$ beyond $\theta=\frac{\pi}{4}$. We observe that aligning the surface magnetization at $\theta=\frac{\pi}{4}$ and simultaneously pushing the system into a fully metallic regime ($m\rightarrow \infty$), will lead to the highest possible value for $\sigma_{xy}^{m.sk1}$/$\sigma_{yy}^{m.s}[\sigma^{m}_{0}]$. In the inset of this panel, this ratio is shown in terms of $m$ for some $\theta$ values. In panel $b$, $\sigma_{yx}^{m.sk1}$/$\sigma_{xx}^{m.s}[\sigma^{m}_{0}]$ is plotted. In general, $\sigma_{yx}^{m.sk1}$/$\sigma_{xx}^{m.s}[\sigma^{m}_{0}]$ follows the same trend as the absolute value of the ratio in panel $a$, only behaving slightly different in the interval $0\leq\theta\leq \frac{\pi}{4}$. This difference is rooted in the fact that here the numerator is again smooth  with respect to $\theta$, but the denominator varies rapidly in this interval. Hence this ratio is not as smooth as in panel $a$. Also, the value of $\theta$ at which these ratios obtain their maximal value is different. In panel $b$ it is reached for $\theta=0$.

Repeating the same calculations for the skew scattering contribution due to scattering of non-magnetic impurities, we obtain the following expression for the conductivity
 \begin{equation}\label{eq:36}
  \sigma^{nm.sk1}_{yx}=-\sigma^{nm.sk}_{xy}=\eta^{nm}\frac{(m^{2}-1)^{2}}{m (m^{2}+3)(5+3m^{2})},
 \end{equation}
with $\eta^{nm}=\frac{\mu}{n_{inm} V_{0}}$. Like all the previously discussed contributions, this contribution is zero for an insulating surface $m=1$. In contrast to $\sigma^{m.sk1}_{ij}$, this contribution is absent in the perfect metallic regime $m\rightarrow  \infty$.

Again these transverse conductivities can be related to the corresponding longitudinal conductivities. We obtain
\begin{equation}\label{eg:40}
\sigma^{m.sk1}_{yx}=-\sigma^{m.sk1}_{xy}=\eta^{m} ~\dfrac{(\tilde{\sigma}^{m.s}_{xx})^{2}\cos\theta}{1+2~\tilde{\sigma}^{m.s}_{xx} \sin^{2}\theta},
\end{equation}
\begin{equation}\label{eg:41}
\sigma^{nm.sk1}_{yx}=-\sigma^{nm.sk1}_{xy}=\eta^{nm}~\dfrac{(\tilde{\sigma}^{nm.s}_{xx})(m^{2}-1)}{m(5+3m^{2})}.
\end{equation}

\begin{figure}
		\centering
		{\includegraphics[width=0.48\textwidth,center]{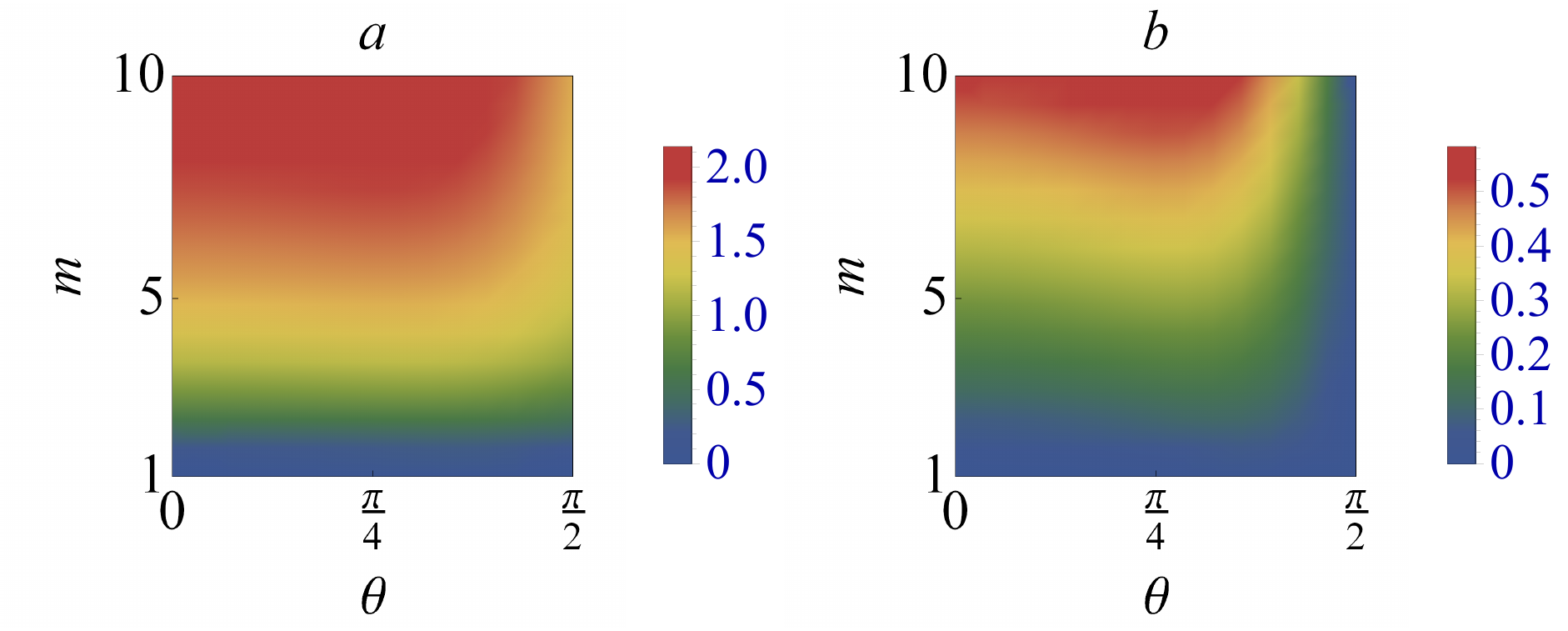}}
	
		\caption{$\sigma_{yx}^{tot.sk1}[\eta^{nm}]$ is plotted in terms of $\theta$ and \textit{m} for $\nu=0.1$ and $\nu=100$ in panel $a$ and $b$, respectively.
		\label{fig:8} }
\end{figure}

The total conventional skew scattering conductivity is now given by
\begin{equation}
	\sigma^{tot.sk1}_{yx}=\eta^{nm}\left[\dfrac{\nu (\tilde{\sigma}^{m.s}_{xx})^{2}\cos\theta}{1+2~\tilde{\sigma}^{m.s}_{xx} \sin^{2}\theta}+ \dfrac{(\tilde{\sigma}^{nm.s}_{xx})(m^{2}-1)}{m(5+3m^{2})}\right],
\end{equation}
in which $\sigma^{tot.sk1}_{ij}=-\sigma^{tot.sk1}_{ji}=\sigma^{m.sk1}_{ij}+\sigma^{nm.sk1}_{ij}$ is expressed (like previous expressions) in unit of $\frac{e^{2}}{h}$ and as a function of  $\nu=\frac{\eta^{m}}{\eta^{nm}}$. Since this contribution is inversely proportional to concentration of present impurities, it will dominate the anomalous hall conductivity in a very dilute regime. However, in a doped system with a low concentration of just magnetic impurities, one can turn off this contribution by simply setting $\theta=\frac{\pi}{2}$. By considering $n_{im}=n_{inm}$, Fig.~\ref{fig:8} shows $\sigma^{ tot.sk1}_{yx}[\eta^{nm}]$ in terms of $\theta$ and $m$ for two different values of $\nu=0.1$ and $\nu=100$ in panels $a$ and $b$, respectively. It shows that increasing $m$ increases $\sigma^{tot.sk1}_{yx}$, while increasing $\nu$ decreases this conductivity. In addition, this contribution becomes insignificant for small values of $m$, independent of the value of $\nu$ and $\theta$. 

Also note that, in contrary to the total contribution of the side jump effect to the AHE, the total contribution of the conventional skew scattering never changes its sign, whatever the value of $\theta$, $m$ or $\bm E$.

\subsubsection{Intrinsic Skew Scattering}
Let us now consider the contribution $\Jv^{sk2}=-e\sum_{\bm k}g^{a2}_{\bm k}  \vv_{0\bm k}$ to the skew scattering. As $g^{a2}_{\bm k} \sim n_{im}^0$, this contribution is independent of the impurity concentration, and therefore also called intrinsic. In order to find the contribution of the intrinsic skew scattering in the conductivity of the system, we first need to calculate $w^{(4)}_{\bm k \bm k'}$ through Eq.~(\ref{w4}) and then by solving Eq.~(\ref{eq:ga2}) we find $g^{a2}_{\bm k}$. Solving these equations separately for electrons scattering off magnetic and non-magnetic impurities, we obtain
\begin{equation}
\begin{split}
w^{(4.m)}_{\bm k \bm k'}=& [w^{(4)}_{0} \gamma  (\gamma^{2}-1)\cos 2\theta (\sin\phi_{\bm k}\cos\phi_{\bm k'}-\sin[\phi_{\bm k'}-\phi_{\bm k}])\\
&-w^{(4)}_{0} \gamma (\gamma^{2}-1)(\sin[\phi_{\bm k'}-\phi_{\bm k}]+\sin\phi_{\bm k'}\cos\phi_{\bm k})+\\w^{(4)}_{0}\gamma^{2} &\sqrt{1-\gamma^{2}}\sin 2\theta(2\sin\phi_{\bm k}-\sin\phi_{\bm k'})]\delta(\varepsilon_{k}-\varepsilon_{k'}),
\end{split}
\end{equation}
\begin{equation}
w^{(4.nm)}_{\bm k\bm k'}=\frac{3 \pi( n_{inm} V_{0}^{2})^{2}}{4 \hbar}\frac{M ~\textit{k}^{2} }{\varepsilon^{3}_{\textit{k}}}\delta(\varepsilon_{\textit{k}}-\varepsilon_{\textit{k}'})\sin(\phi_{\bm k}-\phi_{\bm k'}),
\end{equation}
with $w^{(4)}_{0}=\dfrac{\pi( n_{im} J^{2} S^{2}_{m})^{2}     }{4 \hbar ^{3} v^{2}_{F}}$. Inserting these scattering rates in Eq.~(\ref{eq:ga2}) and using the already found $w^{(2.m)}_{\bm k \bm k '}$ and $g^{s}_{\bm k }$ expressions, we obtain the following transversal conductivities associated to scattering off magnetic and non-magnetic impurities
 \begin{widetext}
 
 \begin{equation}\label{sk2mxy}
\sigma^{m.sk2}_{xy}=\frac{16 (g-2) [m^2-1] \left(-\cos 4 \theta +(m^2+2) \cos 2 \theta +2 m^2+2\right)}{m \left([g+2] \cos 2 \theta +[g+4] m^2-2\right) \left(g \cos 4 \theta+2 [g-2] m^2 \cos 2 \theta +g-4\right)},
\end{equation}

\begin{equation}\label{sk2myx}
\sigma^{m.sk2}_{yx}=\frac{4 (g-2) \left(b \cos 2 \theta -g [2 (g+1) m^4+g-2]-[g-2] [g+2 m^2] \cos 4 \theta +8 m^4-4 m^2+8\right)}{g~ m \left([g+2 m^2] \cos 2 \theta +[g+2] m^2\right) \left([g-2][ \cos 2 \theta + m^2]+\cos 4 \theta -1\right)},
\end{equation}

\begin{equation}\label{sk2nm}
\sigma^{nm.sk2}_{xy}=-\sigma^{nm.a2}_{yx}=-\frac{3}{2}\frac{(m^{2}-1)^{2}}{m(m^{2}+3)^{2}},
\end{equation}
\end{widetext}
with $b=2 [(-2 g^2+g+8) m^2-2 g m^4+g]$. Note that $\sigma^{m.sk2}_{ij}=0$ for $m=1$ (the insulating regime) or in the limit of large $m$ (the perfect metallic regime). If we align all the surface magnetic impurities perpendicular to the surface of the TI ($\theta=0$), the corresponding conductivity is isotropic (ignoring the difference in sign) $\sigma^{m.sk2}_{xy}[\theta=0]=-\sigma^{m.sk2}_{yx}[\theta=0]=\dfrac{12 \left(m^2-1\right)^2}{m \left(3 m^2+1\right)^2}$. Also in case of  $\theta=\dfrac{\pi}{2}$ this intrinsic contribution of the skew scattering is isotropic  $\sigma^{m.sk2}_{xy}[\theta=\dfrac{\pi}{2}]=\sigma^{m.sk2}_{yx}[\theta=\dfrac{\pi}{2}]=\dfrac{4}{3~ m}$.

\begin{figure}
	\subfloat{%
		\includegraphics[width=0.46\textwidth,center]{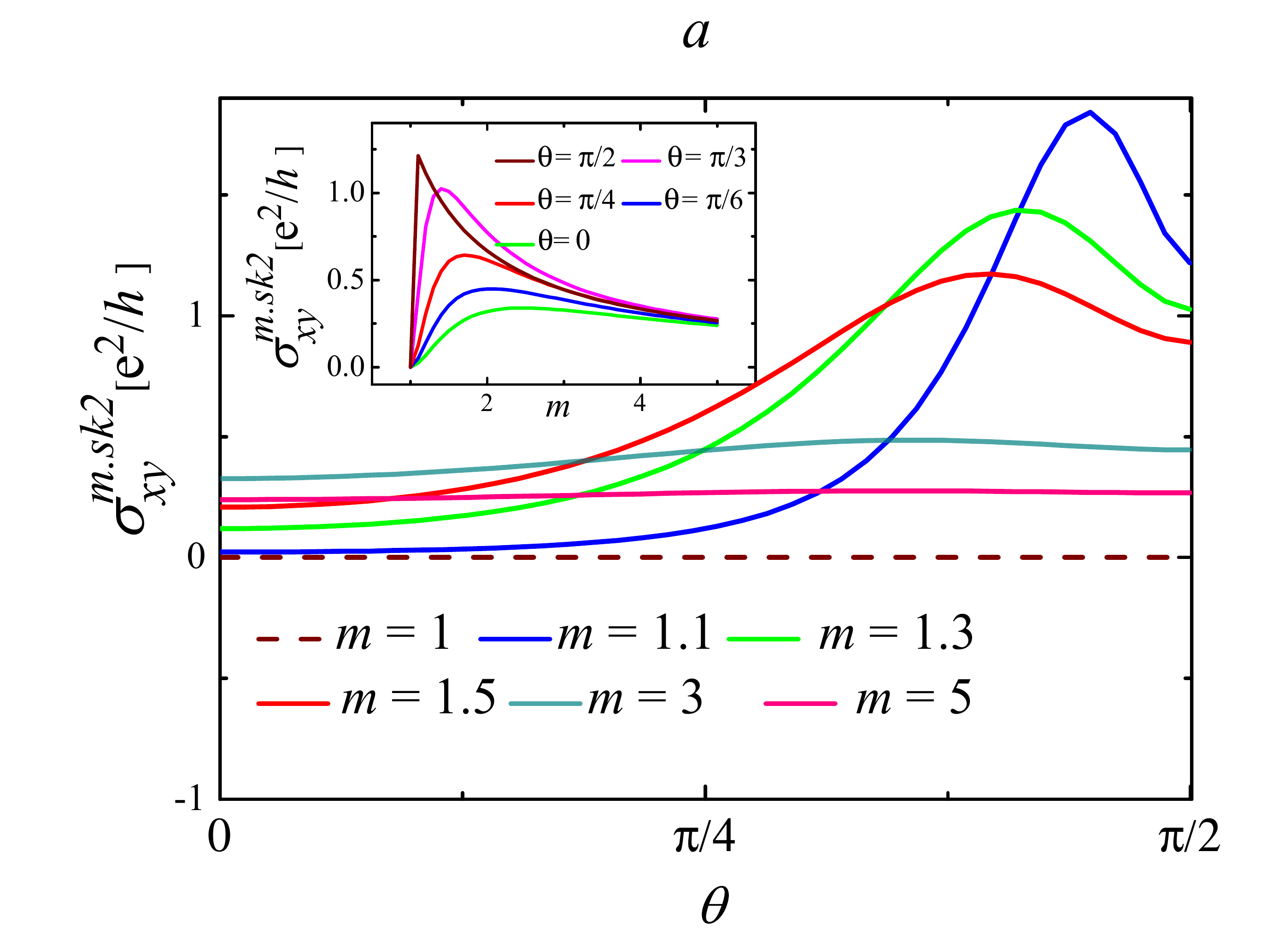}}\\
		\subfloat{%
		\includegraphics[width=0.436\textwidth,center]{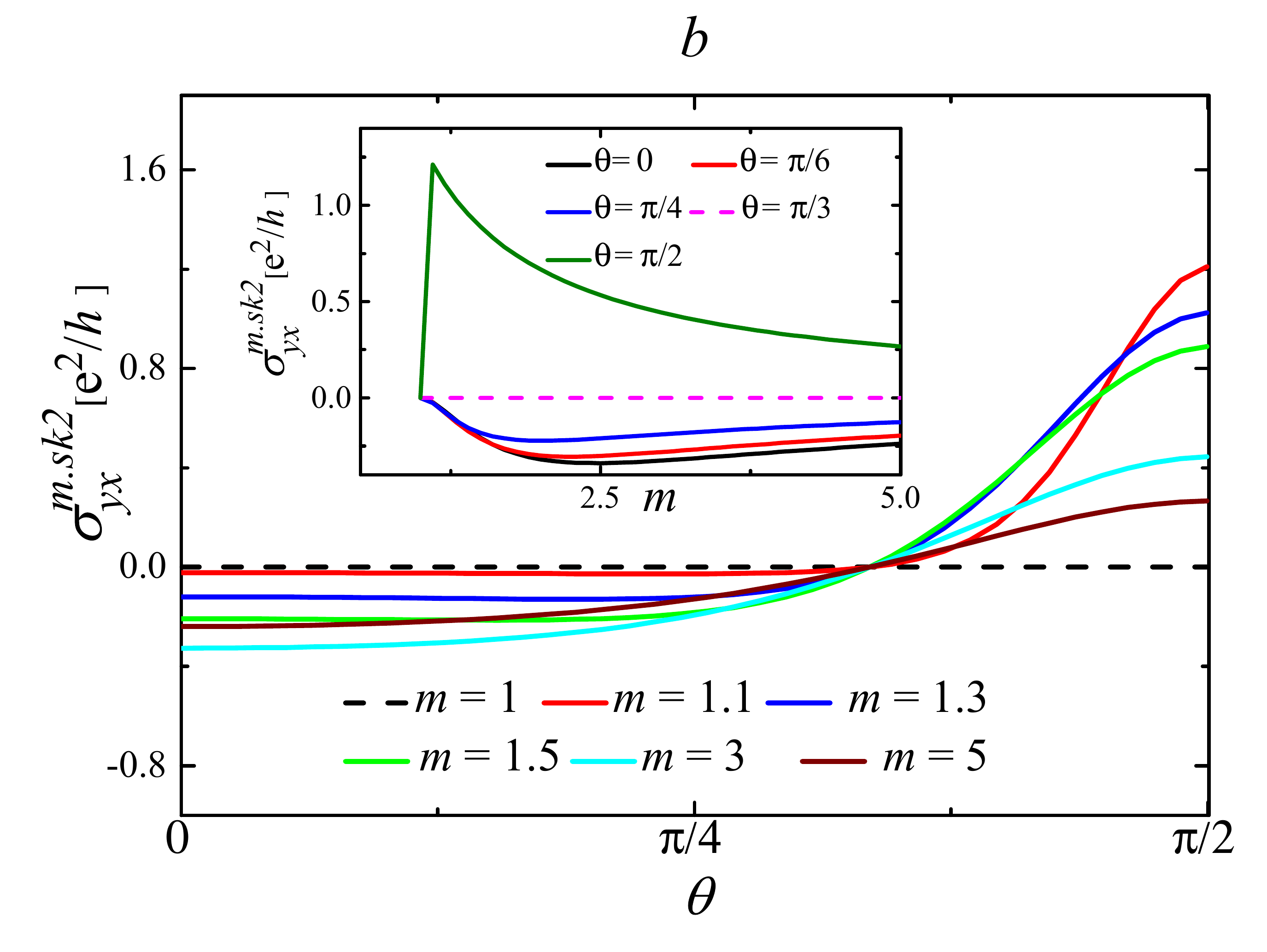}}
	\caption{$\sigma_{xy}^{m.sk2}$ is plotted in terms of $\theta$ for some different values of $m$ in the main window of panel $a$, and against $m$ for some different values of $\theta$ in the inset of this panel.  $\sigma_{yx}^{m.sk2}$ is plotted in panel $b$ for the same choice of parameters $\theta$ and \textit{m}. \label{fig:sk2}} 
\end{figure}

To study $\sigma^{m.sk2}_{ij}$, Eq.~(\ref{sk2mxy}) and Eq.~(\ref{sk2myx}) are illustrated in panel $a$ and $b$ of Fig.~\ref{fig:sk2}, respectively. As it is clear from panel $a$, $\sigma^{m.sk2}_{xy}$ increases with increasing $\theta$ for all values of $m$, and reaches a maximum in the interval $ \frac{\pi}{4}<\theta<\frac{\pi}{2}$, just like the first contribution to the skew scattering $\sigma^{m.sk1}_{xy}$. However, in contrast to this first contribution, $\sigma^{m.sk2}_{xy}$ does not turn off at $\theta=\frac{\pi}{2}$. The inset of panel $a$ shows $\sigma^{m.sk2}_{xy}$ as function of $m$ for some values of $\theta$. Each curve in this inset shows a maximum value in the interval $1\leq\textit{m}\leq 2$, followed by a sharp decrease. In contrast to the first skew scattering contribution $\sigma^{m.sk1}_{xy}$ which does not disappear at large values of $m$, all curves in the inset of panel $a$ approach zero in the limit $m\rightarrow\infty$. Panel $b$ of this figure shows  $\sigma^{m.sk2}_{yx}$ as function of $\theta$ for some values of $\textit{m}$. The inset of this panel shows $\sigma^{m.sk2}_{yx}$ as function of $\textit{m}$ for some values of $\theta$. Surprisingly $\sigma^{m.sk2}_{yx}$ is negative for $0 \leq\theta < \frac{\pi}{3}$ and has a positive value for $\frac{\pi}{3}<\theta\leq  \frac{\pi}{2}$. In contrast to first contribution of skew scattering to the AHE, this contribution changes sign if one changes the spatial orientation of the surface magnetization from $0$ to $\frac{\pi}{2}$. The inset of this figure shows that, for all given values of $\theta$, $|\sigma^{m.sk2}_{yx}|$ starts from zero at $m=1$, then increases till it reaches a maximum value within the interval $1\leq m \leq 2$, after which it decreases. Also in agreement with the main window of panel $b$ which shows that the conductivity at $\theta=\frac{\pi}{2}$ is positive for all values of $m$, the green curve in the inset is positive for all values of $\textit{m}$.

As we indicated in previous section, the conductivity $\sigma^{m.sk1}_{ij}$ can be turned off by rotating the surface magnetization to lie on the surface of the TI. However, this orientation of the magnetization cannot turn off $\sigma^{m.sk2}_{xy}$, but it can turned off by increasing the Fermi level. Furthermore, in case the external electric field is directed in the $\hat{x}$ direction, $\sigma^{m.sk2}_{yx}$ can be turned off for $\theta=\frac{\pi}{3}$. 

\begin{figure}
	\centering
	\subfloat{\includegraphics[width=0.4\textwidth,center]{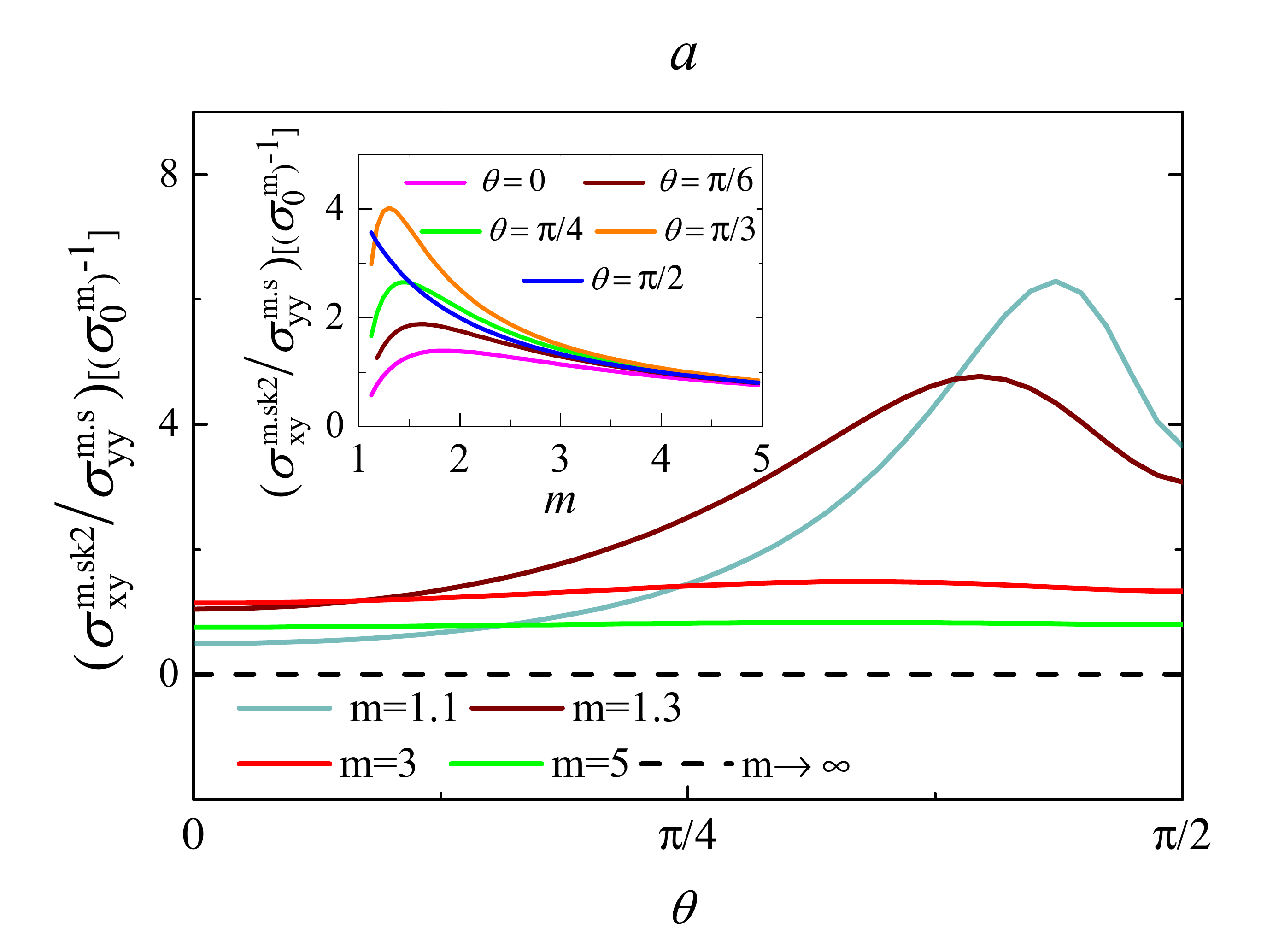}	\label{fig:msk2xyyy} }\\
	\subfloat{\includegraphics[width=0.42\textwidth,center]{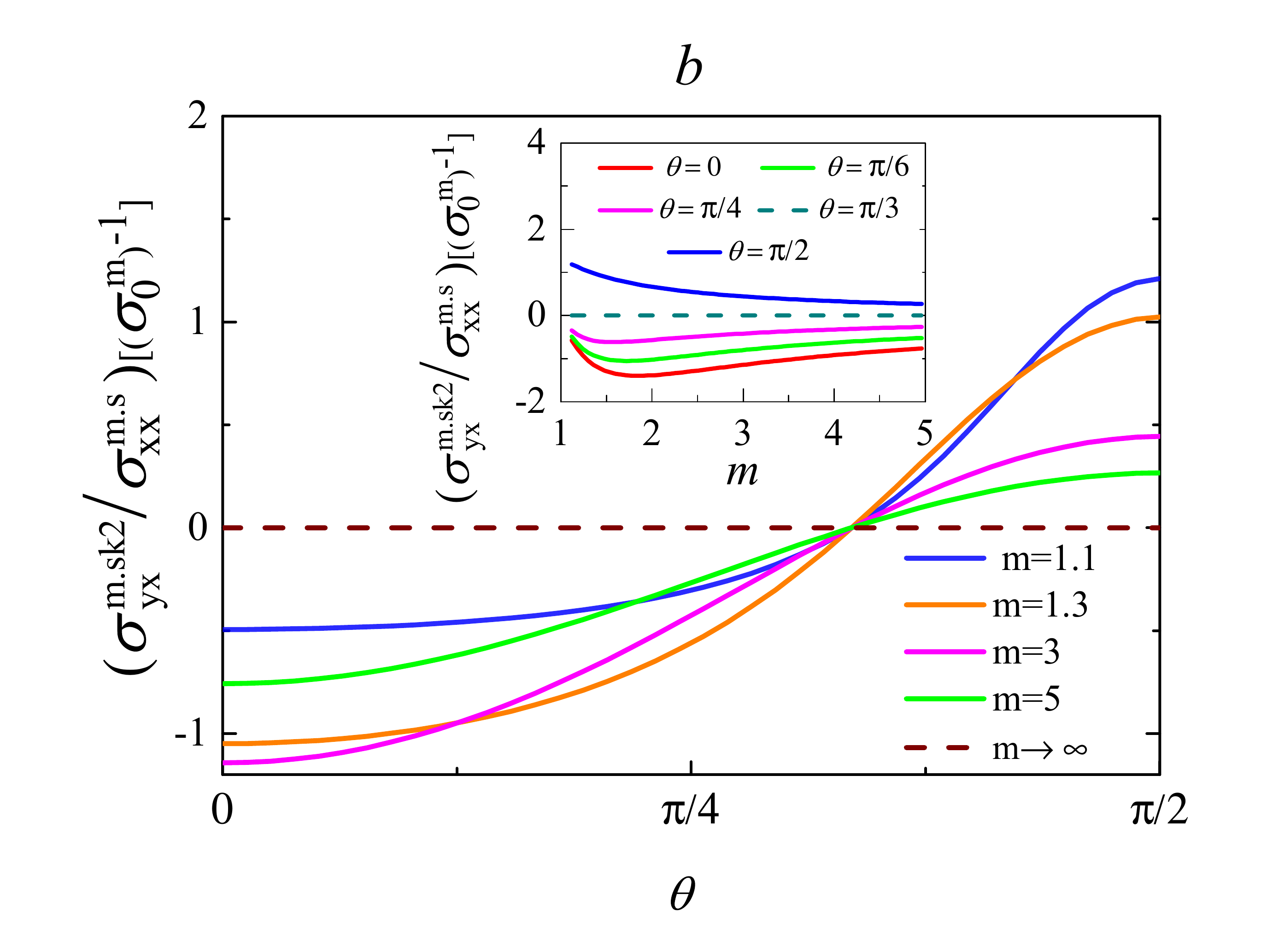}\label{fig:msk2yxxx}}
	\caption{$\sigma_{xy}^{m.sk2}$ / $\sigma_{yy}^{m.s}[(\sigma^{m.s}_{0})^{-1}]$ is plotted in terms of $\theta$ for some different values of $m$ in the main window of panel $a$, and against $m$ for some different values of $\theta$ in the inset of this panel.  $\sigma_{yx}^{m.sk2}$ / $\sigma_{xx}^{to.m.s}[(\sigma^{m.s}_{0})^{-1}]$ is plotted in panel $b$ for the same choice of parameters $\theta$ and \textit{m}.  \label{fig:sjlong}} 
\end{figure}

In the following, in order to compare this contribution with the corresponding longitudinal conductivity, the two ratios $\sigma^{m.sk2}_{xy}$/$\sigma^{m.s}_{yy}$ and $\sigma^{m.sk2}_{yx}$/$\sigma^{m.s}_{xx}$ are plotted in Fig.~\ref{fig:msk2xyyy} and Fig.~\ref{fig:msk2yxxx}, respectively. In panel $a$, $\sigma^{m.sk2}_{xy}$ / $\sigma^{m.s}_{yy}$ shows a negligible change within $0\leq\theta\leq\frac{\pi}{2}$, for $m\geq3$. However, putting the chemical potential just above the bottom of the conduction band ($m \sim$ 1.1) and aligning the surface magnetization around $\theta=0.4 \pi$ realizes the maximum value for this ratio. In panel $b$, $\sigma^{m.sk2}_{yx}$ /$\sigma^{m.s}_{xx}$ is plotted versus $\theta$ and $m$. As expected this ratio changes its sign if we tune the surface magnetization around $\theta=\frac{\pi}{3}$. By altering the spatial orientation of the surface magnetization from $\theta=0$ to $\theta=\frac{\pi}{3}$ the absolute value of this ratio decreases, while after passing through $\theta=\frac{\pi}{3}$ it grows. For all $\theta$ values, this ratio reaches its maximum value by tunning the chemical potential just above the bottom of the conduction band. Besides, by pushing the system into the fully metallic regime, this ratio goes to zero. In contrary to the already plotted ratios associated to the first contribution of the skew scattering, these two plotted ratios clearly follow different trends. This distinction originates from the anisotropy in $\sigma^{m.sk2}_{ij}$ and $\sigma^{m.s}_{ii}$.

Finally, we consider the contribution due to the spin independent intrinsic skew scattering to the AHE, given in Eq.~(\ref{sk2nm}). This expression indicates that this conductivity is isotropic and also like $\sigma^{m.sk2}_{ij}$ it disappears for large values of $m$.
By obtaining the two components of the intrinsic skew scattering related condunctivitis $\sigma_{ij}^{m.sk2}$ and $\sigma_{ij}^{nm.sk2}$, we have found $\sigma_{ij}^{tot.sk2}=\sigma^{m.sk2}_{ij}+\sigma^{nm.sk2}_{ij}$. 

In case that the external electric field is exerted along the $\hat{y}$-direction and the surface magnetization is aligned perpendicular to the surface of the TI, we find  $\sigma_{xy}^{tot.sk2}(\theta=0)=-\frac{3 \left(m^2-1\right)^2 \left(m^4-42 m^2-71\right)}{2 m \left(3 m^4+10 m^2+3\right)^2}$. In the other interesting situation that the external electric field is along the same direction but the surface magnetization lies on the surface of the TI, one can show  that $\sigma_{xy}^{tot.sk2}(\theta=\frac{\pi}{2})=\frac{-m^4+66 m^2+63}{6 m \left(m^2+3\right)^2} $. In addition, one can show that $\sigma_{xy}^{tot.sk2}(\textit{m}=1)=0$ and $\sigma_{xy}^{tot.sk2}(\textit{m}\rightarrow\infty)=0$.

Now, if one changes the direction of the external electric field from $\hat{y}$ to $\hat{x}$ and aligns the surface magnetization again perpendicular to the surface of the TI, one can show that $\sigma_{yx}^{tot.sk2}(\theta=0)= \frac{3 \left(m^2-1\right)^2 \left(m^4-42 m^2-71\right)}{2 m \left(m^2+3\right)^2 \left(3 m^2+1\right)^2}$. In the other interesting situation that the external electric field is still in the $\hat{x}$-direction, but the surface magnetization lies on the surface of the TI, $\sigma_{yx}^{tot.sk2}(\theta=\frac{\pi}{2})= \frac{17 m^4+30 m^2+81}{6 m \left(m^2+3\right)^2}$. Also, like $\sigma_{xy}^{tot.sk2}$,  the conductivity $\sigma_{yx}^{tot.sk2}$ vanishes at the insulator and perfect metallic condition. While $\sigma^{m.sk2}_{ij}$ behaves isotropically with respect to the external electric field direction, not only at $\theta=0$ but also at $\theta=\frac{\pi}{2}$, $\sigma_{ij}^{tot.sk2}$ is isotropic just at $\theta=0$. 
  
\begin{figure}

	\includegraphics[width=0.34\textwidth,center]{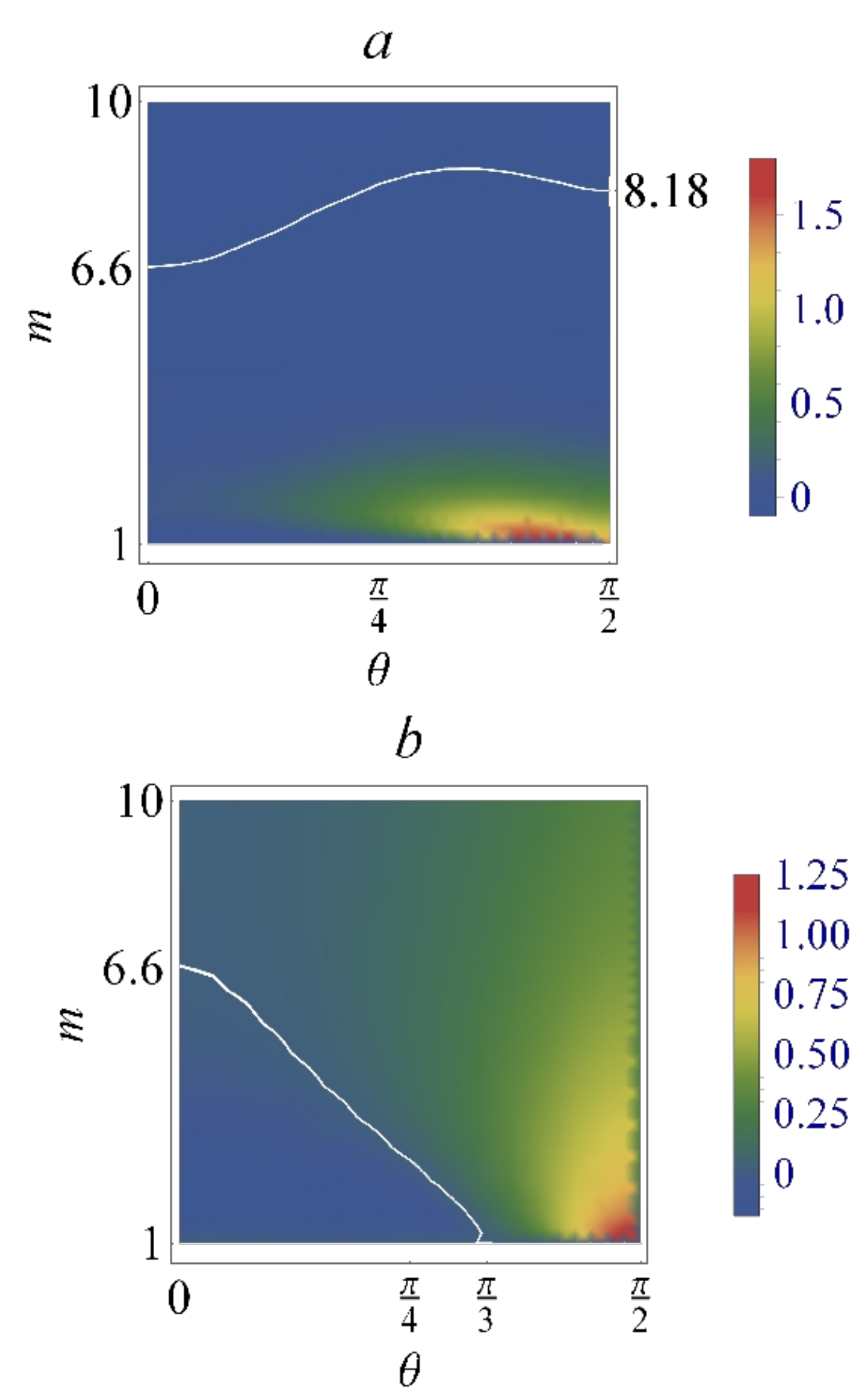}
	\caption{$\sigma_{xy}^{tot.sk2}$ and $\sigma_{yx}^{tot.sk2}$ as functions of $\theta$ and $m$, are plotted in panel $a$ and $b$,  respectively. 
		\label{fig:10} }
\end{figure}

$\sigma_{xy}^{tot.sk2}$ and $\sigma_{yx}^{tot.sk2}$ are illustrated respectively in panel $a$ and $b$ of Fig.~\ref{fig:10}, in terms of $\theta$ and $m$. The white lines in these two panels specify the $(\theta, m)$ combinations for which the corresponding conductivity is zero. %As it can be seen from panel $a$ of this figure, $\sigma^{tot.sk2}_{xy}$ is positive \textcolor{green}{for all those ($\theta$, $m$) located }below the red line, and has a negative value \textcolor{green}{for those ($\theta$,m) which are placed beyond} the red line. In panel b, \textcolor{green}{as the legend of this plot indicates, like panel a there is a change of sign from positive values of $\sigma^{tot.sk2}_{yx}$ to negative values, if we approach the red line from large value of $m$ and cross the red line.}
Panel $a$ shows that, for $1 <m \lesssim 2$,  $\sigma^{tot.sk2}_{xy}$ undergoes a considerable change with respect to $\theta$. Putting the chemical potential just above the gap and also aligning the surface magnetization close to the surface of the TI ($\frac{\pi}{3} \lesssim \theta  \lesssim \frac{\pi}{2} $) causes the system to reach its maximum value for $\sigma^{tot.sk2}_{xy}$. %Surprisingly there are two ways to turn off $\sigma^{tot.sk2}_{xy}$ through adjusting $\theta$ and $m$. Firstly, one can  place surface magnetization at $\theta=0$, then tune chemical potential to reach $\textit{m}\sim 6.6$. Secondly via changing spatial orientation of surface magnetization to lie at $\theta=\frac{\pi}{2}$ and thereafter alter chemical potential to reach $m=8.18$. About $\sigma^{tot.sk2}_{yx}$, also by choosing ($\theta=0$, m=6.6) or ($\theta=\frac{\pi}{3}$, m=1) one can easily turn off this contribution. 
It is also clear from panel $a$ that there is just a small $(\theta, m)$ region with a significant  $\sigma^{tot.sk2}_{xy}$, while in panel $b$ a broad region shows a significant conductivity $\sigma^{tot.sk2}_{yx}$. %In panel $a$, we consider the conductivity of a system in which the external electric field is applied along the $\hat{y}$ direction. So the electrons move approximately parallel to the $\hat{y}$ direction. Consequently, due to presence of the spin orbit coupling, the averaged spin of the conductive electrons in this system would be along the $\hat{x}$ direction. Hence, the spin of these electrons is always perpendicular to the plane of surface magnetization (which lies in the $z-y$ plane). \textcolor{red}{[B: The followingsentence is not complete.]} \textcolor{blue}{[A:done]} But the conductivity $\sigma^{tot.sk2}_{yx}$ shown in panel $b$, corresponds to the system in which the spin of the conducting electrons lies in the plane of the surface magnetization. Due to this underlying difference, $\sigma^{tot.sk2}_{xy}$ differs so strongly from $\sigma^{tot.sk2}_{yx}$.%
   
\subsection{Total conductivity}\label{subsection:total}

To obtain the total value of the anomalous hall conductivity in this system, we need to add up all contributions derived in the previous sections. The total conductivity can be written as an intrinsic part and an extrinsic part 
\begin{equation}
 \sigma^{tot}_{ij}=\sigma^{tot.int}_{ij}+\sigma^{tot.ext}_{ij}
 \end{equation}
where $\sigma^{tot.ext}_{ij}$ is extrinsic and depends on the concentration of impurities, and the intrinsic part $\sigma^{tot.int}_{ij}$ that is independent of the concentration of presentt impurities.   Based on our results we can write $\sigma^{tot.ext}_{ij}=\sigma^{tot.sk1}_{ij}$ and $\sigma^{tot.int}_{ij}= \sigma^{in}_{ij}+\sigma^{tot.sj}_{ij}+\sigma^{tot.sk2}_{ij}$.
In this section we will discuss the AHE in three distinct regimes. First, we assume that the surface of the TI is lightly doped and thus that the intrinsic term dominates the AHE. Secondly, we assume that the system is completely out of the first regime, hence that the extrinsic term dominates the anomalous hall conductivity. Finally, we assume that the system is in the intermediate regime, where we need both the intrinsic and extrinsic terms to properly study the AHE on the surfsce of the system. 
  
\subsubsection{The extrinsic regime}
  
\begin{figure}[htb]
		\includegraphics[height=0.3\textwidth ,width=0.4\textwidth, center]{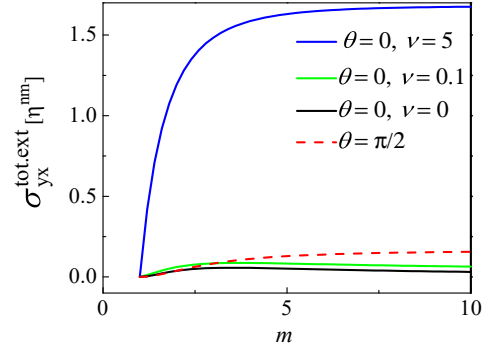}
		\caption{$\sigma_{yx}^{tot.ext}[\eta _{nm}]$ is plotted at $\theta=0$ (solid lines) and $\frac{\pi}{2}$(dashed line) as function of $m$ for some values of $\nu$. 
		\label{fig:144} }
\end{figure}

As we already indicated, the only contribution in the extrinsic regime is $\sigma^{tot.sk1}_{ij}$ which is inversely proportional to the concentration of impurities. Since we already devoted Sec.~\ref{subsec:conventional} to this contribution, we skip over the details and just describe some general important observations. In the absence of the non-magnetic impurity, the resultant magnetic conventional skew scattering varies between 
$0~\lesssim \lvert \sigma^{m.sk1}_{ij}[m, \theta]\lvert \lesssim\frac{  0.12 ~\mu}{J S_{m} n_{im}}$.
The minimum value occurs for systems which have a chemical potential very close to the bottom of the conduction band, regardless of $\theta$, or have an in-plane magnetization, regardless of $m$. The maximum value of this magnetic contribution is reached in those systems which are in a fully metallic regime with $\theta=\frac{\pi}{4}$. In the absence of magnetic impurities, this term changes within $0\lesssim\lvert \sigma^{nm.sk1}_{ij}[m]\lvert \lesssim \frac{ 0.056~\mu}{V_{0} n_{inm}}~$. The minimum value of this non-magnetic contribution of the conventional skew scattering term is obtained in those systems with the chemical potential just above the bottom of the conduction band, and the maximum value is reached in a system with $m=3.5$. In presence of all types of impurities, the total value of this contribution depends not only on the mass of the Dirac fermions, the spatial orientation of the surface magnetization and the concentration of the present impurities, but also on the strength of the magnetic and non-magnetic scattering of itinerant electrons during their skew scattering. There is some experimental and theoretical evidences that the surface magnetization of a TI is preferentially orientated in the plane of the surface or perpendicular to it~\cite{signchangeex, rosenberg_prb_2012}. Therefore we continue our discussion by focusing our attention to the cases $\theta=0$ and $\frac{\pi}{2}$. For these orientations, the total extrinsic anomalous hall conductivity is given by

\begin{equation}
\sigma^{tot.ext}_{xy}[\eta^{nm}]_{\theta=0}=\frac{\nu-\nu m^2}{3 m^2+1}-\frac{m^4-2 m^2+1}{3 m^5+14m^3+15 m},
\end{equation}

\begin{equation}
\sigma^{tot.ext}_{xy}[\eta^{nm}]_{\theta=\frac{\pi}{2}}=\frac{-\left(m^2-1\right)^2}{2 \left(3 m^4+14 m^2+15\right)}.
\end{equation}
Fig.~\ref{fig:144} illustrates above expressions (solid lines correspond to $\theta=0$ and dashed line to $\theta=\frac{\pi}{2}$) for different values of $\nu=\frac{\eta^{m}}{\eta^{nm}}$. As it is clear from this figure, for a system with a magnetization perpendicular to the surface of Ti, $\sigma_{yx}^{tot.ext}$ does not vary much against $m$ for small values of $\nu$. Howvever, for large $\nu$ values, it undergoes a large change with respect to $m$.  

\subsubsection{The intrinsic regime}

In this regime the total anomalous hall conductivity does not depend on the concentration of the impurities and contains three contributions: the intrinsic conductivity arising from the non-zero Berry curvature, the total side jump conductivity and the conductivity produced by intrinsic skew scattering. Our results show that the first intrinsic contribution $\sigma^{in}_{yx}=-\sigma^{in}_{xy}$ changes within $0\leq \lvert \sigma^{in}_{ij}[m]\lvert\leq 0.5 ~$, if $1\leq m\leq\infty$.  Since this term is inversely proportional to $m=\frac{\mu}{M}$, it obviously turns off in the gaples regime ($M=0$). However, the second contribution $\sigma^{tot.sj}_{ij}=\sigma^{tot.m.sj}_{ij}+\sigma^{tot.nm.sj}_{ij}$ is a consequence of the presence of impurities and definitely does not appear in a pure system. In the absence of non-magnetic impurities, by changing the spatial orientation of the surface magnetization within $0 \leq \theta \leq \frac{\pi}{2}$ and also tuning the mass of the Dirac fermions, this contribution varies within $0~\leq\sigma^{tot.m.sj}_{ij}[ m, \theta] <~0.07$. The small value of this contribution is caused by the fact that $\sigma^{m.sj}_{ij}$ and $\sigma^{m.ad}_{ij}$ partially cancel each other in $\sigma^{tot.m.sj}_{ij}$. For example, at $\theta=0$ and $\theta=\frac{\pi}{2}$, this contribution turns off. In the presence of only non-magnetic impurities the absolute value of the corresponding conductivity is limited within $0\leq \lvert \sigma^{tot.nm.sj}_{xy}[m] \lvert \leq 0.45 $. Therefore, we can say that in the presence of both magnetic and non-magnetic impurities, we can ignore the contribution of the magnetic impurities in $\sigma^{tot.sj}_{ij}$. 

We also want to stress that, although concentration of impurities does not appear in the final expression for $\sigma^{tot.sj}_{ij}$, this contribution does originate from the present impurities. There are two parameters which play an important role in the transport of electrons during their side jump, the so called side jump relaxation times and the side jump velocity. The first one is inversely proportional to the concentration of impurities, though the second one is directly proportional to the concentration of impurities. Therefore, interestingly, their product is independent on the concentration of impurities.

The last effect which contributes to the AHE in this regime originates from the intrinsic skew scattering. In the absence of non-magnetic impurities, this term varies as $0 \leq\sigma^{m.sk2}_{xy}[m, \theta]\leq 2.2 $, which minimum value occurs for a system with $(m, \theta)=(1,\theta)$, and the maximum value for a system with $($m$, \theta)=(1.04,0.48 \pi)$. In addition, $-0.34 \leq \sigma^{m.sk2}_{yx}[m, \theta]\leq 1.33 ~$, where the minimum and maximum values are obtained in systems with $($m$, \theta)=(2.5,\simeq 0)$ and $($m$, \theta)=(\simeq1.01, \simeq \frac{\pi}{2})$, respectively. 
In the absence of magnetic impurities, the conductivity corresponding to non-magnetic intrinsic skew scattering changes as $0~\lesssim \lvert \sigma^{nm.sk2}_{ij}[m]\lvert \lesssim 0.23~$, with $(m_{min}, m_{max})=(1, \simeq 3.8)$. If both kinds of impurities are present, we obtain 
$-0.035 ~\lesssim\sigma^{tot.sk2}_{xy}[m, \theta] \lesssim 2.1 $. Pushing the system into the metallic regime and setting $\theta$ to $\frac{\pi}{2}$ leads to the minimum value of $\sigma^{tot.sk2}_{xy}$. Putting the chemical potential just above the bottom of the conduction band along with turning slightly the magnetization within $\frac{\pi}{3}\leq\theta< \frac{\pi}{2}$  leads to the maximum value for this contribution. In addition, we show that 
$-0.18 ~ \lesssim\sigma^{tot.sk2}_{yx}[m, \theta] \lesssim 1.3 $, where setting ($m$, $\theta$) to ($\simeq 1.8$, 0) and ($\simeq 1$, $\simeq \frac{\pi}{2}$) lead to the minimum and maximum values, respectively.

\begin{figure}[htb]
		\includegraphics[width=0.32\textwidth,center]{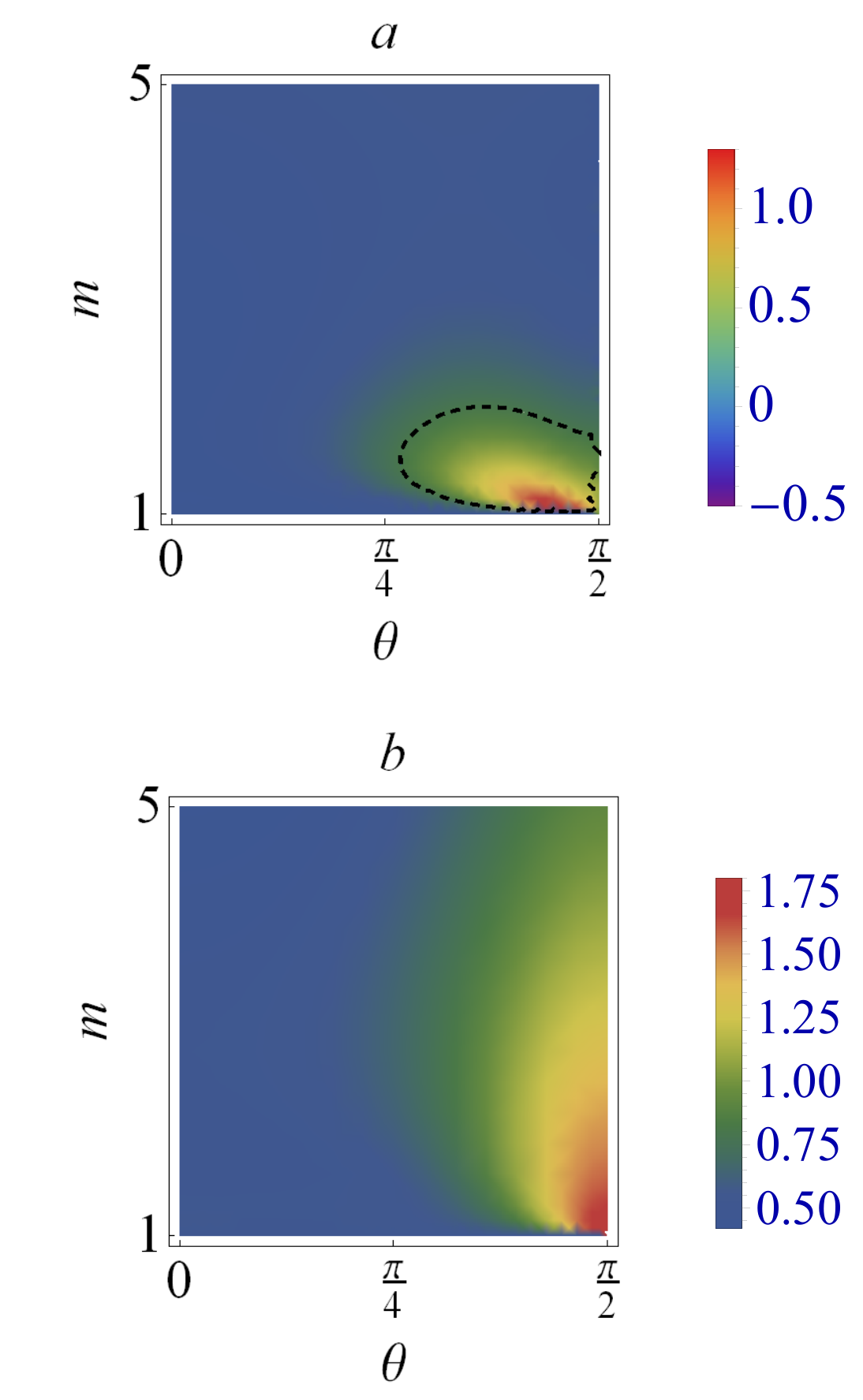}
		\caption{$\sigma_{xy}^{tot.int}[\theta, m]$ and $\sigma_{yx}^{tot.int}[\theta, m]$ are plotted in panel $a$ and $b$, respectively. 
		\label{fig:15} }
\end{figure}

Adding up all these contributions provides us with the total anomalous conductivity in the intrinsic regime. Based on Fig.~\ref{fig:15} the $\sigma^{tot.int}_{ij}$ is anisotropic and and its component varies within $-0.6 ~\lesssim \sigma^{tot.int}_{xy}[m, \theta]\lesssim 1.35 ~$, and also $0.25\lesssim \sigma^{tot.int}_{yx}[m, \theta]\lesssim 1.83~$. The black dashed line in panel $a$ of Fig.~\ref{fig:15} indicates the $(m, \theta)$ combinations for which $\sigma^{tot.int}_{xy}[m, \theta]=0$. Accordingly, tuning $(m, \theta)$ around this dashed line leads to a sign change in $\sigma^{tot.int}_{xy}$.
\\

At the end of this section, we consider again the two special magnetization orientations $\theta=0$ and $\frac{\pi}{2}$. The total intrinsic anomalous hall conductivities corresponding to these two important cases are
 \begin{align}
 &\sigma^{tot.int}_{xy}[\theta=0]=\frac{m^6+95 m^4+79 m^2-207}{2 \left(m^2+3\right)^2 \left(3 m^3+m\right)},\\
  &\sigma^{tot.int}_{yx}[\theta=0]=\frac{m^6-81 m^4-49 m^2+225}{2 \left(m^2+3\right)^2 \left(3 m^3+m\right)},
  \end{align} 
     \begin{align}
 &\sigma^{tot.int}_{xy}[\theta=\frac{\pi}{2}]=-\frac{29 m^4+54 m^2+45}{6 m \left(m^2+3\right)^2},\\
    &\sigma^{tot.int}_{yx}[\theta=\frac{\pi}{2}]=  \frac{17 m^4+62 m^2+81}{2 m \left(m^2+3\right)^2}.
    \end{align} 

\begin{figure}
		\includegraphics[width=0.45\textwidth,center]{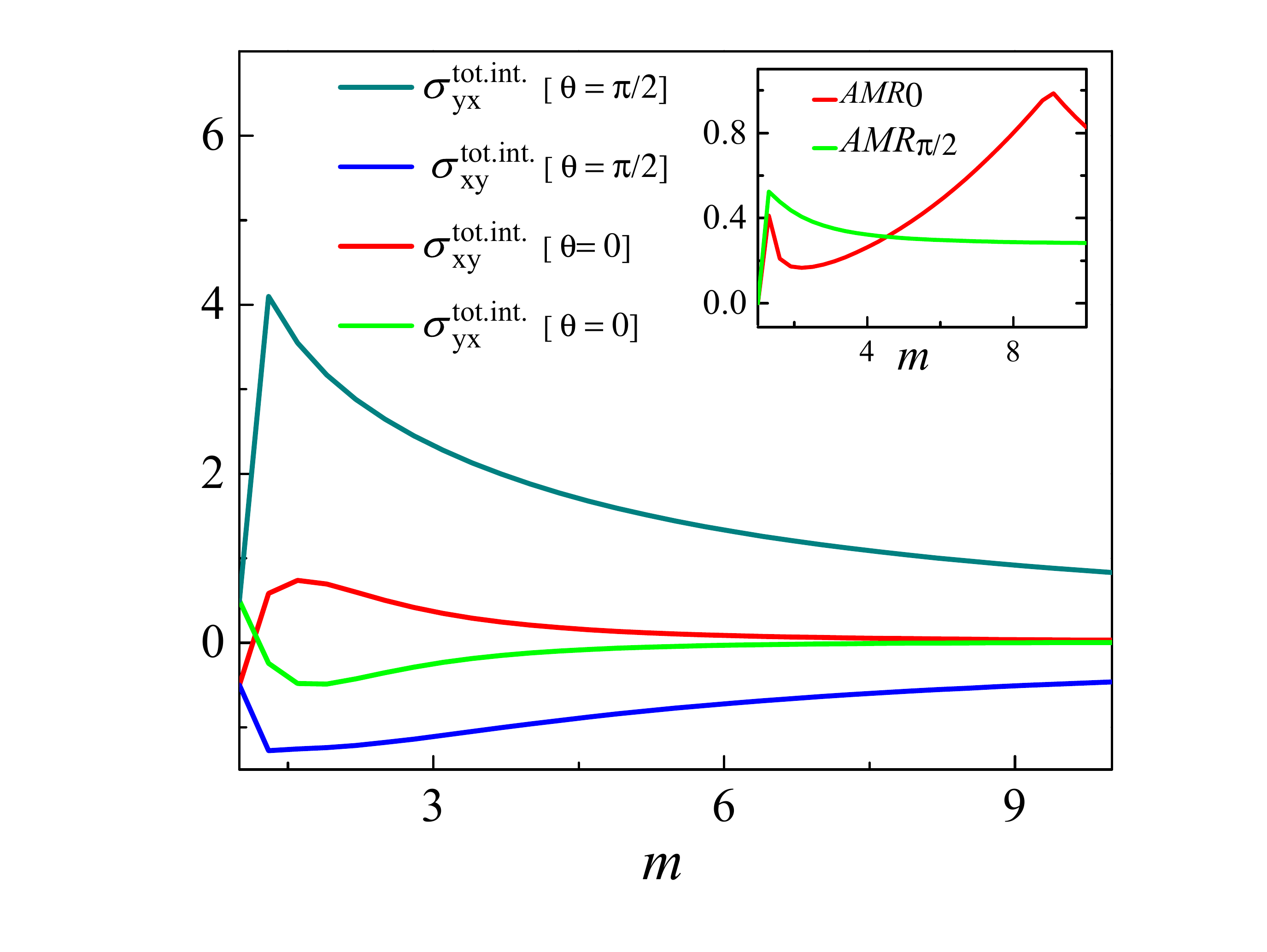}
			\caption{ $\sigma_{ij}^{tot.int.}$ is plotted as function of $m$ for two values of $\theta=0$ and $\theta=\frac{\pi}{2}$. The inset shows $AMR_{0}$ and $AMR_{\pi/2}$ as function of $m$. 
		\label{fig:11} }
\end{figure}

Fig.~\ref{fig:11} illustrates how above expressions behave with respect to $m$. Also, in the inset of this figure 
\begin{equation}
AMR_{\theta}[m]=\dfrac{\left|~\left|\sigma^{tot.int}_{yx}[\theta, m]\right| -\left| \sigma^{tot.int}_{xy}[\theta, m]\right|~ \right| }{~ \left|\sigma^{tot.int}_{yx}[\theta, m]\right| +\left|\sigma^{tot.int}_{xy}[\theta, m]\right|~ },
\end{equation}
is shown to measure the anisotropy of this contribution.
$\sigma^{tot.int}_{xy}[\theta=0]$ starts from $-0.5 $ at $m=1$, then increases until a maximum positive value is reached around $m=1.8$, after which it continually decreases till it approaches zero for large values of $m$. From all contributions to this total intrinsic anomalous hall conductivity, only $\sigma^{in}_{xy}$ has a non-zero value at $m=1$. In contrary to the other contributions, $\sigma^{in}_{ij}$ contains not only information of the conduction band but also of the valence band. Hence at $m=1$, the contribution of the valence band in $\sigma^{in}_{ij}$ leads to a non-zero value for $\sigma^{tot.int}_{ij}$. $\sigma^{tot.int}_{yx}[\theta=0]$ starts from $0.5 $ at $m=1$, then decreases till it reaches its minimum value around $m=1.8$, and finally follows the same trend as $\sigma^{tot.int}_{xy}[\theta=0]$ at large values of $m$.  As it is clear from the figure, the conductivities for the in plane magnetization are larger than for the out-of-plane case. By increasing $m$, $AMR_{0}[m]$ acquires a local maximum close to $m=1$, and another local maximum around $m=10$. In addition, $AMR_{\frac{\pi}{2}}[m]$ does not change as much against $m$. 

\subsubsection{The intermediate regime}
\begin{figure}
	
		\includegraphics[width=0.45\textwidth, center]{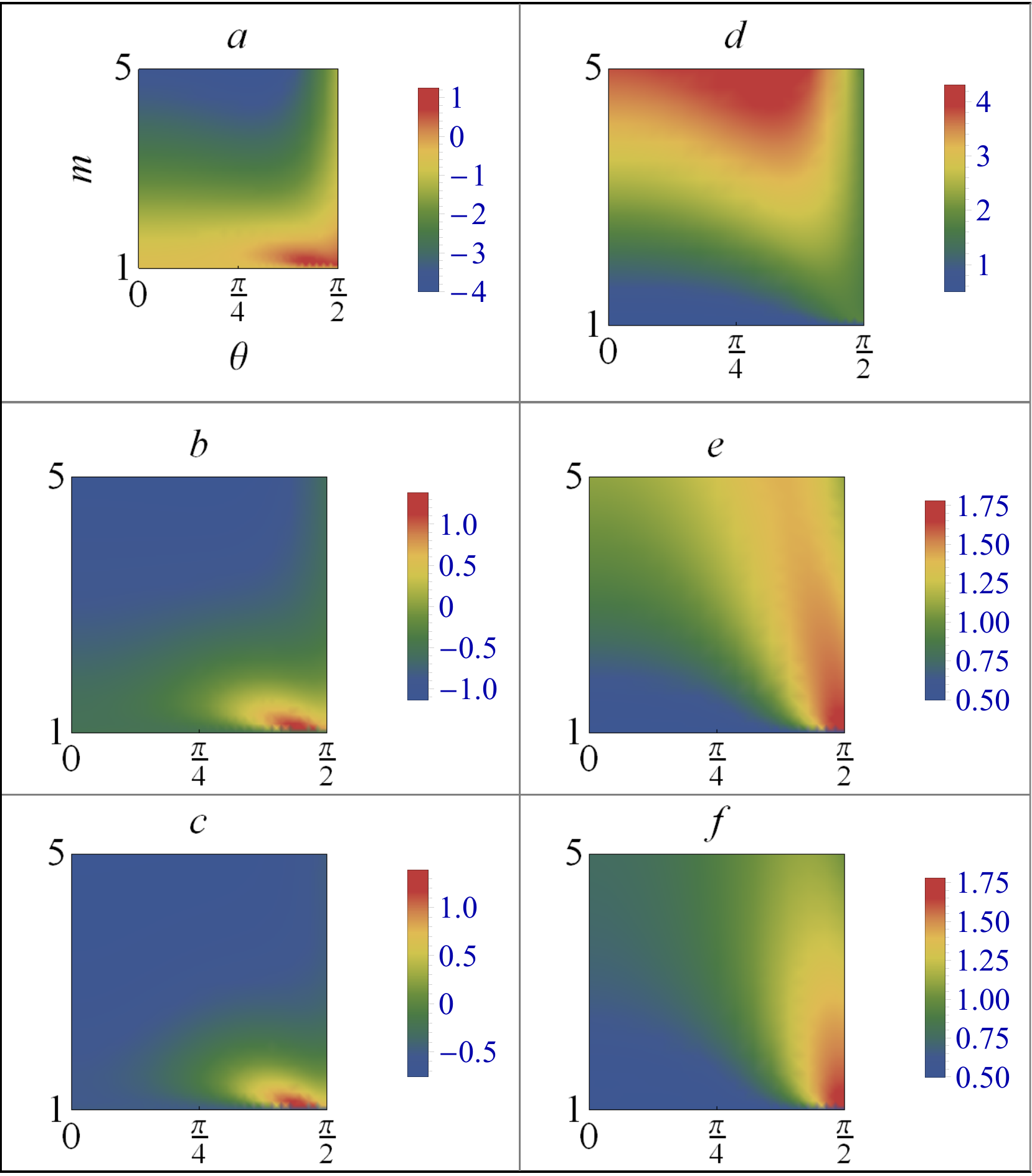}
		\caption{$\sigma_{xy}^{tot}$ and $\sigma_{yx}^{tot}$ are plotted respectively in the first and second column, in terms of $\theta$ and $m$, for some values of $n_{im}$. First, second and third row correspond to $n_{im}=0.1$, $0.5$ and $1$, respectively. 
		\label{fig:interxyyx} }
\end{figure}

In this regime, extrinsic and intrinsic terms both contribute to the AHE with comparable sizes. Fig.~\ref{fig:interxyyx} is shown to discuss the behavior of the conductivity $\sigma^{tot}_{ij}$ in terms of $\theta$ and $m$ for some values of $n_{im}(=n_{inm})$ at an arbitrary $\nu=\frac{V_{0}}{J S_{m}}=1.4$. The first column of this figure illustrates how $\sigma^{tot}_{xy}$ behaves against $\theta$ and $m$ for $n_{im}=0.1$, $0.5$ and $1$ in panels $a$, $b$ and $c$, respectively. In the second column, $\sigma^{tot}_{yx}$ is shown for the same choices of $n_{im}$. As it is shown in panels $a$ and $b$, there is a crossover from a positive value to a negative value for the corresponding conductivity, if we increase $m$ from $1$ to $5$, regardless of $\theta$. In panel $c$, this crossover happens if $\theta$ is larger then $\theta=\frac{\pi}{4}$. In addition, the maximum value of $\sigma^{tot}_{xy}$ occurs if we place the chemical potential close to the bottom of the conduction band and also adjust the surface magnetization close to $\frac{\pi}{2}$. Surprisingly, in contrast to $\sigma^{tot}_{xy}$, the second column illustrates that $\sigma^{tot}_{yx}$ is  positive. This term in panel $d$ gets its maximum value for large values of $m$ (close to $5$) and a magnetization away from $\theta=\frac{\pi}{2}$. However, by increasing the concentration of the impurities till $n_{im}=1$ in panel $f$, $\sigma^{tot}_{yx}$ acquires its maximum value in a system with $\theta \simeq \frac{\pi}{2}$ and a small value of $m \simeq 1$. Then, although these two components of $\sigma^{tot}_{ij}$ behave differently respect to given parameters, they share this feature that at $n_{im}=1$ they get their maximum value if $m \simeq 1$ and $\theta \simeq \frac{\pi}{2}$.

\begin{figure}
 	\centering{}
		\includegraphics[width=0.44\textwidth,center]{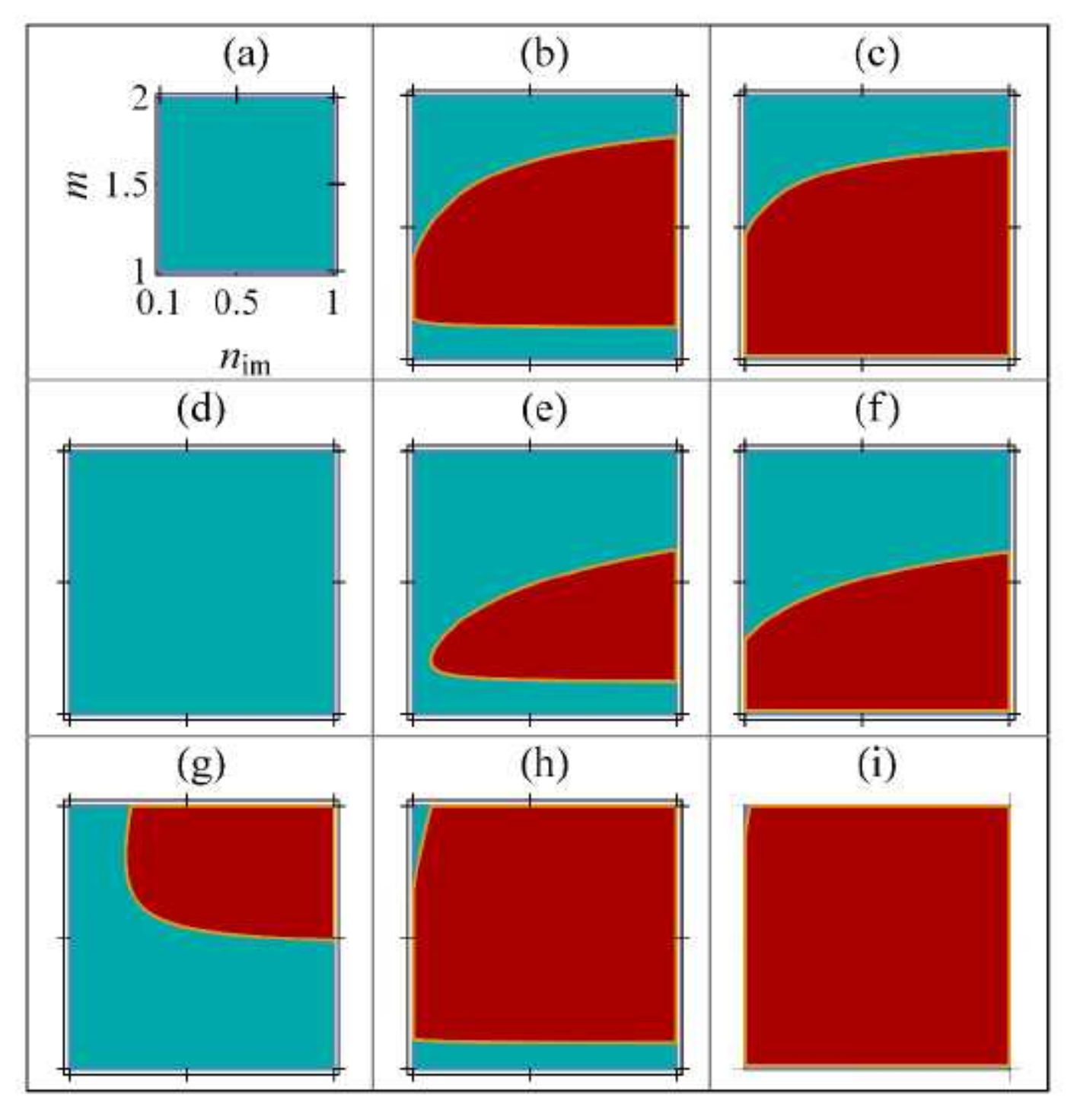}
		\caption{$\sigma_{xy}^{tot}$ is plotted in terms of $n_{im}$ and $m$ for some values of $\theta$ and $\nu$. The blue part corresponds to negative values of $\sigma_{xy}^{tot}$ and the red part corresponds to positive values. The first, seconds and third row correspond to $\nu=10$, $0.1$ and $0$, respectively. Also, first, second and third column correspond to  $\theta=\frac{\pi}{6}$, $\frac{\pi}{3}$ and $0.44 \pi$, respectively.} 
	\label{fig:interxy} 
\end{figure}

In addition, there is a region in panel $a$ with a positive conductivity, while it has the opposite sign in panel $c$. Therefore, it can be inferred that if we exert the external electric field parallel to the in-plane component of the surface magnetization, one can change the sign of anomalous conductivity in the intermediate regime via changing the concentration of the present impurities. To go further into the detail of this sign change via the change in concentration of impurities,  Fig.~\ref{fig:interxy} is shown. Except for panel $a$ and $d$, $\sigma^{tot.}_{xy}$ undergoes a sign change by tunning the impurity concentration. 
This sign change in the anomalous Hall conductivity has recently been observed experimentally~\cite{signchangeex}. 

\begin{figure}
	\includegraphics[width=0.42\textwidth,center]{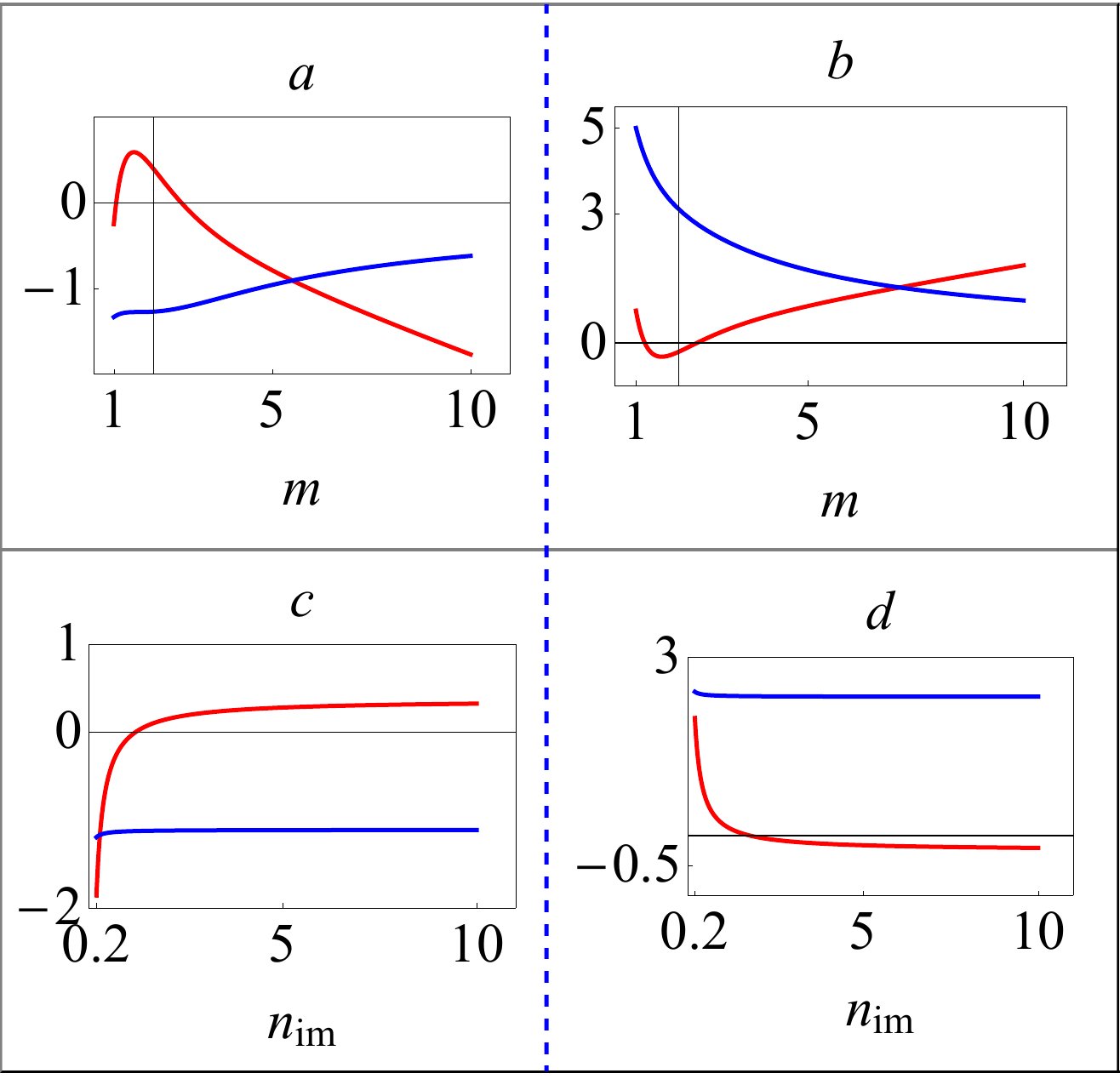}
		\caption{$\sigma_{xy}^{tot}$ and $\sigma_{yx}^{tot}$ are plotted in the first and the second column, respectively. In the first row corresponding conductivity is plotted against $m$ and in the second row against $n_{im}$. The red and blue curves show corresponding conductivity at $\theta$=0 and $\frac{\pi}{2}$, respectively.\label{fig:intermatends} }
\end{figure}

Finally, like in previous sections, we briefly discuss the two cases $\sigma^{tot.}_{xy}[\theta=0]$ and $\sigma^{tot.}_{xy}[\theta=\frac{\pi}{2}]$. As all red curves in Fig.~\ref{fig:intermatends} show, $\sigma^{tot.}_{ij} [\theta=0]$ undergoe a sign change via changing $m$ or $n_{im}$, in contrary to $\sigma^{tot.}_{ij}[\frac{\pi}{2}]$(blue curve) which does not show such a sign change. Therefore, since the AHE in this regime is very anisotropic, observing its sign change with respect to $m$ or $n_{im}$, requires properly adjusting all the involved parameters, the direction of the external electric field, the orientation of the surface magnetization, the position of the chemical potential, the concentration of the impurities and also the ratio of the non-magnetic scattering potential to the magnetic scattering potential.    

\section{Summary}\label{sec:summary}
In this work, the anomalous hall conductivity of a 3D TI is investigated using the semi-classical Boltzmann approach along with a modified relaxation time scheme, in terms of the Fermi level and the band gap, the spatial orientation of the present surface magnetization $\theta$ (an orientation perpendicular to the surface corresponds to $\theta=0$, an orientation in the $\hat{y}$-direction to $\theta=\frac{\pi}{2}$) and also the concentration of magnetic and non-magnetic impurities.
There are three contributions to the AHE, namely the intrinsic effect (arising from nonzero Berry curvature), the side jump effect and the skew scattering effect. They are competing to dominate the anomalous hall conductivity of the system. In this work by applying a fully analytical method we investigate how the spatial orientation of the surface magnetization and also the value of $m$ influence the transport of the massive Dirac Fermions on the surface of a 3D TI, doped with point like, randomly placed, magnetic and no-magnetic impurities. Since the 
contribution of non-magnetic impurities to the AHE has been investigated by others before~\cite{mac}, here we mainly focus on the effect of magnetic impurities to the AHE.

Concerning the side jump contribution, we discuss all important transport parameters, such as the electron deflection, side jump velocity of the electrons, the side jump associated mean free paths and finally the corresponding charge conductivity. We also extract an analytical expression for the conductivity due to the magnetic side jump effect as a function of the longitudinal conductivity. Moreover, we show that in the absence of non-magnetic impurities, the total magnetic side jump contribution is isotropic versus the direction of the exerted external electric field. Interestingly, our results open up some possibilities for engineering the total magnetic side jump contribution. For example, setting $\theta$ to either $0$ or $\frac{\pi}{2}$, turns off the magnetic side jump contribution, whatever is $m$. We demonstrate that in the presence of non-magnetic impurities, the total side jump contribution is anisotropic. Furthermore, by tuning the surface magnetization near the surface of the TI and also putting the chemical potential just above the bottom of the conduction band, one can again turn off and on the total side jump contribution in the presence of both kinds of impurities. 

Next, the contributions coming from conventional and intrinsic skew scattering are investigated. Our results show that in the absence of non-magnetic impurities, the conventional skew scattering is isotropic. In contrary to the total magnetic side jump contribution, which vanishes in a fully metallic regime, the conductivity corresponding to magnetic conventional skew scattering surprisingly gets its maximal value in this regime. However, similar to the side jump magnetic contribution, it disappears in a system with in-plane magnetization. In the presence of non-magnetic impurities, the total contribution of the conventional skew scattering is still isotropic, and in the metallic regime it reaches a significant value if the magnetization is out of plane. In addition, the skew scattering effect ontributes to the AHE through an additional correction, called the intrinsic term. Our results show that the correction arising from the magnetic intrinsic skew scattering is anisotropic. And in contrary to the previous conventional contribution does not vanish at $\theta=\frac{\pi}{2}$. Besides, this intrinsic contribution disappears at the fully metallic regime, just  like the side jump contribution. Remarkably, by applying an external electric field perpendicular to the in-plane component of the magnetization, and setting $\theta=\frac{\pi}{3}$, one can turns off this term. 

The presence of non-magnetic impurities allows one to have more degrees of freedom in engineering the total contribution of the intrinsic skew scattering. For example, if we exert an electric field along the in-plane component of the magnetization, the total intrinsic conductivity arising from skew scattering can always be turned off, regardless of $\theta$. If we exert the electric field perpendicular to the in-plane component of the magnetization, we can still turn off this contribution, but just for those systems with $\theta \leq\frac{\pi}{3}$. 

Therefore, by considering all these observations, we come to the conclusion that in the metallic regime (or the gapless regime) the conventional skew scattering dominates the AHE of a system with a low concentration of magnetic impurities. Out of this very dilute regime, by tuning $\theta$ around $\frac{\pi}{2}$ and $m$ around $1$ and also exerting an external electric field perpendicular to the in-plane component of the magnetization, the AHE gets its maximum reachable value. If the extrinsic and intrinsic terms both contribute to the AHE with comparable sizes (the intermediate regime), one can observe a sign change in the anomalous Hall conductivity not only via tuning the Fermi level or the spatial orientation of the surface magnetization, but surprisingly also via tuning the concentration of the impurities, for a certain range of the other parameters.

 \clearpage
\appendix
\newpage
\begin{widetext}
\numberwithin{equation}{section}

\section{Longitudinal magnetic mean free paths $\lambda^{m.s}_{i}$ }\label{sec:Longitudinal mean free path}

To calculate the current density of the system $\textbf{J}=-e \sum_{\bm k} \textbf{\textbf{v}}_{\bm k} f_{\bm k}$, we need to find the following three terms in the velocity of the electrons $\textbf{v}_{\bm k}=\textbf{v}_{0\bm k}+\textbf{v}^{an}_{\bm k}+\textbf{v}^{sj}_{\bm k}$ and also the following five terms in the distribution function of the Dirac Fermions $f_{\bm k}= f^{0}+g^{s}_{\bm k}+g^{a1}_{\bm k}+g^{a2}_{\bm k}+g^{ad}_{\bm k}$. These terms in the velocity of the electrons are calculated in the main text of this work. The next step is to calculate the equilibrium distribution function of the Dirac Fermions. As the dynamics of these Fermions during scattering off non-magnetic impurities is isotropic, it can be treated by the widely used relaxation time scheme. Here we go through the details of the calculation of all introduced corrections to the conductivity of the system that arise from scattering by magnetic impurities. Based on Eq.~(\ref{eq:generalg}), in order to find $g^{p}_{\bm k}$, first the associated mean free paths $\lambda^{p}_{i}$ have to be calculated. In this section we first clarify the procedure to obtain $\lambda^{m.s}_{i}$. In Sec.~\ref{sec:side mean free path}, Sec.~\ref{sec:conventional skew} and finally in Sec.~\ref{sec:intrinsic skew}, we go through the calculation of $\lambda^{m.ad}_{i}$, $\lambda^{m.sk1}_{i}$, $\lambda^{m.sk2}_{i}$, respectively. 
 
To find the longitudinal conductivity of the system, we need to find the distribution function of the electrons during their conventional scattering off magnetic impurities. According to Eq.~(\ref{eq:generalg}) and Eq.~(\ref{eq:gs}), we arrive at
\begin{equation}
\vv_{0\bm k}\cdot \hat{x_{i}} =\sum_{\bm k '} w^{(2. m)}_{\bm k \bm k' } \left[\lambda^{m.s}_{i} (\bm k ) - \lambda^{m.s}_{i} (\bm k ')\right],
\end{equation}
where $w^{(2.m)}_{\bm k \bm k'}$ is transition rate of magnetic scatterings. Replacing the mean free path with its Fourier expansion $\lambda^{m.s}_{i}(\bm k,\theta)=\sum_{n=1}^{\infty}[\lambda^{0}_{i}(\textit{k})+ {\lambda^{m.s, c}}_{i,n} (\textit{k})\cos n\phi_{\bm k}+{\lambda}^{m.s, s}_{i,n} (\textit{k})\sin n\phi_{\bm k}]$ leads to  
\begin{equation}\label{appenxix2}
\vv_{0\bm k}\cdot \hat{x_{i}} =\sum_{\bm k '}\sum_{n=1}^{\infty} w^{(2.m)}_{\bm k \bm k'}(~\lambda^{0}_{i}+\lambda^{m.s,c}_{i,n}[\cos n\phi_{\bm k } -\cos n\phi_{\bm k'}] + \lambda^{m.s,s}_{i,n}[\sin n\phi_{\bm k } -\sin n\phi_{\bm k'}]).
\end{equation}
where $\hat{x}_{1}= \hat{x}$ and $\hat{x}_{2}=\hat{y}$. Before continuing our discussion, note that conservation of the number of particles imposes 
\begin{equation}
N=\sum_{\bm k}f_{\bm k}^{0}=\sum_{\bm k}(f_{\bm k}^{0}+ g^{s}_{\bm k}+g^{ad}_{\bm k}+g^{a1}_{\bm k}+g^{a2}_{\bm k}).
\end{equation}
Therefore, $\sum_{\bm k}~ (g^{s}_{\bm k}+g^{ad.}_{\bm k}+g^{a1}_{\bm k}+g^{a2}_{\bm k})=0$, and hence all the constant terms in the Fourier expansions for $\lambda^{0}_{i}$ are zero. 
After some calculations we obtain
\begin{equation}\label{appendix3}
\lambda^{m.s}_{i}(\bm k,\theta)=\dfrac{\alpha^{m.s,0}_{i}+(\alpha^{m.s,c}_{i}+\delta_{i,1})\cos\phi_{\bm k}+(\alpha^{m.s,s}_{i}+\delta_{i,2})\sin\phi_{\bm k}}{2(\beta_{k} \cos \phi_{\bm k} +1) \left(\gamma_{k}^2 \cos 2 \theta +1\right)}\lambda^{m.s}_{0},
\end{equation}
 where $\beta_{k} =\dfrac{\gamma_{k} \sqrt{1-\gamma_{k}^2} \sin 2 \theta }{\gamma_{k}^2 \cos 2 \theta +1}$, $\alpha^{m.s,0}_{i}=\gamma_{k}\sqrt{1-\gamma_{k}^2}\sin 2 \theta~\dfrac{\lambda^{m.s,c}_{i,1}}{\lambda^{m.s}_{0}}$, $\alpha^{m.s,c}_{i}=\left(1-\gamma_{k}^2\right)\left(2 \sin ^2\theta -1\right)\dfrac{\lambda^{m.s,c}_{i,1}}{\lambda^{m.s}_{0}}$, $\alpha^{m.s,s}_{i}=(\gamma^{2}_{k}-1)\dfrac{\lambda^{m.s,s}_{i,1}}{\lambda^{m.s}_{0}}$ and $\lambda^{m.s}_{0} =\frac{4\hbar^{3} v^{3}_{F}\sqrt{1-\gamma^{2}_{k}}}{n_{im} J^{2}S^{2}_{m}\varepsilon_{k}} $. 
As it is clear from the equations above, $\lambda^{m.s,c}_{i,1}$ and $\lambda^{m.s,s}_{i,1}$ are the only two required Fourier coefficients of $\lambda^{m.s}_{i}$, as the other higher order Fourier coefficients ($n >1$) are a function of these two primary coefficients. These crucial Fourier coefficients can be obtained straightforwardly as follows:
\begin{equation}\label{appendix4}
 \left\{
 \begin{array}{c}
 \pi \lambda^{m.s,c}_{i,1}-\int^{2\pi}_{0} \dfrac{\alpha^{m.s,0}_{i}+(\alpha^{m.s,c}_{i}+\delta_{i,1})\cos\phi_{\bm k}+(\alpha^{m.s,s}_{i}+\delta_{i,2})\sin\phi_{\bm k}}{2(\beta_{k} \cos \phi_{\bm k} +1) \left(\gamma_{k}^2 \cos 2 \theta +1\right)}\lambda^{m.s}_{0}\cos\phi_{\bm k} d\phi_{\bm k}=0\\ \\
  \pi \lambda^{m.s,s}_{i,1}-\int_{0}^{2\pi}\dfrac{\alpha^{m.s,0}_{i}+(\alpha^{m.s,c}_{i}+\delta_{i,1})\cos\phi_{\bm k}+(\alpha^{m.s,s}_{i}+\delta_{i,2})\sin\phi_{\bm k}}{2(\beta_{k} \cos \phi_{\bm k} +1) \left(\gamma_{k}^2 \cos 2 \theta +1\right)}\lambda^{m.s}_{0}\sin\phi_{\bm k}d\phi_{\bm k}=0
 \end{array}
 \right.~.
 \end{equation}
After solving the above set of integral equations, we arrive at:
\begin{equation}\label{appendix5}
\lambda^{m.s,c}_{1,1}= \frac{4~\hbar ^{3} v^{3}_{F}}{n_{im} J^{2} S_{m}^{2}\varepsilon_{k}}\frac{\sqrt{1-\gamma_{k}^{2}} }{1+\Gamma_{k}+\left(1+\Gamma_{k}\gamma_k^2\right)\cos2\theta}~, \lambda^{m.s,s}_{1,1}=0,
\end{equation}
\begin{equation}\label{appendix6}
\lambda^{m.s,c}_{2,1}=0,~~~~~ \lambda^{m.s,s}_{2,1}= \frac{4~\hbar ^{3} v^{3}_{F}}{n_{im} J^{2} S_{m}^{2}\varepsilon_{k}}\frac{\sqrt{1-\gamma_{k}^{2}} }{1-\gamma^{2}_{k}+(1+\Gamma_{k})(1+\gamma^{2}_{k}\cos2\theta)}~,
\end{equation}
where $\Gamma=\left(1-\dfrac{\gamma^{2}_{k}(1-\gamma^{2}_{k})\sin^{2}2\theta}{(1+\gamma_{k}^{2}\cos2\theta)^{2}}\right)^{1/2}$. Putting the found nonzero Fourier coefficient $\lambda^{m.s,c}_{1,1}$ and  $\lambda^{m.s,s}_{2,1}$ in Eq.~(\ref{appendix3}) gives
\begin{equation}\label{appendix7}
\lambda^{m.s}_{1}=\dfrac{A_{k}\cos\phi_{\bm k}+\gamma_{k}\sin 2\theta \sqrt{1-\gamma^{2}_{k}}}{[1+\beta_{k}\cos\phi_{\bm k}][1+\gamma^{2}_{k}\cos2\theta][A+[1-\gamma^{2}_{k}]\cos2\theta]}\dfrac{\lambda^{m.s}_{0}}{2},
\end{equation}
\begin{equation}\label{appendix8}
\lambda^{m.s}_{2}=\dfrac{  (1+\Gamma_{k})
 \sin\phi_{\bm k}}{[1+ \beta_{k} \cos\phi_{\bm k}][1+A_{k}-\gamma^{2}_{k}]}\dfrac{\lambda^{m.s}_{0}}{2},
\end{equation}
where $A_{k}=(1+\Gamma_{k})(1+\gamma^{2}_{k}\cos2\theta)$. Therefore, the resultant correction to the distribution function of the electrons due to the conventional scattering of electrons from magnetic impurities is
\begin{equation}
g^{m.s}_{\bm k}=e E \left[\dfrac{A_{k}\cos\phi_{\bm k}+\gamma_{k}\sin 2\theta \sqrt{1-\gamma^{2}_{k}}}{[1+\beta_{k}\cos\phi_{\bm k}][1+\gamma^{2}_{k}\cos2\theta][A_{k}+[1-\gamma^{2}_{k}]\cos2\theta]}\dfrac{\lambda^{m.s}_{0}}{2} \cos\chi+\dfrac{  (1+\Gamma_{k})
 \sin\phi_{\bm k}}{[1+ \beta_{k} \cos\phi_{\bm k}][1+A_{k}-\gamma^{2}_{k}]}\dfrac{\lambda^{m.s}_{0}}{2} \sin\chi \right]
\partial_{\varepsilon_k} f^0~.
\end{equation}

\section{Side jump associated mean free paths $\lambda^{m.ad}_{i}$}\label{sec:side mean free path} 
As it was indicated in the main text, electrons during scattering off magnetic impurities undergo a side jump which changes the velocity of the electrons and also their distribution function. This leads to the following two corrections to the conductivity $\sigma^{m.sj}_{ij}$, $\sigma^{m.ad}_{ij}$. Since we have already found the associated distribution function $g^{m.s}_{\bm k}$, we needn't to calculate $\sigma^{m.sj}_{ij}$, thus in this section we just present the details of calculating $\sigma^{m.ad}_{ij}$. As before, we ignore the side jump of the electrons during their skew scattering. As we did to calculate $g^{m.ad}_{\bm k}$, we replace $\lambda^{m.ad}_{i}$ in Eq.~\ref{eq:gadist} with their Fourier expansions  $\lambda^{m.ad}_{i}(\bm k,\theta)=\sum_{n=1}^{\infty}[\lambda^{m.ad,c}_{i,n}\cos n\phi_{\bm k}+ \lambda^{m.ad,s}_{i,n}\sin n\phi_{\bm k}]$. By assuming that the external electric field is exerted along $\hat{x_{i}}$, Eq.~(\ref{eq:gadist}) is converted into 

\begin{equation}\label{appenxix9}
\vv^{m.sj}_{\bm k}\cdot \hat{x_{i}} =\sum_{\bm k'}\sum_{n=1}^{\infty} w^{(2.m)}_{\bm k \bm k'}(~\lambda^{m.ad,c}_{i,n}  [\cos n\phi_{\bm k} -\cos n\phi_{\bm k'}] + \lambda^{m.ad,s}_{i,n}[\sin n\phi_{\bm k} -\sin n\phi_{\bm k'}]).
\end{equation}
Using the already found functions $\mathbf{v}^{m.sj}_{\bm k}$ in Eq.~(\ref{eq:vsidm}) and using Eq.~(\ref{eq:w2}), we arrive at
\begin{equation}\label{appendix10}
	\lambda^{m.ad}_{i}(\bm k,\theta)=\frac{\alpha^{m.ad,0}_{i}+[\alpha^{m.ad, c}_{i} -\delta_{i,2}]\cos \phi_{\bm k }+(\alpha^{m.ad, s}_{i}+ [2 -\cos2\theta]~\delta_{i,1})\sin \phi_{\bm k}}{ 2 [\beta _{k} \cos \phi_{\bm k} +1][\gamma_{k}^2 \cos 2 \theta +1]}\lambda^{m.ad}_{0},
\end{equation}
where $\alpha^{m.ad,0}_{i}=\gamma_{k}\sqrt{1-\gamma_{k}^2}\sin 2 \theta~\dfrac{\lambda^{m.ad, c}_{i,1}}{\lambda^{m.ad}_{0}}$, $\alpha^{m.ad,c}_{i}=\left(1-\gamma_{k}^2\right)\left(2 \sin ^2\theta -1\right)\dfrac{\lambda^{m.ad,c}_{i,1}}{\lambda^{m.ad}_{0}}$, $\alpha^{m.ad,s}_{i}=(\gamma^{2}_{k}-1)\dfrac{\lambda^{m.ad,s}_{i,1}}{\lambda^{m.ad}_{0}}$ and $\lambda^{m.ad}_{0}=\frac{\hbar  v_F}{2 \varepsilon_{k} }\gamma_{k} \sqrt{1-\gamma_{k}^2} $. 
Finally we have to solve the set of equations
\begin{equation}\label{appendix11}
\left\{
\begin{array}{c}
\pi \lambda^{m.ad,c}_{i,1}-\int^{2\pi}_{0} \dfrac{\alpha^{m.ad,0}_{i}+[\alpha^{m.ad, c}_{i} -\delta_{i,2}]\cos \phi_{\bm k}+(\alpha^{m.ad, s}_{i}+ [2 -\cos2\theta]~\delta_{i,1})\sin \phi_{\bm k}}{ 2 [\beta _{k} \cos \phi_{\bm k} +1][\gamma_{k}^2 \cos 2 \theta +1]}\lambda^{m.ad}_{0}\cos\phi_{\bm k} ~d\phi_{\bm k}=0
\\
\pi \lambda^{m.ad,s}_{i,1}-\int^{2\pi}_{0} \dfrac{\alpha^{m.ad,0}_{i}+[\alpha^{m.ad, c}_{i} -\delta_{i,2}]\cos \phi_{\bm k}+(\alpha^{m.ad, s}_{i}+ [2 -\cos2\theta]~\delta_{i,1})\sin \phi_{\bm k}}{ 2 [\beta _{k} \cos \phi_{\bm k} +1][\gamma_{k}^2 \cos 2 \theta +1]}\lambda^{m.ad}_{0}\sin\phi_{\bm k}~d\phi_{\bm k}=0
\end{array}
\right.~.
\end{equation}
Their solution is 
\begin{equation}\label{appendix12}
\lambda^{m.ad,s}_{1,1}=\frac{[\sqrt{1-\beta^{2}_{k}}-1] [\cos 2 \theta -2 ] }{\beta_{k}^2[1+\gamma^{2}_{k}\cos2\theta] +[\sqrt{1-\beta_{k} ^2}-1][\gamma_{k}^2-1]} \lambda^{m.ad}_{0}, ~~~ \lambda^{m.ad,c}_{1,1}=0,
	\end{equation}
	
\begin{equation}\label{appendix13}
\lambda^{m.ad,c}_{2,1}=-\frac{ 1}{[~[\sqrt{1-\beta_{k}^2}-\beta_{k}^2]\gamma_{k}^2+1]\cos 2 \theta -\beta_{k}^2+\sqrt{1-\beta_{k}^2}+\beta _{k}\gamma_{k}  \sqrt{1-\gamma_{k}^2} \sin 2 \theta +1} \lambda^{m.ad}_{0}, ~~~ \lambda^{m.ad,s}_{2,1}=0.
\end{equation}
Inserting $\lambda^{m.ad,s}_{1,1}$ and $\lambda^{m.ad,c}_{2,1}$ in Eq.~(\ref{appendix10}) leads to

\begin{equation}\label{appendix14}
\lambda^{m.ad} _{1}(\bm k,\theta)=\dfrac{(2 - \cos 2\theta) \sin \phi_{\bm k}}{~\left( 1+\gamma_{k}^{2}\cos 2 \theta +  \beta_{k}^{-1}[\sqrt{1-\beta_{k}^{2}}-1][\gamma_{k}^{2}-1]\right)~[1+\beta_{k}  \cos \phi _{\bm k}]}\lambda^{m.ad}_{0},
\end{equation}

\begin{equation}\label{appendix15}
\lambda^{m.ad}_{2}(\bm k, \theta)=- \dfrac{\gamma_{k}\sqrt{1-\gamma^{2}_{k}}\sin2\theta + c_{k}~ \cos \phi_{\bm k}}{2[1+\beta_{k}\cos \phi_{\bm k}][1+ \gamma_{k}^2\cos 2 \theta][c_{k}+(1-\gamma_{k}^{2}) \cos 2 \theta]}\lambda^{m.ad}_{0},
\end{equation}
where $c_{k}=(1-\beta_{k}^2+\sqrt{1-\beta_{k}^{2}})(1+\gamma_{k}^2 \cos 2\theta) +\beta_{k}\gamma_{k} \sqrt{1-\gamma_{k}^2} \sin2 \theta$.

Finally, the associated correction to the distribution function of the electrons arising from the side jump can be written as, based on Eq.~(\ref{eq:generalg}),

\begin{equation}
g^{m.ad}_{\bm k}=\frac{e E \lambda^{m.ad}_{0}\partial_{\varepsilon_k} f^0}{[1+\beta_{k}  \cos \phi _{\bm k}]} \left[ \dfrac{(2 - \cos 2\theta) \sin \phi_{\bm k} \cos\chi}{~\left( 1+\gamma_{k}^{2}\cos 2 \theta +  \beta_{k}^{-1}[\sqrt{1-\beta_{k}^{2}}-1][\gamma_{k}^{2}-1]]~\right)}  - \dfrac{\gamma_{k}\sqrt{1-\gamma^{2}_{k}}\sin2\theta + c_{k}~ \cos \phi_{\bm k}}{2[1+ \gamma_{k}^2\cos 2 \theta][c_{k}+(1-\gamma_{k}^{2}) \cos 2 \theta]}\sin\chi \right]
~.
\end{equation}

\section{Conventional skew scattering associated mean free path  $\lambda^{m.sk1}_{i}$} \label{sec:conventional skew}
 
The conventional and intrinsic skew scattering contribute to the conductivity of the system via changing the distribution function of the electrons, as the velocity of the electrons does not change, in contrary to the side jump effect. Based on Eq.~(\ref{eq:skew1conduc})

\begin{align}\label{appendix17}
g^{m.a1}_{\bm k }=\frac{\sum_{\bm k'} w^{(2.m)}_{\bm k \bm k'} g^{m.a1}_{\bm k'}+\sum_{\bm k'} w^{(3a.m)}_{\bm k \bm k'}\left(g^{m.s}_{\bm k'}-g^{m.s}_{\bm k}\right)} {\sum_{\bm k'} w^{(2.m)}_{\bm k \bm k'}}.
\end{align}
Since $g^{m.s}_{\bm k}$ has been already found, we just need to find $g^{m.a1}_{\bm k}$. By using Eq.~(\ref{eq:w3am}), it can be straightforwardly proven that $\sum_{\bm k'}w^{(3a.m)}_{\bm k \bm k'}=0$. Therefore Eq.~(\ref{appendix17}) can be rewritten in terms of the mean free paths as

\begin{equation}\label{appendix18}
\lambda^{m.a1}_{i}(\bm k,\theta)=\dfrac{\sum_{\bm k' } w^{(3a.m)}_{\bm k \bm k'}\lambda^{m.s}_{i}(\bm k',\theta)+\sum _{\bm k'} w^{(2.m)}_{\bm k \bm k'}\lambda^{m.a1}_{i}(\bm k', \theta)}{\sum_{\bm k'} w^{(2.m)}_{\bm k \bm k'}}.
\end{equation}
Using the Fourier expansions of these mean free paths $\lambda^{m.a1}_{i}(\bm k,\theta)=\sum_{n=1}^{\infty}[{\lambda^{m.a1, c}}_{i,n}(\textit{k}) \cos n\phi_{\bm k}+{\lambda}^{m.a1,s}_{i,n} (\textit{k})\sin n\phi_{\bm k}]$ leads to  

\begin{equation}\label{appendix19}
\lambda^{m.a1}_{i}(\bm k,\theta)=\frac{\alpha^{m.a1,0}_{i}+[\alpha^{m.a1,c}_{i}-\delta_{i,2}\frac{J s_{m} \cos\theta~ k^{2}}{2 \varepsilon_{k}}\lambda^{m.s,s}_{i,1}]\cos\phi_{\bm k}+ [\alpha^{m.a1, s}_{i}+\delta_{i,1}\frac{J s_{m} \cos\theta~ k^{2}}{2 \varepsilon_{k}}\lambda^{m.s,c}_{i,1}]\sin \phi_{\bm k}}{2(1+\gamma^{2}_{k}\cos2\theta)(1+\beta_{k}\cos\phi_{\bm k}]},
\end{equation}
where $\alpha^{m.a1,0}_{i}=\gamma_{k}\sqrt{1-\gamma_{k}^2}\sin 2 \theta~\lambda^{m.a1, c}_{i,1}$, $\alpha^{m.a1,c}_{i}=\left(1-\gamma_{k}^2\right)\left(2 \sin^{2}\theta -1\right)\lambda^{m.a1,c}_{i,1}$, $\alpha^{m.a1,s}_{i}=(\gamma^{2}_{k}-1)\lambda^{m.a1,s}_{i,1}$.
Four unknown crucial Fourier coefficients $\lambda^{m.a, c}_{i,1}$ and $\lambda^{m.a1, s}_{i,1}$ can be found through solving a set of equations such Eq.~(\ref{appendix4}) and Eq.~(\ref{appendix11}). After solving this set of equations, we arrive at

\begin{equation}\label{appendix20}
\lambda^{m.a1, s}_{1,1}=\frac{\hbar v_{F}}{J n_{im} S_{m}}\frac{2 (1-\gamma^{2}_{k})\sqrt{1-\gamma^{2}_{k}}~\cos\theta}{[1+\Gamma_{k}+\cos 2\theta+\gamma_{k}^{2}~\Gamma_{k} \cos 2 \theta][2+\Gamma_{k}+\gamma_{k}^{2}~\Gamma_{k} \cos 2\theta-2\gamma^{2} \sin^2\theta ]},
\end{equation}
\begin{equation}\label{appendix21}
\lambda^{m.a1,c}_{2,1}=\frac{\hbar v_{F}}{J n_{im} S_{m}}\frac{4 (1-\Gamma_{k})\sqrt{1-\gamma^{2}_{k}}(1+\gamma_{k}^2 \cos 2 \theta)~\cos\theta}{[\gamma_{k}^2 (\Gamma_{k}+1) \cos 2 \theta -\gamma_{k}^2+\Gamma_{k}+2] [\gamma_{k}^2 (\Gamma_{k}\cos 4 \theta +\Gamma_{k}-2)+2 (\Gamma_{k}-1) \cos 2 \theta ]},
\end{equation}
and $\lambda^{m.a1,c}_{1,1}=\lambda^{m.a1,s}_{2,1}=0.$
Inserting the resultant non-zero Fourier coefficients in Eq.~(\ref{appendix19}), one obtains the mean free paths of the electrons during magnetic conventional skew scattering:
\begin{equation}\label{appendix22}
\lambda^{m.a1}_{1}(\bm k,\theta)=-\dfrac{\hbar v_{F}}{J S_{m} n_{im} }\dfrac{(1+\Gamma_{k})~\cos\theta ~\sin\phi_{\bm k}}{[(\Gamma_{k}\gamma_{k}^{2}+1)\cos2\theta+\Gamma_{k}+1][p_{k}+p_{k}\beta_{k} \cos\phi_{\bm k}]},
\end{equation}
\begin{equation}\label{appendix23}
\lambda^{m.a1}_{2}(\bm k,\theta)=\dfrac{\hbar v_{F}}{J S_{m} n_{im} }\dfrac{\cos\theta \left \{ [(4-4\Gamma_{k})\cos2\theta +(2\Gamma_{k}-3)\gamma_{k}^{2}+ (2\Gamma_{k}-1)\gamma_{k}^{2}\cos 4\theta ]\cos\phi_{\bm k}+ B_{k}\right \}}{[1+\gamma_{k}^{2}\cos2\theta]~[\Gamma_{k} \gamma_{k}^{2}\cos4\theta+ (2\Gamma_{k}-2)\cos 2\theta+ (\Gamma_{k}-2) \gamma_{k}^{2} ][p_{k}+p_{k}\beta \cos\phi_{\bm k}]},
\end{equation}
$p_{k}= \{\gamma ^2 [(\Gamma_{k} +1) \cos 2 \theta -1]+\Gamma_{k} +2\} (1-\gamma_{k}^{2})^{-3/2}$, 
$B_{k}=2 (1-\gamma_{k}^{2})^{-1/2}\gamma_{k} (\Gamma_{k}-1)(1+\gamma_{k}^{2}\cos2\theta)\sin2\theta$.

Finally, the associated correction to the distribution function of the electrons arising from the side jump can be written, based on Eq.~(\ref{eq:generalg}), as 

\begin{equation}
g^{m.a1}_{\bm k}=\alpha^{m.a1}\left(\dfrac{[(4-4\Gamma_{k})\cos2\theta-2\gamma^{2}_{k}+ (2\Gamma_{k}-1)\gamma_{k}^{2}(\cos 4\theta+1)]\cos\phi_{\bm k}+B_{k}}{(1+\gamma^{2}_{k}\cos2\theta)~(\Gamma_{k} \gamma^{2}_{k}\cos4\theta+ (2\Gamma_{k}-2)\cos 2\theta+ (\Gamma_{k}-2) \gamma^{2}_{k})} \sin\chi -\dfrac{(1+\Gamma_{k})~~\sin\phi_{\bm k}  \cos\chi }{(\Gamma_{k}\gamma_{k}^{2}+1)\cos2\theta+\Gamma_{k}+1}\right), 
\end{equation}  
with
$\alpha^{m.a1}=\dfrac{e E \partial_{\varepsilon_k}f^0}{p_{k}+p_{k}\beta_{k}\cos\phi_{\bm k}}\dfrac{\hbar v_{F}}{J S_{m} n_{im} }. $

\section{Intrinsic skew scattering associated mean free path  $\lambda^{m.a2}_{i}$} \label{sec:intrinsic skew} 

Like the conventional skew scattering, this contribution to the skew scattering just alters the distribution function of the electrons and leaves the velocity of the electrons unchanged. Using Eq.~(\ref{eq:skew2conduc}), which connects the conventional magnetic scattering of electrons to their intrinsic magnetic skew scattering, we arrive at

 \begin{equation}
 \sum_{\bm k'} w^{(4m)}_{\bm k \bm k'}\left[\lambda^{m.s}_{i}(\bm k)-\lambda^{m.s}_{i}(\bm k')\right]+
 \sum_{\bm k'}w^{(2.m)}_{\bm k\bm k'}\left[\lambda^{m.a2}_{i}(\bm k)-\lambda^{m.a2}_{i}(\bm k')\right]=0, 
  \end{equation}
which after rewriting gives
\begin{align}\label{appendix17}
\lambda^{m.a2}_{i}=\frac{\sum_{\bm k'} w^{(2.m)}_{\bm k \bm k'} \lambda^{m.a2}_{i}(\bm k') + w^{(4.m)}_{\bm k \bm k'}\left[\lambda^{m.s}_{i}(\bm k')-\lambda^{m.s}_{i}(\bm k)\right]} {\sum_{\bm k'} w^{(2.m)}_{\bm k \bm k'}}.
\end{align}

Applying the Fourier expansions of the mean free paths $\lambda^{m.a2}_{i}$ as $\lambda^{m.a2}_{i}(\bm k)=\sum_{n=1}^{\infty}[{\lambda_{i,n}^{m.a2, c}}\cos n\phi_{\bm k}+{\lambda}_{i,n}^{m.a2,s}\sin n\phi_{\bm k}]$ leads to 

\begin{align}\label{appendix17}
\lambda^{m.a2}_{i}=\frac{\sum_{\bm k', n} w^{(2.m)}_{\bm k \bm k'} [{\lambda_{i,n}^{m.a2, c}}\cos n \phi_{\bm k'}+{\lambda}^{m.a2,s}_{i,n}\sin n\phi_{\bm k'}] +\sum_{k'} w^{(4.m)}_{\bm k \bm k'}\left[\lambda^{m.s}_{i}(\bm k')-\lambda^{m.s}_{i}(\bm k)\right]} {\sum_{\bm k '} w^{(2.m)}_{\bm k \bm k'}}.
\end{align}
Using the already found $ w^{(2.m)}_{\bm k \bm k'}$, $w^{(4.m)}_{\bm k \bm k'}$ and $\lambda^{m.s}_{i}(\bm k')$, we arrive at
\begin{equation}\label{eq:lambdaa2}
\begin{split}
\lambda^{m.a2}_{i}~=&\left([\gamma_{k}^{3}-\gamma_{k}]\left([2 \cos 2\theta + 1]\lambda^{m.s.c}_{i,1} \sin\phi_{\bm k}-\lambda^{m.s.s}_{i,1}~[\cos 2\theta + 2]\cos\phi_{\bm k}\right)- \gamma^{2}_{k} \sqrt{1-\gamma_{k}^{2}}~\lambda^{m.s.s}_{i,1}~ \sin 2\theta \right)\alpha_{1}^{m.a2} \\
& +\left(\frac{1}{2}[\gamma_{k}^{2}-1] ~\lambda^{m.a2.s}_{i,1} \sin\phi_{\bm k}+\frac{1}{2}[\gamma_{k}^{2}-1]~\lambda^{m.a2.c}_{i,1}~\cos 2\theta \cos\phi_{\bm k}+ \frac{1}{2}\gamma_{k} \sqrt {1-\gamma_{k}^{2}}~\lambda^{m.a2.c}_{i,1}~\sin 2\theta \right) \alpha_{0}^{m.a2}\\
&-4~\lambda^{m.s}_{i} \gamma_{k} \sqrt{1-\gamma_{k}^{2}}\sin\phi_{\bm k} \sin2 \theta ~\alpha_{1}^{m.a2},
\end{split}
\end{equation}
where $\alpha_{0}^{m.a2}=\dfrac{ 1  \ }{	
		\gamma_{k}^{2} \cos 2 \theta +\gamma_{k} \sqrt{1-\gamma_{k}^{2}}   \sin 2 \theta  \cos \phi_{\bm k} +1 }$, and
		 $\alpha_{1}^{m.a2}=\dfrac{ n_{im}J^{2} S_{m}^{2} \ }{8 \hbar^{2}v^{2}_{F}} \dfrac{ 1  \ }{	
		\gamma_{k}^{2} \cos 2 \theta +\gamma_{k} \sqrt{1-\gamma_{k}^{2}}   \sin 2 \theta  \cos \phi_{\bm k} +1 }$.
	These two crucial non-zero Fourier coefficients $\lambda^{m.s,c}_{1,1}$ and $\lambda^{m.s,s}_{2,1}$ are given in Eq.~(\ref{appendix5}) and Eq.~(\ref{appendix6}). Four unknown crucial Fourier coefficients $\lambda^{m.a2, c}_{i,1}$ and $\lambda^{m.a2, s}_{i,1}$ can be obtained by solving a set of equations as we did in Eq.~(\ref{appendix4}) or Eq.~(\ref{appendix11}). After solving this set of equations, we arrive at
\begin{equation}
\begin{split}
\lambda^{m.a2.s}_{1,1}[\frac{\hbar v_{F}}{\varepsilon_{k}}]=&
\frac{8(\gamma_{k}-\Gamma_{k}\gamma_{k}) \left(\gamma_{k}^4 (\Gamma_{k}-2) (\Gamma_{k} +1)+\gamma_{k}^2 (\Gamma_{k} -1) \left(\gamma_{k}^2 \Gamma_{k} +1\right) \cos 4 \theta -\xi_{k}\cos 2 \theta +\gamma_{k}^2+2 \Gamma_{k}^2+\Gamma_{k}-2\right)}{\Gamma_{k}~\sqrt{1-\gamma_{k}^2}   \left([\gamma_{k}^2 \Gamma_{k}+1]\cos 2 \theta +\Gamma_{k}+1\right)\left(\left(\cos 2 \theta [\Gamma_{k} -1] -\sin^2 2 \theta ~\right)\gamma_{k}^2+\Gamma_{k} -1\right)},
\end{split}
\end{equation}
\begin{equation}
\lambda^{m.a2.c}_{2,1}[\frac{\hbar v_{F}}{\varepsilon_{k}}]=\frac{16\gamma_{k} \sqrt{1-\gamma^{2}_{k}}[\Gamma_{k} -1] \left(-\gamma_{k}^2\cos 4 \theta +[2 \gamma_{k}^2+1]\cos2 \theta+2\gamma_{k}^2+2\right)}{\left(\gamma_{k}^2 (\Gamma_{k}+1) \cos2 \theta -\gamma_{k}^2+\Gamma_{k}+2\right) \left(\gamma_{k}^2 (\Gamma_{k} \cos 4 \theta +\Gamma_{k}-2)+2 (\Gamma_{k}-1) \cos 2 \theta \right)},
\end{equation}
where $\xi_{k}=\gamma_{k}^2 \left(\Gamma_{k}  [\gamma_{k}^2-4 \Gamma_{k}+1]+4\right)-2 \Gamma_{k} $, $\lambda^{m.a2.c}_{1,1}=\lambda^{m.a2.s}_{2,1}=0$. We can express the correction to the distribution function of the electrons due to the magnetic intrinsic skew scattering as
\begin{equation}\label{eq:lambdaa2}
\begin{split}
g^{m.a2}_{k}~=&\left([\gamma_{k}^{3}-\gamma_{k}]\left([2 \cos 2\theta + 1]\lambda^{m.s.c}_{1,1} \sin\phi_{\bm k}-\lambda^{m.s.s}_{1,1}~[\cos 2\theta + 2]\cos\phi_{\bm k}\right)- \gamma^{2}_{k} \sqrt{1-\gamma_{k}^{2}}~\lambda^{m.s.s}_{1,1}~ \sin 2\theta \right)\alpha_{2}^{m.a2} \\
& +\left(\frac{1}{2}[\gamma_{k}^{2}-1] ~\lambda^{m.a2.s}_{1,1} \sin\phi_{\bm k}+\frac{1}{2}[\gamma_{k}^{2}-1]~\lambda^{m.a2.c}_{1,1}~\cos 2\theta \cos\phi_{\bm k}+ \frac{1}{2}\gamma_{k} \sqrt {1-\gamma_{k}^{2}}~\lambda^{m.a2.c}_{1,1}~\sin 2\theta \right) \alpha_{3}^{m.a2}\\
&-4~\lambda^{m.s}_{1} \gamma_{k} \sqrt{1-\gamma_{k}^{2}}\sin\phi_{\bm k} \sin2 \theta ~\alpha_{2}^{m.a2}\\
&+\left([\gamma_{k}^{3}-\gamma_{k}]\left([2 \cos 2\theta + 1]\lambda^{m.s.c}_{2,1} \sin\phi_{\bm k}-\lambda^{m.s.s}_{2,1}~[\cos 2\theta + 2]\cos\phi_{\bm k}\right)- \gamma^{2}_{k} \sqrt{1-\gamma_{k}^{2}}~\lambda^{m.s.s}_{2,1}~ \sin 2\theta \right)\alpha_{4}^{m.a2} \\
& +\left(\frac{1}{2}[\gamma_{k}^{2}-1] ~\lambda^{m.a2.s}_{2,1} \sin\phi_{\bm k}+\frac{1}{2}[\gamma_{k}^{2}-1]~\lambda^{m.a2.c}_{2,1}~\cos 2\theta \cos\phi_{\bm k}+ \frac{1}{2}\gamma_{k} \sqrt {1-\gamma_{k}^{2}}~\lambda^{m.a2.c}_{2,1}~\sin 2\theta \right) \alpha_{5}^{m.a2}\\
&-4~\lambda^{m.s}_{2} \gamma_{k} \sqrt{1-\gamma_{k}^{2}}\sin\phi_{\bm k} \sin2 \theta ~\alpha_{4}^{m.a2},
\end{split}
\end{equation}
$\alpha_{2}^{m.a2}= e E \partial_{\varepsilon_{k}} f^0 \alpha_{1}^{m.a2} \cos\chi$, $\alpha_{3}^{m.a2}= e E \partial_{\varepsilon_{k}} f^0 \alpha_{0}^{m.a2} \cos\chi$, $\alpha_{4}^{m.a2}= e E \partial_{\varepsilon_{k}} f^0 \alpha_{1}^{m.a2} \sin\chi$, $\alpha_{5}^{m.a2}=  e E \partial_{\varepsilon_{k}} f^0 \alpha_{0}^{m.a2} \sin\chi$. 

As we already proved $\lambda^{m.s,s}_{1,1}=0$ and $\lambda^{m.s,c}_{2,1}=0$, finally we can write the final expression for the correction to the distribution function of the electrons as
\begin{equation}\label{eq:lambdaa22}
\begin{split}
g^{m.a2}_{k}~=&\sin\phi_{\bm k}\left( [\gamma_{k}^{3}-\gamma_{k}][2 \cos 2\theta + 1]\lambda^{m.s.c}_{1,1}\alpha_{2}^{m.a2}  +\frac{1}{2}[\gamma^{2}_{k}-1] ~\lambda^{m.a2.s}_{1,1}\alpha^{m.a2}_{3}-4~\lambda^{m.s}_{1} \gamma_{k} \sqrt{1-\gamma^{2}_{k}} \sin2 \theta ~\alpha^{m.a2}_{2}\right)\\
& +\frac{1}{2}[\gamma_{k}^{2}-1]~\lambda^{m.a2.c}_{1,1}~\cos 2\theta ~\alpha^{m.a2}_{3}\cos\phi_{\bm k}+ \frac{1}{2}\gamma_{k} \sqrt {1-\gamma_{k}^{2}}~\lambda^{m.a2.c}_{1,1}~\sin 2\theta \alpha_{3}^{m.a2}\\
&+ \sin\phi_{\bm k}  \left([\gamma_{k}^{3}-\gamma_{k}] [2 \cos 2\theta + 1]\lambda^{m.s.c}_{2,1} \alpha^{m.a2}_{4}
 +\frac{1}{2}[\gamma_{k}^{2}-1] ~\lambda^{m.a2.s}_{2,1}\alpha^{m.a2}_{5}-4~\lambda^{m.s}_{2} \gamma_{k} \sqrt{1-\gamma_{k}^{2}}\sin2 \theta ~\alpha_{4}^{m.a2} \right)\\
&+\left(\lambda^{m.s.s}_{2,1}~[\cos 2\theta + 2][\gamma_{k}-\gamma^{3}_{k}]\alpha^{m.a2}_{4}+\frac{1}{2} \alpha^{m.a2}_{5}[\gamma_{k}^{2}-1]~\lambda^{m.a2.c}_{2,1}~\cos 2\theta \right)\cos\phi_{\bm k }\\
& - \frac{1}{2}\gamma_{k} \sqrt{1-\gamma_{k}^{2}}~\sin 2\theta (~2 \alpha^{m.a2}_{4} \gamma_{k}\lambda^{m.s.s}_{2,1}- \lambda^{m.a2.c}_{2,1} ~\alpha^{m.a2}_{5}).\\
\end{split}
\end{equation}

\end{widetext}


\begin{thebibliography}{2}

\bibitem{kane_1} 
	L. Fu, \textit{et al.}, Mele, Phys. Rev. Lett. \textbf{98}, 106803 (2007).
%	
     \bibitem{zhang_1} 
    X.-L. Qi and S.-C. Zhang, Rev. Mod. Phys.  \textbf{ 83}, 1057 (2011).
%
     \bibitem{kane_2} 
	C. L. Kane and E. J. Mele, Phys. Rev. Lett. \textbf{95}, 146802 (2005).
	%
\bibitem{bernevig_1} 
	B. A. Bernevig, \textit{et al.}, Science \textbf{314}, 1757 (2006).
	%
   \bibitem{Hassan_m}
	M. Z. Hasan and C. L. Kane, Rev. Mod. Phys. \textbf{82}, 3045 (2010).
	%
\bibitem{fu_tci}
	L. Fu, Phy. Rev. Lett. \textbf{106}, 106802 (2011).
	%
    \bibitem{TI_book}
	S.-Q. Shen, \textit{Topological Insulators: Dirac Equation in Condensed Matters} (Springer-Verlag, Berlin 2012).
	%
\bibitem{roushan_nat_2009}
	P. Roushan \textit{et al.}, Nature \textbf{460}, 1106 (2009).
	%
\bibitem{weak_1}

	M. Liu,\textit{et al.}, Phy. Rev. Lett. \textbf{108},  036805 (2012).
\bibitem{moore_1}	
	J. E. Moore and L. Balents, Phys. Rev. B \textbf{75}, 121306 (2007).
	%
\bibitem{Qi_1}
	X.-L. Qi \textit{et al.}, Phys. Rev. B \textbf{78}, 195424 (2008).
	%
\bibitem{hish_1}	
	D. Hsieh  \textit{et al.}, Phy. Rev. Lett. \textbf{103}, 146401 (2009).
	%	

\bibitem{culcer_prb_2010}
	D. Culcer, E. H. Hwang, T. D. Stanescu, and S. Das Sarma, Phys. Rev. B \textbf{82}, 155457 (2010).
	%
\bibitem{culcer_review}
	D. Culcer, Physica E \textbf{44}, 860  (2012).
	
\bibitem{rui} 
	R. Yu \textit{et. al.}, Science \textbf{329}, 61(2010).
	

\bibitem{scalQAHE} 
	X. Kou \textit{et. al.}, Phys. Rev. Lett. Science \textbf{113}, 137201(2014).	
	
	\bibitem{beyond2d} 
	Y. Xing \textit{et. al.}, New J. Phys. \textbf{20}, 043011(2018). 
	
	  \bibitem{signchangeex}
	N. Liu,\textit{et al.} Nat. Commun. \textbf{9}, 1282(2018).
	
		\bibitem{sensitivity} 
     Y. Ni \textit{et. al.}, IEEE Transactions on Magnetics \textbf{52}, 4002304(2016).
	
 
	\bibitem{firstcoordinate}	
N. A. Sinitsyn \textit{et al.} Phys. Rev. B	\textbf{73}, 075318(2006).
	%
		
	\bibitem{lutinger1}
	W. Kohn and J. M. Luttinger, Phys. Rev. \textbf{108}, 590 (1957).
	
\bibitem{lutinger2} 
	R. karplus and J. M. Lutinger, phys. Rev. \textbf{95}, 1154(1954). 
	%
\bibitem{lutinger3}              
	J. M. Luttinger, Phys. Rev. \textbf{112}, 739(1958).
	%
\bibitem{topical}
N. A. Sinitsyn, J. Phys.: Condens. Matter \textbf{20}, 023201 (2008) .

	\bibitem{Jairo_method}
	K. V'yborny, \textit{et al.}, Phys. Rev. B \textbf{79}, 045427 (2009).
	
	
	\bibitem{fourth}	
C. Xiao \textit{et a} Phys. Rev.B \textbf{95}, 035426 (2017).
		
	\bibitem{mac}
	N. A. Sinitsyn \textit{et al.}  Phys. Rev. B \textbf{75} 045315 (2007).	
	%	
		\bibitem{mahan_nutshell}
	G. D. Mahan, \textit{Condensed Matter in a Nutshell} (Princeton University Press, 2011).
	%
	
		\bibitem{size of gap}
	Y. L. Chen  \textit{et al.}, Science \textbf{329}, 659 (2010).
	
		\bibitem{amir_2015}
	A. Sabalipour \textit{et al.}  J. Phys.: Condens. Matter \textbf{27} 115301(2015).
	
		\bibitem{rosenberg_prb_2012}
	G. Rosenberg and M. Franz, Phys. Rev. B \textbf{85}, 195119 (2012).
	%
	
	
	
	
	
	
	
	
\end{thebibliography}
\end{document}